\documentclass[aps,reprint,preprintnumbers,showpacs,superscriptaddress]{revtex4-2}
\usepackage[english]{babel}
\usepackage{amsmath,amssymb,bbm,graphicx,color,comment,txfonts}
\usepackage[bookmarks=true,colorlinks,citecolor=blue,urlcolor=blue]{hyperref}
\usepackage{dsfont}
\usepackage{soul}

\usepackage{bbm}
\usepackage{float}

\usepackage{dcolumn}   
\usepackage{bm}        
\usepackage{filecontents}
\usepackage{lineno}

\usepackage{mathtools}
\usepackage[usenames,dvipsnames]{xcolor} 

\usepackage[export]{adjustbox}
\usepackage{adjustbox}

\usepackage{braket}
\usepackage{tikz}
\usepackage{bbold}
\def\ketbra#1{\ket{#1}\!\!\bra{#1}}
\def\Tr{\mathrm{Tr}}
\let\baraccent=\= \renewcommand{\=}[1]{\stackrel{#1}{=}}
\newcommand{\eq}[1]{Eq.\thinspace(\ref{#1})}
\newcommand{\eqs}[2]{Eqs.\thinspace(\ref{#1},\ref{#2})}

\newcommand{\fig}[1]{Fig.\thinspace{}\ref{#1}}
\newcommand{\fc}[1]{({#1})}
\newcommand{\figc}[2]{Fig.\thinspace{}\ref{#1}\thinspace{}\fc{#2}}

\usetikzlibrary{arrows,shapes,positioning}
\usetikzlibrary{decorations.markings}
\tikzstyle arrowstyle=[scale=1]
\tikzstyle directed=[postaction={decorate,decoration={markings,
    mark=at position 0.8 with {\arrow[arrowstyle]{stealth}}}}]
\tikzstyle reverse directed=[postaction={decorate,decoration={markings,
    mark=at position 0.2 with {\arrowreversed[arrowstyle]{stealth};}}}]

\tikzstyle directeds=[postaction={decorate,decoration={markings,
    mark=at position 0.5 with {\arrow[arrowstyle]{stealth}}}}]
\tikzstyle reverse directeds=[postaction={decorate,decoration={markings,
    mark=at position 0.5 with {\arrowreversed[arrowstyle]{stealth};}}}]

\newcommand{\plaqv}{
\tikz[baseline=-0.62ex]{
\draw [gray] (0,-0.19) -- (0.3,-0.19);
\draw [gray] (0,0.19) -- (0.3,0.19);
\filldraw [color=black, fill=black!65] (0,0) ellipse (0.06 and 0.23); 
\filldraw [color=black, fill=black!65] (0.3,0) ellipse (0.06 and 0.23); }
}

\newcommand{\plaqh}{
\tikz[baseline=-0.62ex]{
\draw [gray] (-0.19,-0.19) -- (-0.19,0.19);
\draw [gray] (0.2,-0.19) -- (0.2,0.19);
\filldraw [color=black, fill=black!65] (0,-0.15) ellipse (0.23 and 0.058); 
\filldraw [color=black, fill=black!65] (0,0.15) ellipse (0.23 and 0.058); }
}

\usepackage[normalem]{ulem}

\begin{document}
\preprint{MIT-CTP/6013}
\title{
Digital dissipative state preparation for frustration-free gapless quantum systems
}
\author{Johannes Feldmeier}
\thanks{These authors contributed equally.\\}
\affiliation{Physics Department, Harvard University, 17 Oxford St, Cambridge MA 02138, USA}

\author{Yu-Jie Liu}
\thanks{These authors contributed equally.\\}
\affiliation{Center for Theoretical Physics - a Leinweber Institute, Massachusetts Institute of Technology, Cambridge, MA 02139, USA}

\author{Mikhail D. Lukin}
\affiliation{Physics Department, Harvard University, 17 Oxford St, Cambridge MA 02138, USA}

\author{Soonwon Choi}
\affiliation{Center for Theoretical Physics - a Leinweber Institute, Massachusetts Institute of Technology, Cambridge, MA 02139, USA}

\date{\today}

\begin{abstract}
Preparing algebraically correlated ground states of quantum many-body systems is an important, yet challenging task for quantum simulation.
We introduce a protocol that employs local projective measurements and unitary feedback for frustration-free gapless systems. Our approach prepares \emph{a priori} unknown ground states in time that scales polynomially with system size.
We analytically show the performance of our protocol for the dynamics of a single particle; we argue the same mechanism generalizes to many-body systems based on the physics of quasiparticles. Our theory predicts that the transient cooling dynamics directly reveals the system's universal critical properties. 
In particular, the state preparation time is linear in the inverse of the finite-size gap (up to log correction) when the system's dynamical critical exponent is larger than or equal to the effective spatial dimension explored by the quasiparticles. 
We verify these predictions in numerical simulations of ferromagnetic Heisenberg models in one- and two dimensions, a Fredkin spin chain, and a two-dimensional model of resonating valence bond states.
Our protocol stabilizes gapless many-body ground states fully digitally without requiring analog rotations, enabling access to high-fidelity states beyond conventional adiabatic approaches in near-term experiments.
\end{abstract}

\maketitle
\begin{figure}[t]
\centering
\includegraphics[trim={0cm 0cm 0cm 0cm},clip,width=0.99\linewidth]{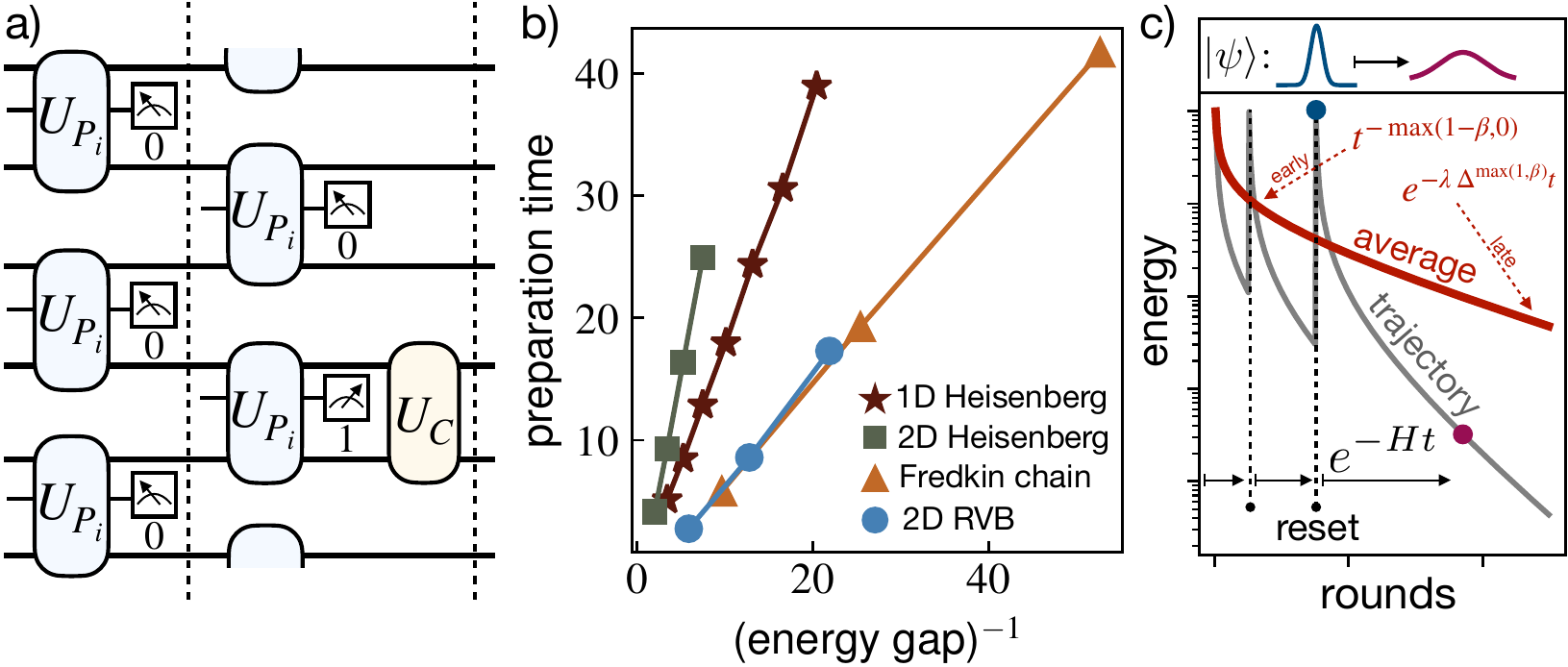}
\caption{\textbf{Protocol.} 
\textbf{a)} 
Our protocol consists of sequentially measuring the projectors of the Hamiltonian \eq{eq:frustration_free_H} and applying local corrections $U_C$ for outcomes $P_i=1$.
\textbf{b)} 
For a variety of \textit{gapless} many-body systems, the average time to prepare the ground state up to a fixed many-body infidelity, $\epsilon$, has inverse-linear scaling with the finite-size energy gap $\Delta\sim N^{-z/d}$, as shown by our numerical simulation with $\epsilon =0.2$.
\textbf{c)}
This performance scaling can be exactly shown for a solvable model of effective single-particle dynamics: A localized excitation undergoes imaginary time evolution with stochastic resets to the initial state (gray).
The solution reveals the universal scaling of the average dynamics of the energy (red).
}
\label{fig:1}
\end{figure}

\textbf{\textit{Introduction.}}--
Simulating correlated quantum systems on large-scale quantum computers requires high-fidelity preparation of entangled many-body  states. As increasingly accurate digital and early error-corrected quantum hardware continues to improve, the design of efficient  protocols becomes increasingly important.
A particular challenge is the fast, scalable, and hardware-efficient preparation of gapless quantum many-body ground states, which play a key role near quantum critical points~\cite{Sachdev_2011}.
Frustration-free systems constitute an important class of relevant problems, as they not only realize many gapped quantum phases of matter~\cite{schuch:2011}, but can also host paradigmatic examples of gapless systems~\cite{RK1988,verstraete:2006,Fradkin:2006,castelnovo:2008,Salberger:2016,dellAnna2016_fredkin,movassagh2016_supercrit,movassagh2018_fredkin,setqpt_tns}. As such, they can serve as valuable starting points for engineering and probing gapless quantum systems. The Hamiltonian for such systems can be represented as a sum over (non-commuting) local projectors $P_i$,
\begin{equation} \label{eq:frustration_free_H}
H = \sum_i P_i \,, \quad P_i^2 = P_i, \quad P_i\ket{\Omega} = 0 \;\forall \;i,
\end{equation}
where $\ket{\Omega}$ is the simultaneous ground state of all $P_i$. Here, we are interested in situations where only the Hamiltonian $H$ is given and $\ket{\Omega}$ is not necessarily known.
Existing methods to prepare the unknown state $\ket{\Omega}$ include adiabatic preparation~\cite{adiabaticQC2018,farhi2000_adiabatic} or Lindblad-based open system dynamics and other protocols of engineered dissipation
~\cite{verstraete2009_dissip,diehl2008_dissip,Kraus2008markov,ticozzi2012stabilize,google2024_dissip,lloyd2025_cooling,Gilyen2017_lovasz,Cubitt2023_dissip,polla2021_cooling,roy2020_steering,Matthies2024_cooling,molpeceres2025_cooling,Herasymenko2023steering}. These methods, however, often incur algorithmic overhead, for example due to implementing continuous evolution on digital devices; filtering-based methods additionally require large initial ground state overlap~\cite{gilyen2019_qsvt,Lin2020_nearoptimal,Thibodeau2023_frust}, rendering them limited for practical implementation.

In this work, we introduce a simple digital cooling protocol based on measurements of the local projectors $P_i$ and subsequent correction operations  (\figc{fig:1}{a}). While  this procedure is straightforward to analyze when the projectors commute~\cite{verstraete2009_dissip}, for realizing gapless spectra, the projectors must be non-commuting, giving rise to complicated, generally intractable dynamics. 
From numerical simulations, we find efficient ground state preparation in a time linear in the inverse of the finite-size gap for a variety of many-body systems, \figc{fig:1}{b}.
This performance can be understood by modeling the protocol as an effective cooling process of quasiparticles; they undergo a combination of imaginary time evolution and stochastic resets (\figc{fig:1}{c}). On average, this leads to depletion of quasiparticles at a rate $\Delta$, the energy gap of $H$.
We present exact solutions for the single-particle problem and argue its generalization to the many-body settings, leading to the state preparation time $T_c =O (\Delta^{-\max(1,\, \beta)} |\log (\epsilon \, \Delta^{\min(1,\beta)} / N)|)$, with target average global infidelity $\epsilon$, system size $N$, and the ratio $\beta$ of effective spatial dimension explored by the quasiparticles and dynamical critical exponent. Our effective model also predicts the full scaling form of the transient dynamics toward the ground state, which we verify through extensive numerical simulations.
Our protocol can be understood as digital dissipative dynamics, under which arbitrary initial states are cooled down to a pure dark state. In contrast to other dissipative dynamics in quantum optics such as coherent population trapping~\cite{alzetta1976experimental,Gray:78}, the present approach concerns \emph{a priori} unknown, highly-entangled target many-body state and displays exponential convergence over time.

\textbf{\textit{Protocol.}}--
We consider gapless instances of \eq{eq:frustration_free_H} with finite-size gap $\Delta = \Delta(N) \sim N^{-z/d}$ decreasing with system size $N$, where $d$ is the spatial dimension and $z$ the dynamical critical exponent. In local frustration-free systems, $z\geq2$ is generally expected~\cite{gosset2016_gap,ogunnaike2023_lindblad,masaoka2024_gap,masaoka2025oral}.
Our algorithm consists of repeated, sequential applications of measurement-feedback layers. 
To do this, we organize the non-commuting $P_i$ into groups $B_{a=1,\ldots,\mathcal{A}}$, such that projectors within a group have non-overlapping support and therefore commute.
Each group $B_a$ is associated with a single layer: 
all projectors $P_j\in B_a$ are measured, and, if the outcome is $P_i=1$, a unitary correction $U_C(i)$ is applied.
The feedback $U_C(i)$ is a finite-depth circuit chosen to lower the energy of the local $P_i$~\footnote{In our examples, the projectors $P_i$ are of rank 1, such that we can find local feedback that corrects $P_i=1$ to $P_i=0$. We note that general projectors $P_i$ can be written as a sum of rank 1 projectors, and the corresponding perfect feedback can be constructed accordingly. In general, we only require that $U_C(i)$ does not commute with $P_i$. We expect the protocol to still behave the same qualitatively, which we verified in the examples studied in this work.}.
We may restrict $U_C(i)$ to preserve the symmetries of $H$.
A single round of our protocol consists of $\mathcal{A}$ layers.
We denote the state after $t$ rounds by $\ket{\psi(t)}$. 
The success of our algorithm is heralded by the absence of $P_i=1$ measurements in sufficiently many rounds.

\textit{\textbf{Guaranteed convergence and empirical performance.}}-- 
This simple algorithm defines driven-dissipative dynamics, whose analysis is intractable in general.
Nevertheless, we present several rigorous results.
First, we bound its state-preparation time by considering the \emph{reset-free trajectory}, where every measurement is assumed to yield $P_i=0$. After a single round of $P_i=0$ outcomes, $\ket{\psi(t)}$ is updated according to
\begin{equation} \label{eq:projectionRound}
\ket{\psi(t+1)} =\frac{\mathcal{P} \ket{\psi(t)}}{\|\mathcal{P} \ket{\psi(t)}\|}, \;\; \mathcal{P} = \prod_{a=1}^{\mathcal{A}} \mathcal{P}_a, \;\; \mathcal{P}_a = \prod_{i \in B_a} (1 - P_i).
\end{equation}
We call $\mathcal{P}$ a \textit{projection-round operator}. Using the detectability lemma (DL)~\cite{aharonov2009_DL,anshu2016_DL}, every state $\ket{\psi_\perp}$ orthogonal to the ground state satisfies 
$1 - 4 \braket{\psi_\perp|H|\psi_\perp} \leq \| \mathcal{P} \ket{\psi_\perp} \|^2 \leq (1 + \Delta/\mathcal{A}^2)^{-1} $.
Then, for generic states $\ket{\psi(t)}$ that
typically have non-zero overlap with the low-lying excited states of $H$, the DL implies the minimal timescale $\sim \Delta^{-1} |\log\epsilon|$ required to  reach arbitrarily small infidelity $\epsilon$ in the reset-free trajectory.
Since the open system dynamics in our algorithm enter the reset-free trajectory at sufficiently late times, the state-preparation time cannot be parametrically faster than $\Delta^{-1}$
~\footnote{Via the DL, the spectral gap $\Delta$ of the frustration-free Hamiltonian is equivalent (up to constants) to the gap of $\mathcal{P}$; in particular, the slow modes of $\mathcal{P}$ are supported on low-lying excited states of $H$. We assume that the states $\ket{\psi(t)}$ along the cooling dynamics generically carry a nonzero weight in this low-energy sector that is not parametrically smaller in system size than their weight in the ground-state subspace. We expect that these weights, which come from gapless excitations, cannot be removed by any finite-depth quantum circuits. Therefore, most trajectories which eventually reach the ground state must, at late times, enter a reset-free trajectory. The infidelity under a reset-free trajectory cannot decay faster than $\sim e^{-c t \Delta}$; this implies a convergence time $t \gtrsim \Delta^{-1} |\log\epsilon|$ to reach small infidelity $\epsilon$ (if we only assume that the weight in the low-energy sector is non-zero, as done in the main text, that sets the minimal timescale to reach arbitrarily small infidelity). 
Under these assumptions, this also sets a minimal timescale for the full protocol to converge for any choices of $U_C(i)$.}.

The actual convergence time $T_c$ to reach the ground state can in principle greatly exceed the DL bound owing to the early time dynamics when $P_i=1$ outcomes are obtained frequently. In fact, the protocol might even fail to converge, never reaching the reset-free trajectory.
As our second result, we show that the protocol is guaranteed to converge eventually when using random Pauli corrections, with an upper bound on $T_c$ that grows exponentially in system size, see Supplemental Material (SM)~\cite{SM}.
As our protocol also applies to computationally hard Hamiltonians, such as those encoding SAT problems~\cite{cook:1971,Lichtenstein:1982}, a rigorous polynomial upper bound on $T_c$ is not expected without further assumptions. Instead, we aim to determine the actual scaling of $T_c$ for physical many-body models.

To this end, we numerically simulate the protocol for a variety of gapless many-body spin Hamiltonians, starting from product initial states and employing deterministic local unitary corrections. The different systems and their dynamics under the protocol are discussed in detail below. Crucially, in all examples, the time $T_c$ to obtain the ground state (at infidelity $\epsilon = 0.2$) is consistent with $T_c = \tilde{O}(1/\Delta)$ (\figc{fig:1}{b}).
Surprisingly, this scaling matches that of reset-free trajectories, even though such trajectories are exponentially rare and frequent resets may, naively, induce longer convergence times. 
To understand these empirical observations, in what follows, we present a solvable model of the dynamics and argue its generalization.

\begin{figure}[t]
\centering
\includegraphics[trim={0cm 0cm 0cm 0cm},clip,width=0.99\linewidth]{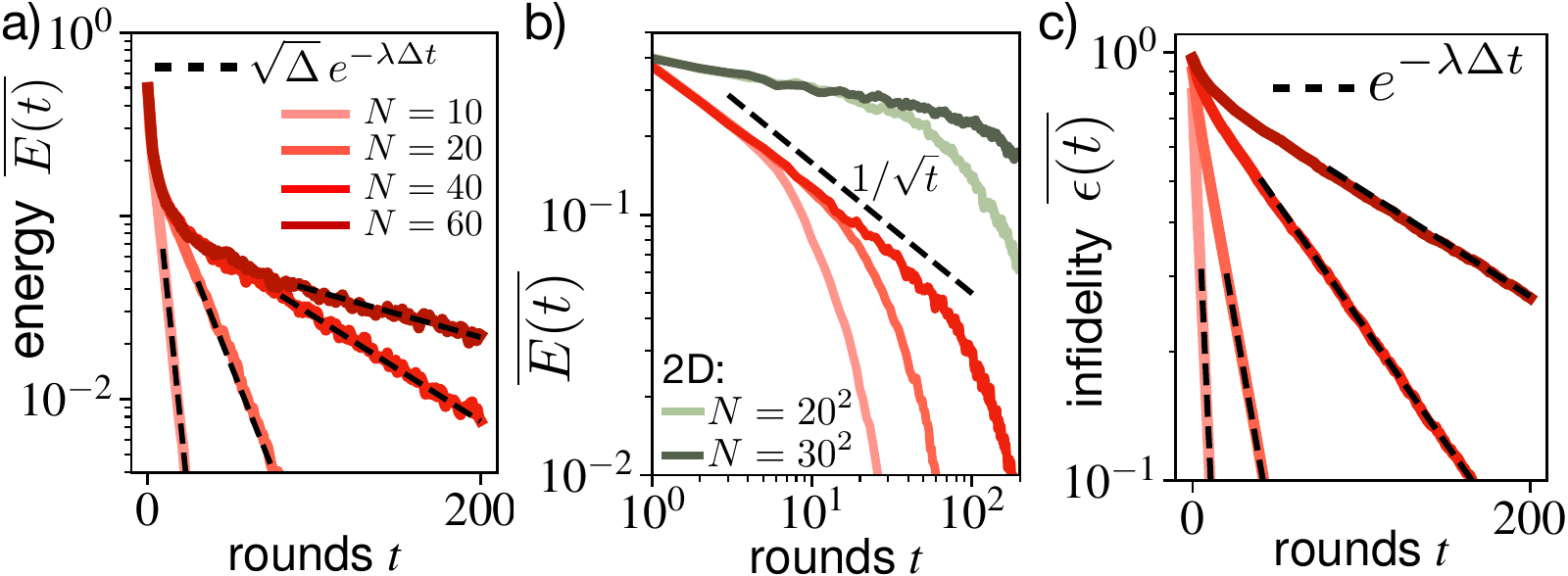}
\caption{
\textbf{Single-particle dynamics.} 
\textbf{a)} The trajectory-averaged energy $\overline{E}(t)$ in the single-particle Heisenberg chain. $\overline{E}(t)$ decays with rate linear in the energy gap $\Delta(N)$, following the prediction of \eq{eq:avg_energy} (with $\beta = 1/2$). $\lambda \approx 1.0$ is a system-size-independent constant. 
\textbf{b)} At early times, $\overline{E(t)} \sim 1/\sqrt{t}$ decays algebraically, in agreement with \eq{eq:avg_energy} for $\beta=1/2$. In contrast, dynamics in 2D, where $\beta=1$, shows no clear regime with algebraic decay (discernible from a predicted slower logarithmic decay at early times for $\beta=1$, see SM~\cite{SM}).
\textbf{c)} Average ground state infidelity $\overline{\epsilon(t)}$, decaying with the same rate $\lambda \Delta$ as the energy.
}
\label{fig:2}
\end{figure}

\textbf{\textit{Single-particle system.}}--
We solve the cooling dynamics within the single-particle sector of the nearest-neighbor ferromagnetic (FM) Heisenberg model in $d$ dimensions, $H_{\mathrm{Heis}} = -\sum_{\braket{i_1 i_2}} \bigl(\boldsymbol{S}_{i_1}\cdot \boldsymbol{S}_{i_2} - 1/4\bigr)$ with periodic boundaries. The key idea is to map the dynamics to an exactly solvable process of imaginary time evolution with stochastic resets. 
We may first write $H_{\mathrm{Heis}} = \sum_{j=1}^d H^{(j)}$, where $H^{(j)}$ takes the form of \eq{eq:frustration_free_H} via
\begin{equation} \label{eq:fmHeisenberg}
P_{i} =\frac{1}{2}\bigl(\ket{01}-\ket{10}\bigr)_{i,i+\hat{e}_j}\bigl(\bra{01}-\bra{10}\bigr),
\end{equation}
local singlet projectors on bonds $i,i+\hat{e}_j$ along the $\hat{e}_j$-direction.
We restrict to the subspace $Z_{\mathrm{total}} = \sum_i Z_i=N-2$, spanned by states $\ket{i}$ with a single spin at site $i$ in state $\ket{1}$ and all others in $\ket{0}$. The Hamiltonian is diagonalized in momentum space, $H_{\mathrm{Heis}}=\sum_k \varepsilon(k) \ket{k}\bra{k}$, with dispersion $\varepsilon(k) \sim k^2$ for small momenta $k\ll 1$, energy gap $\Delta \sim N^{-2/d}$, and ground state $\ket{\Omega}=\frac{1}{\sqrt{N}}\sum_i \ket{i}$.
Our protocol measures $P_{i}$ and applies $U_C(i) = Z_i$ if $P_{i}=1$, correcting a singlet into a triplet.
One round consists of $\mathcal{A}=2d$ layers, corresponding to even and odd bonds along each direction, and we initialize the particle in a triplet state $\ket{\psi(0)}=\frac{1}{\sqrt{2}}(\ket{i} + \ket{i+\hat{e}_1})$.
Key to understanding this process is realizing that every round corresponds to one of two possible events:

\textit{1) Projection (reset-free) evolution:}
If all $P_{i}=0$ in a round, $\ket{\psi(t)}$ evolves according to \eq{eq:projectionRound}, which we may intuitively think of as imaginary time evolution.
Indeed, in the SM~\cite{SM} we show analytically that within the single-particle sector, 
\begin{equation} \label{eq:imagT}
\mathcal{P}^\tau \ket{k} \xrightarrow{\tau \gg 1, \; k\ll 1} \exp\bigl\{ -2\varepsilon(k) \tau \bigr\} \ket{k}.
\end{equation}
Inserting a general dispersion $\varepsilon(k) \sim k^z$, the energy $e(\tau) \equiv \braket{\psi(0)|(\mathcal{P}^\dagger)^\tau H_{\mathrm{Heis}}\mathcal{P}^\tau|\psi(0)} \, / \, ||\mathcal{P}^\tau \ket{\psi(0)} ||^2$ upon applying $\mathcal{P}^\tau$ to $\ket{\psi(0)}$ decreases as (see SM~\cite{SM})
\begin{equation} \label{eq:sp_energy}
e(\tau \Delta \ll 1) = \frac{1}{2}\,\frac{\beta}{\tau}, \quad e(\tau \Delta \gg 1) \sim \beta \Delta \exp\bigl\{-4 \Delta \tau \bigr\}.
\end{equation}
As shown in \figc{fig:1}{c}, the wavefunction delocalizes and $e(\tau)$ initially decays algebraically. When the extent of the wavefunction reaches system size, the decay becomes exponential due to overlap with the lowest excited state.
Moreover, we expect $\beta \sim d/z$, as each dimension contributes equal kinetic energy, smaller for softer modes with large $z$. We show in the SM~\cite{SM} that in fact $\beta = d/z$ must hold \textit{exactly}.

\textit{2) Resets:}
If $P_{i}=1$ for some $i$ during the round, the position of the particle collapses, and, after applying $U_C(i)=Z_i$, we have
$\ket{\psi(t+1)} = \ket{\psi(0)} \;\; \text{(up to spatial translation)}$, 
and the process effectively resets.
The probability for such a reset during the full round is 
\begin{equation} \label{eq:probEnergy}
p(t) \leq 2 E(t) = 2e(\tau(t)),
\end{equation}
where the instantaneous $E(t)$ is determined by the number of rounds since the previous reset, denoted by $\tau(t)$, such that $E(t)=e(\tau(t))$.
Asymptotically, as $E(t)\ll 1$, the inequality in \eq{eq:probEnergy} saturates.
\figc{fig:1}{c} shows a schematic of the single-particle energy dynamics. 
Notably, the last-reset time is crucial in determining the overall convergence timescale.

We provide an analytical solution to the Markov process (\eqs{eq:sp_energy}{eq:probEnergy}) in the SM~\cite{SM} and obtain for the trajectory-averaged energy $\overline{E(t)}$,
\begin{equation} \label{eq:avg_energy}
\begin{split}
\overline{E(t)} \sim
\begin{cases}
1/ t^{\max(1-\beta,0)}, \;\; \Delta t\ll 1 \\
\Delta^{\max(1-\beta,0)}\, \exp\{-\lambda \,\Delta^{\max(1,\beta)}t\}, \;\; \Delta t \gg 1
\end{cases} ,  \;\; (\beta \neq 1).
\end{split}
\end{equation}
Here, $\lambda$ is a \textit{system-size independent} constant related to the quality of the reset step (see SM~\cite{SM}). 
The dependence of \eq{eq:avg_energy} on the parameter $\beta=d/z$ reveals a characteristic competition between geometry and dynamical critical exponent that is independent of microscopic details.
To see this, we note that the time scale associated with the gap is $\Delta^{-1} \sim N^{z/d}$, while the expected average number of resets before converging is given by the initial ground state overlap via $|\braket{\Omega|\psi(0)}|^{-2}\sim N$. Therefore, when $\beta > 1$, the dynamics in \eq{eq:avg_energy} is dominated by the large number of resets and converges with rate $\Delta^\beta \sim N^{-1}$; when $\beta<1$, the convergence time reflects the effective imaginary time evolution and exhibits a convergence rate linear in $\Delta$. The ratio $\beta$ is thus a universal parameter characterizing the speed of the cooling.
We note that the case $\beta = 1$ is marginal and incurs a log-correction compared to \eq{eq:avg_energy}, such that $\overline{E(t)}|_{\beta=1} \sim |\log \Delta|^{-1} \, \exp\{-\lambda \Delta t / |\log \Delta|\}$.

With a unique ground state at zero energy, the infidelity $\epsilon(t)$ is bounded by $\epsilon(t) \leq E(t)/\Delta$, and hence $\overline{\epsilon(t)} =O( \Delta^{-\min(1,\beta)} e^{-\lambda \,\Delta^{\max(1,\beta)}t})$.
Thus, when $\beta\leq 1$, the convergence time $T_c = O( \Delta^{-\max(1,\beta)} |\log( \overline{\epsilon} \,\Delta^{\min(1,\beta)})|)$ using the local recovery nearly yields the best achievable scaling (over different recovery strategies) suggested by the reset-free trajectories. 
We numerically verify the decay of energy and infidelity for the single-particle model (where $z=2$) in \fig{fig:2}. In particular, \figc{fig:2}{b} confirms the algebraic decay $\overline{E(t)}\sim 1/\sqrt{t}$ at early times for $d=1$ and the absence of a clear algebraic regime for $d=2$ as predicted by \eq{eq:avg_energy}.

\begin{figure}[t]
\centering
\includegraphics[trim={0cm 0cm 0cm 0cm},clip,width=0.99\linewidth]{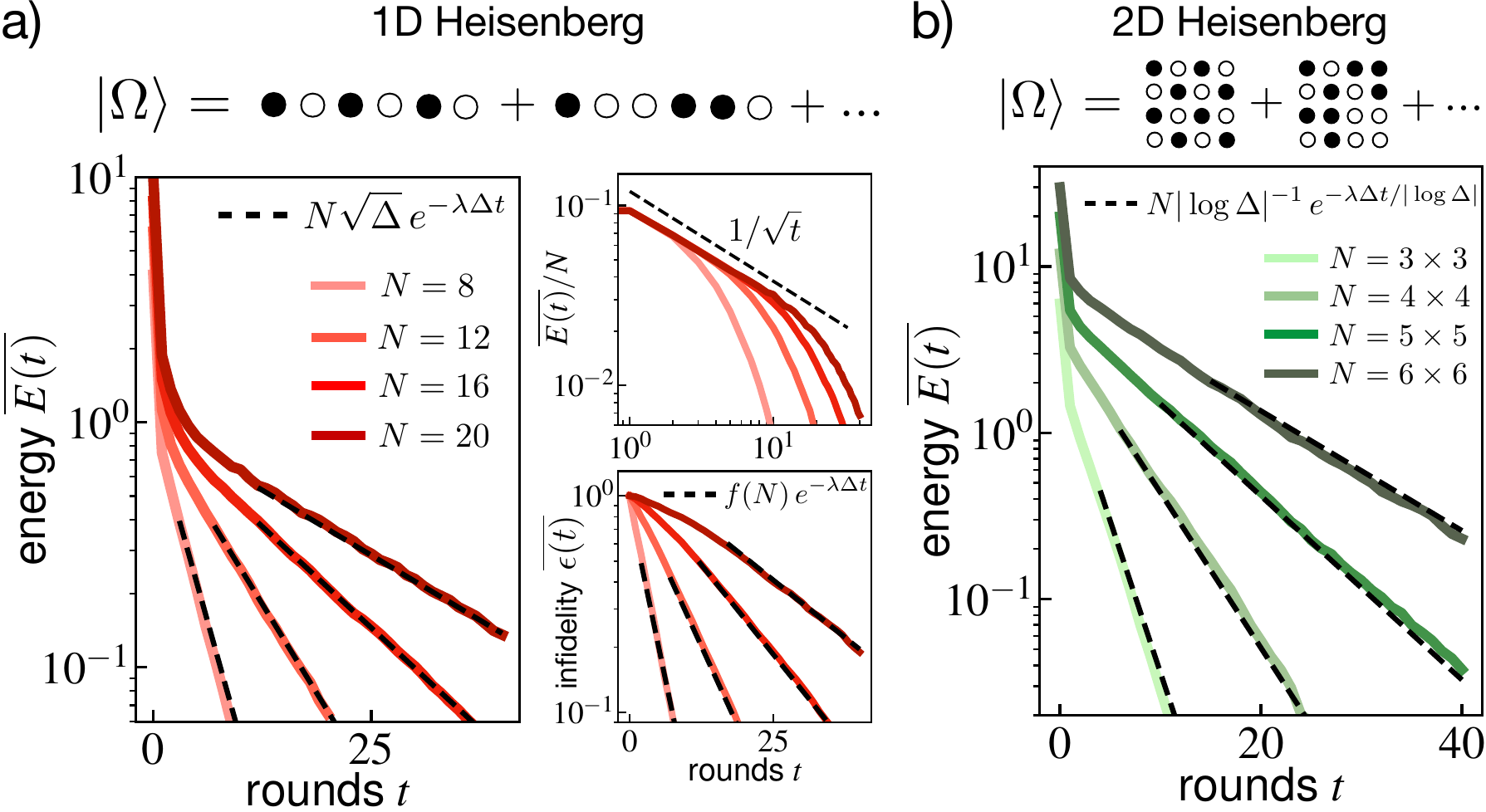}
\caption{\textbf{Many-body dynamics.} 
\textbf{a)} \textit{Left:} Average energy in the Heisenberg chain starting from a Néel product state. $\overline{E(t)}$ decays with rate linear in $\Delta(N)$, following our expectation for $\beta = 1/2$ (black dashed lines with $\lambda \approx 1.0$). \textit{Upper right:} At early times, $\overline{E(t)} / N \sim 1/\sqrt{t}$ decays algebraically.
\textit{Lower right:} Dynamics of the average infidelity. The asymptotic decay $\overline{\epsilon(t)} \sim f(N) \, e^{-\lambda\Delta t}$, with $f(N) \lesssim N / \sqrt{\Delta}$, implies an efficient preparation time. Black dashed lines show $f(N)\sim \sqrt{N}$.
\textbf{b)} Average energy in the 2D Heisenberg model starting from a Néel state, consistent with a log-correction to the decay rate as expected for $\beta = 1$ (black dashed lines with $\lambda \approx 3.1$). There is no discernible early time algebraic decay.
}
\label{fig:3}
\end{figure}

\textbf{\textit{Many-body systems.}}--
To generalize the single-particle scaling \eq{eq:avg_energy} to the many-body setting, we note that a measurement $P_i=1$ with subsequent correction results in a local energy bump around position $i$. 
Heuristically, we expect this local excitation to overlap with the system's quasiparticles with dispersion $\varepsilon(k) \sim k^z$, and thus behave akin to a single particle of the previous section.
For initial states with extensive energy $E(0) \sim N$, such as product states, a simple mean-field picture suggests that the protocol performs $\mathcal{O}(N)$ single-particle cooling processes in parallel. Under this phenomenological assumption, we expect an energy $\overline{E(t)} \sim N \, \overline{E_{\mathrm{sp}}(t)}$, where $\overline{E_{\mathrm{sp}}(t)}$ is the single-particle result of \eq{eq:avg_energy}, which provides a bound $\overline{\epsilon(t)} \leq \frac{N}{\Delta} \; \overline{E_{\mathrm{sp}}(t)}$ on the global many-body infidelity. As we will see below, the dimension $d$ in $\beta = d/z$ corresponds to an effective spatial dimension explored by the quasiparticles. The resulting preparation time $T_c =O( \Delta^{-\max(1,\beta)} |\log (\overline{\epsilon} \,\Delta^{\min(1,\beta)}/N) |)$ for $\beta\neq 1$ is consistent with our numerical results of \figc{fig:1}{b}. Beyond the convergence time, the theory makes precise predictions on the scaling form of the transient cooling dynamics, which we verify across those physical models in the following.

We first consider the $Z_{\mathrm{total}}=0$ sector of the \textit{1D Heisenberg chain} of \eq{eq:fmHeisenberg} with periodic boundaries (and $z=2$~\cite{auerbach2012_interacting}), where $\ket{\Omega}$ is the half-filled Dicke state, see \figc{fig:3}{a}. The protocol starts from a Néel state with $|\braket{\Omega|\psi(0)}|^2 = O(2^{-N})$ and performs layered singlet measurements and local $Z$-corrections. 
\figc{fig:3}{a} demonstrates numerically that the energy decays asymptotically as $\overline{E(t)} \sim N\sqrt{\Delta} \exp(-\lambda\Delta t)$, in excellent agreement with the mean field picture and the single-particle solution \eq{eq:avg_energy} for $\beta = d/z = 1/2$.
We emphasize that this scaling form for $\overline{E(t)}$ captures the full system size dependence of the decay rate $\sim\Delta(N)$ and the prefactor $\sim N\sqrt{\Delta(N)}$.
\figc{fig:3}{a} further confirms that $\overline{E(t)}/N \sim 1/\sqrt{t}$ decays algebraically at early times, and that the average global infidelity follows $\overline{\epsilon(t)}\sim f(N) \exp(-\lambda\Delta t)$. For accessible $N$, we find the prefactor $f(N)$ consistent with $f(N)\sim \sqrt{N}$, well below the loose infidelity upper bound 
$f(N) = O(N/\sqrt{\Delta})$ obtained from the energy dynamics.

Next, we utilize matrix-product states (MPS) to simulate the dynamics of the \textit{2D Heisenberg model} ($z=2$) up to system sizes $N=6 \times 6$ with open boundary conditions. \figc{fig:3}{b} demonstrates that $\overline{E(t)}\sim N |\log \Delta|^{-1} \, e^{-\lambda \Delta t /|\log \Delta|}$ is consistent with our parallel single-particle cooling picture for $\beta = d/z = 1$. Notably, in contrast to the 1D chain, $\overline{E(t)}$ shows no early time algebraic decay.

\begin{figure}[t]
\centering
\includegraphics[trim={0cm 0cm 0cm 0cm},clip,width=0.99\linewidth]{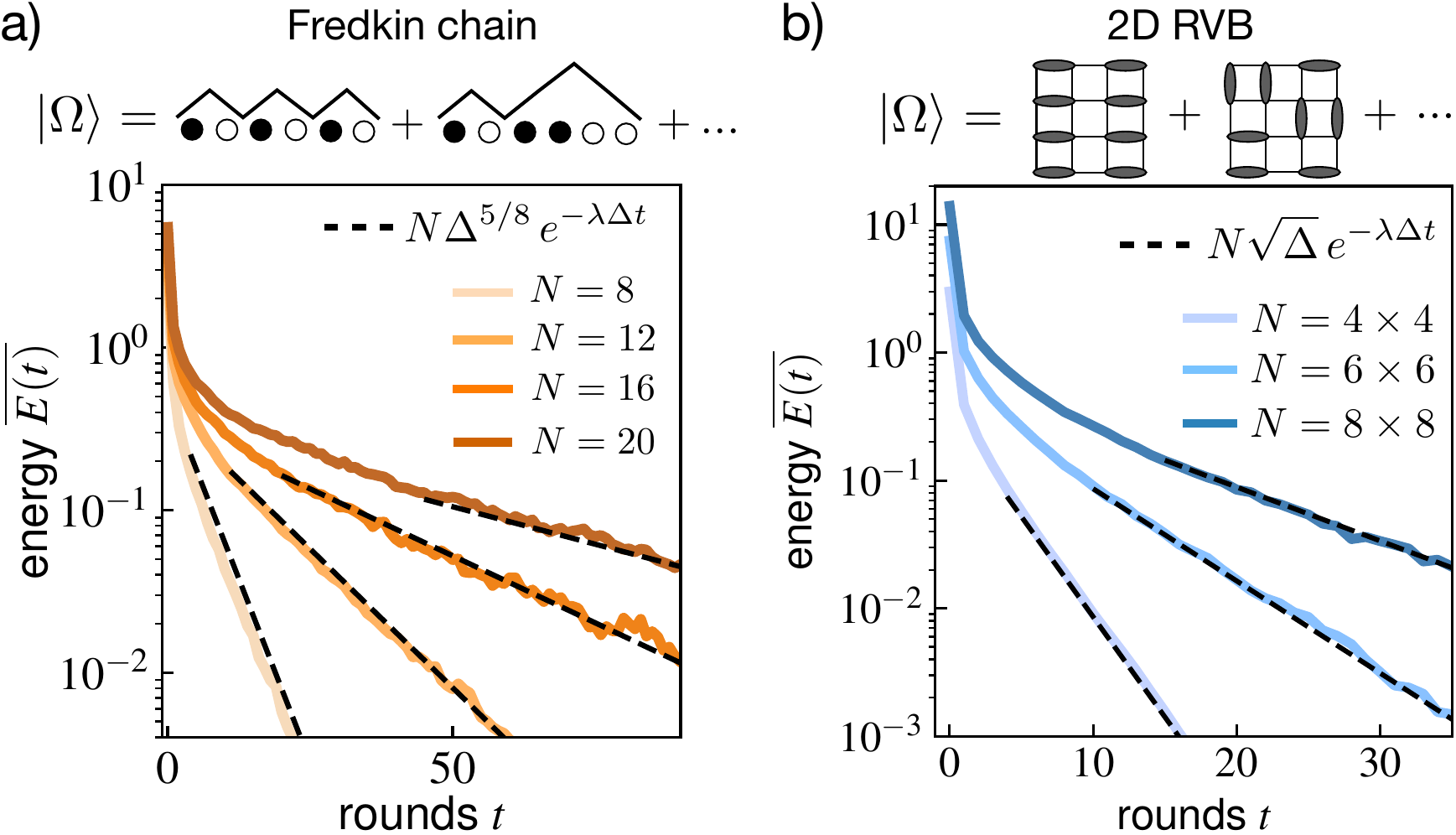}
\caption{\textbf{Algebraically correlated systems.} 
\textbf{a)} Average energy starting from a Néel state in the Fredkin spin chain with $z\approx 8/3$.
The dynamics is consistent with the scaling form expected for $\beta = d/z = 3/8$ (black dashed lines with $\lambda \approx 1.9$).
\textbf{b)} Average energy starting from a columnar product state of a 2D square lattice quantum dimer model (first state depicted in the RVB superposition). $\overline{E(t)}$ decays with rate linear in the gap $\Delta(N)$ and follows our prediction for $\beta = 1/2$ (black dashed lines with $\lambda \approx 2.1$).
Dynamics was simulated with MPS at bond dimension $\chi=512$, averaging over at least $10^3$ trajectories for each system size; gaps $\Delta(N)$ were obtained using exact diagonalization. We further found a decay of the infidelity consistent with $\overline{\epsilon(t)} \sim \sqrt{N} e^{-\lambda\Delta t}$ (see SM).
}
\label{fig:4}
\end{figure}
\textbf{\textit{Algebraically correlated ground states.}}-- 
While the above examples exhibit dynamical exponent $z=2$, general frustration-free systems can have $z>2$, such as the \textit{Fredkin spin chain}~\cite{Salberger:2016} $H_{\mathrm{Fred}} = \sum_{i=1}^{N-1} P_i$, where
\begin{equation}
P_i = \frac{1}{2}\bigl( \ket{01}-\ket{10} \bigr)_{i,i+1} \bigl( \bra{01}-\bra{10} \bigr) \otimes \bigl( \mathbb{1} - \ket{10}\bra{10}_{i-1,i+2}  \bigr).
\end{equation}
$P_i$ projects onto singlets at sites $i,i+1$, conditioned on sites $i-1,i+2$ being orthogonal to the state $\ket{10}$~\cite{dellAnna2016_fredkin}. Within the Krylov sector connected to a Néel initial state, the ground state of $H_{\mathrm{Fred}}$ is an equal superposition of so-called \textit{Dyck-paths}, characterized by a non-negative cumulative charge $\sum_{i=0}^{x} Z_i \geq 0$ for all $x$, see \figc{fig:4}{a}. 
The ground state exhibits algebraic (boundary-bulk) correlations, logarithmic entanglement entropy and a gap $\Delta(N) \sim N^{-z}$ with $z \approx 8/3$~\cite{Chen2017_fredkin,DellAnna:2019}. Our protocol measures $P_i$ and applies $U_C = Z_i$ if $P_i=1$, such that the corrected state exhibits $P_i=0$.
\figc{fig:4}{a} shows the dynamics of the average energy, consistent with $\overline{E(t)} \sim N \Delta^{5/8} \exp(-\lambda \Delta t)$, and thus $\beta = d/z = 3/8$ even for $z \neq 2$. 
We further found early time algebraic decay consistent with $1/t^{5/8}$ (see SM~\cite{SM}).

More generally, the spatial dimensions explored by the quasiparticles in constrained systems may differ from the spatial dimension $d$ of the lattice, giving rise to an \textit{effective} $\beta_{\mathrm{eff}}$. A trivial example is a $d=2$ system of decoupled 1D chains, with $\beta_{\mathrm{eff}}=1/z$ despite $d=2$. 
Intriguingly, we find such behavior in the \textit{2D quantum dimer model} on the square lattice, described by the canonical Rokhsar-Kivelson (RK) Hamiltonian $H_{\mathrm{QDM}} = \sum_i P_i$ with
\begin{equation} \label{eq:QDM_hamiltonian}
P_i = \frac{1}{2}\Bigl(\Ket{\plaqh} - \Ket{\plaqv}\Bigr)_i \Bigl( \Bra{\plaqh} - \Bra{\plaqv} \Bigr),
\end{equation}
where $i$ runs over the plaquettes of the lattice.
The ground state $\ket{\Omega}$ is a resonating valence bond (RVB) state with excitation gap $\Delta(N) \sim N^{-z/2}$ with $z=2$~\cite{RK1988} and equal time correlations described by 2D conformal field theory~\cite{Fradkin:2006}.
We simulate systems up to $N=8 \times 8$ sites with open boundary conditions using MPS, starting from an initial columnar product state depicted in \fig{fig:4}(b). For measurement outcomes $P_i=1$ we apply a local $Z$-correction on a spin of the corresponding plaquette, which leads to $P_i=0$ in the corrected state.
Our results in \fig{fig:4}(b) are in excellent agreement with $\overline{E(t)} \sim N \sqrt{\Delta} \exp(-\lambda\Delta t)$ and therefore $\beta_{\mathrm{eff}} = 1/2$. 
Remarkably, this suggests that excitations of $H_{\mathrm{QDM}}$ may exhibit effectively one-dimensional features. 
Exploring the origin of this phenomenon, either theoretically or experimentally, is an exciting prospect for future work.

\textbf{\textit{Near-term implementations.}}-- 
Our protocol avoids continuous quantum dynamics, an appealing feature for near-term digital implementations. 
As a concrete example, the number of rounds to achieve global infidelity $\epsilon=0.2$ for an $N=6\times 6$ RVB system (60 qubits) is $T_c \approx 9$.
Each round consists of $25$ four-body projectors $P_i$, each of which requires $36$ CNOTs to implement (using two additional ancillas), yielding an approximate CNOT count of $N^{\mathrm{(cool)}}_{\mathrm{CNOT}}\approx 8.1\times 10^3$ (plus 225 measurements).
For comparison, we have simulated adiabatic preparation from a columnar product state to the RVB state at $\epsilon=0.2$ and find a required ramp time $T_{\mathrm{ramp}}\approx 120$ (in units of $P_i$) for an optimized, smooth ramp profile~\footnote{The adiabatic ramp is $H(t) = (1-r(t))\, H_{\mathrm{col}} + r(t)\, H_{\mathrm{RVB}}$, where $H_{\mathrm{col}}$ has a columnar state as the exact ground state. For a linear ramp, $r(t) = t/ T_{\mathrm{ramp}}$ with $0\leq t \leq T_{\mathrm{ramp}}$. We have tested a smooth profile with $\partial_t r(t) = \frac{1}{\mathcal{N}}e^{-T_{\mathrm{ramp}}\bigl(1/t + 1/(T_{\mathrm{ramp}}-t)\bigr)}$, where $\mathcal{N}$ is such that $r(T_{\mathrm{ramp}})=1$.}. Digital implementation of this ramp requires Trotterization of the evolution under $H_{\mathrm{RVB}}$. Assuming a coarse Trotter step of $0.2$, the resulting CNOT estimate is $N^{\mathrm{(adiab)}}_{\mathrm{CNOT}}\approx 5.4 \times 10^5$, more than an order of magnitude above our protocol's.
The specific adiabatic path may be optimized further; however, the required time for adiabatic preparation is lower bounded by $T_{\mathrm{ramp}} = \Omega( \Delta^{-1})$~\cite{boixo2010_adiab} and $n$-th order Trotterization requires circuit depth $\Omega\bigl((T_{\mathrm{ramp}})^{1+1/n}\bigr)$ to ensure faithful digital implementation~\cite{berry2007_trotter}. As a result, for $\beta<1$, the observed circuit depth $O(\Delta^{-1})$ of our protocol scales better than adiabatic preparation based on a finite-order Trotter scheme.
The absence of continuous dynamics in our protocol is also appealing for error-corrected devices~\cite{google2023_surface,Bluvstein2024_logic,quantinuum2024_qec,google2025_qec}, where synthesis of analog gates carries additional overhead~\cite{dawson2005_solvay,bluvstein2025_architect}. 

Under realistic noise present on current devices, our protocol generally stabilizes non-equilibrium steady states with a finite energy. However, it can still enable a high-quality realization of 2D gapless quantum states.
For instance, we find numerically that both $N=4 \times 4$ RVB states and $N=4\times 4$ half-filled Dicke states have $N^{\mathrm{(cool)}}_{\mathrm{CNOT}}\approx 10^3$ (to reach $\epsilon=0.2$ in the absence of noise), close to the capabilities of state-of-the-art devices.

Furthermore, access to the history of measurement outcomes $P_i$ can be used for postselecting good trajectories with fewer undesired outcomes:
We have simulated our protocol for an $N=4\times 4$ Heisenberg model with a local dephasing channel of strength $p=0.02$ after each layer of operations and postselected trajectories with few undesired measurement outcomes.
We find that a $30\%$ postselection rate improves the achievable average global state infidelity from $\epsilon\approx 0.8$ to $\epsilon_{\mathrm{ps}}\approx 0.55$. 
Finally, instead of restarting the protocol whenever a trajectory is rejected, one can adopt a heralded strategy and continue running until the acceptance criterion is satisfied. In this approach, the quantum state remains stabilized within a relatively low-energy subspace that is much closer to the target ground state than the initial product state, even if some projectors are temporarily violated. From this subspace, the acceptance condition is reached more rapidly than under naive restarting.
The strategy is particularly well suited for next-generation digital quantum platforms, such as continuously operating Rydberg neutral-atom arrays~\cite{manetsch2025tweezer,li2025fast,chiu2025continuous}.

\textbf{\textit{Conclusions.}}-- 
We introduced a cooling protocol for gapless frustration-free systems that efficiently prepares ground states from initial product states.
Further quantitative speed-ups are conceivable either by optimizing the local feedback or introducing non-local controls. However, we find empirically that the performance of the purely local feedback scheme already matches that of reset-free trajectories (which correspond to effective imaginary time evolution) up to log-corrections, for a physically relevant class of systems.
Importantly, no prior knowledge of the ground state is assumed, which, if available, may be used to speed up state preparation~\cite{Piroli2024_Dicke,Yu2024_Dicke}.
It would be interesting to investigate how measurement history and nonlocal feedback can be further leveraged to mitigate experimental errors, such as measurement imperfections, and to accelerate convergence in systems with topological excitations, including gapped systems with a finite correlation length such as non-bipartite RVB models.~\cite{moessner2001_rvb,misguich2002_rvb}.
Our protocol could also be tested in higher-spin Fredkin and Motzkin chains with superlogarithmic ground-state entanglement entropy~\cite{dellAnna2016_fredkin,movassagh2016_supercrit,movassagh2018_fredkin,salberger2018_fredkin}, and in other tensor-network solvable critical states in two and higher dimensions~\cite{verstraete:2006,zhu:2019,zhu:2023,xu:2020, setqpt_tns,set_isotns,Boesl2025,Boesl:2025_stringnetiso}.
We further note that our example systems all featured a $U(1)$ symmetry. In the SM~\cite{SM}, we examine a critical cluster-Ising chain as a gapless uncle Hamiltonian~\cite{uncle_h} of a GHZ ground state. We show that our protocol exhibits dynamics consistent with $\beta = 1/2$ despite the absence of a $U(1)$ symmetry.
Our protocol may potentially be extended by constructing increasingly non-local Hamiltonians with the same ground state but an amplified gap~\cite{somma2013_gapAmp,low2017_gapAmp}.
Finally, while we relied on a purely phenomenological approach to many-body systems, it would be interesting to rigorously construct conditions under which the universal scaling of \eq{eq:avg_energy} holds.
Testing these scaling forms in practice for large systems is a challenging nonequilibrium problem which could be the subject of early useful quantum simulation.

\textbf{\textit{Acknowledgements.}}-- We thank Mohamed Abobeih, Ignacio Cirac, Jinen Guo, Simon Hollerith, Barbara Kraus, Clemens Kuhlenkamp, Frank Pollmann, Luke Stewart, and Elias Trapp for insightful discussions. We are particularly grateful to Simon Hollerith for valuable feedback on the manuscript. 
MPS simulations were performed using the TeNPy library~\cite{hauschild2018_tenpy}. 
We acknowledge financial support by the NSF CAREER Award DMR-2237244, the Alfred P. Sloan Fellowship,
the US Department of Energy (DOE Quantum Systems Accelerator Center, contract number 7568717), the Center for Ultracold Atoms (an NSF Physics Frontier Center, grant number PHY-2317134), and the National Science Foundation (grant numbers PHY-2012023, CCF-2313084, and QLCI grant OMA-2120757).
JF acknowledges support by the Harvard Quantum Initiative.
The authors acknowledge the MIT Office of Research Computing and Data for providing high-performance computing resources that have contributed to the research results reported within this paper.

\textit{Note added:--} Upon completion of the manuscript, we became aware of a recent independent work~\cite{dorstel2025frustration}. Our work considers several additional model systems beyond 1D and provides an extended theoretical analysis including exact solutions.

\let\oldaddcontentsline\addcontentsline
\renewcommand{\addcontentsline}[3]{}
\bibliography{references}

\begin{thebibliography}{78}%
\makeatletter
\providecommand \@ifxundefined [1]{%
 \@ifx{#1\undefined}
}%
\providecommand \@ifnum [1]{%
 \ifnum #1\expandafter \@firstoftwo
 \else \expandafter \@secondoftwo
 \fi
}%
\providecommand \@ifx [1]{%
 \ifx #1\expandafter \@firstoftwo
 \else \expandafter \@secondoftwo
 \fi
}%
\providecommand \natexlab [1]{#1}%
\providecommand \enquote  [1]{``#1''}%
\providecommand \bibnamefont  [1]{#1}%
\providecommand \bibfnamefont [1]{#1}%
\providecommand \citenamefont [1]{#1}%
\providecommand \href@noop [0]{\@secondoftwo}%
\providecommand \href [0]{\begingroup \@sanitize@url \@href}%
\providecommand \@href[1]{\@@startlink{#1}\@@href}%
\providecommand \@@href[1]{\endgroup#1\@@endlink}%
\providecommand \@sanitize@url [0]{\catcode `\\12\catcode `\$12\catcode
  `\&12\catcode `\#12\catcode `\^12\catcode `\_12\catcode `\%12\relax}%
\providecommand \@@startlink[1]{}%
\providecommand \@@endlink[0]{}%
\providecommand \url  [0]{\begingroup\@sanitize@url \@url }%
\providecommand \@url [1]{\endgroup\@href {#1}{\urlprefix }}%
\providecommand \urlprefix  [0]{URL }%
\providecommand \Eprint [0]{\href }%
\providecommand \doibase [0]{https://doi.org/}%
\providecommand \selectlanguage [0]{\@gobble}%
\providecommand \bibinfo  [0]{\@secondoftwo}%
\providecommand \bibfield  [0]{\@secondoftwo}%
\providecommand \translation [1]{[#1]}%
\providecommand \BibitemOpen [0]{}%
\providecommand \bibitemStop [0]{}%
\providecommand \bibitemNoStop [0]{.\EOS\space}%
\providecommand \EOS [0]{\spacefactor3000\relax}%
\providecommand \BibitemShut  [1]{\csname bibitem#1\endcsname}%
\let\auto@bib@innerbib\@empty
\bibitem [{\citenamefont {Sachdev}(2011)}]{Sachdev_2011}%
  \BibitemOpen
  \bibfield  {author} {\bibinfo {author} {\bibfnamefont {S.}~\bibnamefont
  {Sachdev}},\ }\href@noop {} {\emph {\bibinfo {title} {{Quantum Phase
  Transitions}}}},\ \bibinfo {edition} {2nd}\ ed.\ (\bibinfo  {publisher}
  {Cambridge University Press},\ \bibinfo {year} {2011})\BibitemShut {NoStop}%
\bibitem [{\citenamefont {Schuch}\ \emph {et~al.}(2011)\citenamefont {Schuch},
  \citenamefont {P\'erez-Garc\'{\i}a},\ and\ \citenamefont
  {Cirac}}]{schuch:2011}%
  \BibitemOpen
  \bibfield  {author} {\bibinfo {author} {\bibfnamefont {N.}~\bibnamefont
  {Schuch}}, \bibinfo {author} {\bibfnamefont {D.}~\bibnamefont
  {P\'erez-Garc\'{\i}a}},\ and\ \bibinfo {author} {\bibfnamefont
  {I.}~\bibnamefont {Cirac}},\ }\bibfield  {title} {\bibinfo {title}
  {Classifying quantum phases using matrix product states and projected
  entangled pair states},\ }\href {https://doi.org/10.1103/PhysRevB.84.165139}
  {\bibfield  {journal} {\bibinfo  {journal} {Phys. Rev. B}\ }\textbf {\bibinfo
  {volume} {84}},\ \bibinfo {pages} {165139} (\bibinfo {year}
  {2011})}\BibitemShut {NoStop}%
\bibitem [{\citenamefont {Rokhsar}\ and\ \citenamefont
  {Kivelson}(1988)}]{RK1988}%
  \BibitemOpen
  \bibfield  {author} {\bibinfo {author} {\bibfnamefont {D.~S.}\ \bibnamefont
  {Rokhsar}}\ and\ \bibinfo {author} {\bibfnamefont {S.~A.}\ \bibnamefont
  {Kivelson}},\ }\bibfield  {title} {\bibinfo {title} {{Superconductivity and
  the Quantum Hard-Core Dimer Gas}},\ }\href
  {https://doi.org/10.1103/PhysRevLett.61.2376} {\bibfield  {journal} {\bibinfo
   {journal} {Phys. Rev. Lett.}\ }\textbf {\bibinfo {volume} {61}},\ \bibinfo
  {pages} {2376} (\bibinfo {year} {1988})}\BibitemShut {NoStop}%
\bibitem [{\citenamefont {Verstraete}\ \emph {et~al.}(2006)\citenamefont
  {Verstraete}, \citenamefont {Wolf}, \citenamefont {Perez-Garcia},\ and\
  \citenamefont {Cirac}}]{verstraete:2006}%
  \BibitemOpen
  \bibfield  {author} {\bibinfo {author} {\bibfnamefont {F.}~\bibnamefont
  {Verstraete}}, \bibinfo {author} {\bibfnamefont {M.~M.}\ \bibnamefont
  {Wolf}}, \bibinfo {author} {\bibfnamefont {D.}~\bibnamefont {Perez-Garcia}},\
  and\ \bibinfo {author} {\bibfnamefont {J.~I.}\ \bibnamefont {Cirac}},\
  }\bibfield  {title} {\bibinfo {title} {Criticality, the area law, and the
  computational power of projected entangled pair states},\ }\href
  {https://doi.org/10.1103/PhysRevLett.96.220601} {\bibfield  {journal}
  {\bibinfo  {journal} {Phys. Rev. Lett.}\ }\textbf {\bibinfo {volume} {96}},\
  \bibinfo {pages} {220601} (\bibinfo {year} {2006})}\BibitemShut {NoStop}%
\bibitem [{\citenamefont {Fradkin}\ and\ \citenamefont
  {Moore}(2006)}]{Fradkin:2006}%
  \BibitemOpen
  \bibfield  {author} {\bibinfo {author} {\bibfnamefont {E.}~\bibnamefont
  {Fradkin}}\ and\ \bibinfo {author} {\bibfnamefont {J.~E.}\ \bibnamefont
  {Moore}},\ }\bibfield  {title} {\bibinfo {title} {Entanglement entropy of 2d
  conformal quantum critical points: Hearing the shape of a quantum drum},\
  }\href {https://doi.org/10.1103/PhysRevLett.97.050404} {\bibfield  {journal}
  {\bibinfo  {journal} {Phys. Rev. Lett.}\ }\textbf {\bibinfo {volume} {97}},\
  \bibinfo {pages} {050404} (\bibinfo {year} {2006})}\BibitemShut {NoStop}%
\bibitem [{\citenamefont {Castelnovo}\ and\ \citenamefont
  {Chamon}(2008)}]{castelnovo:2008}%
  \BibitemOpen
  \bibfield  {author} {\bibinfo {author} {\bibfnamefont {C.}~\bibnamefont
  {Castelnovo}}\ and\ \bibinfo {author} {\bibfnamefont {C.}~\bibnamefont
  {Chamon}},\ }\bibfield  {title} {\bibinfo {title} {Quantum topological phase
  transition at the microscopic level},\ }\href
  {https://doi.org/10.1103/PhysRevB.77.054433} {\bibfield  {journal} {\bibinfo
  {journal} {Phys. Rev. B}\ }\textbf {\bibinfo {volume} {77}},\ \bibinfo
  {pages} {054433} (\bibinfo {year} {2008})}\BibitemShut {NoStop}%
\bibitem [{\citenamefont {Salberger}\ and\ \citenamefont
  {Korepin}(2017)}]{Salberger:2016}%
  \BibitemOpen
  \bibfield  {author} {\bibinfo {author} {\bibfnamefont {O.}~\bibnamefont
  {Salberger}}\ and\ \bibinfo {author} {\bibfnamefont {V.}~\bibnamefont
  {Korepin}},\ }\bibfield  {title} {\bibinfo {title} {{Entangled spin chain}},\
  }\href {https://doi.org/10.1142/S0129055X17500313} {\bibfield  {journal}
  {\bibinfo  {journal} {Rev. Math. Phys.}\ }\textbf {\bibinfo {volume} {29}},\
  \bibinfo {pages} {1750031} (\bibinfo {year} {2017})},\ \Eprint
  {https://arxiv.org/abs/1605.03842} {arXiv:1605.03842 [quant-ph]} \BibitemShut
  {NoStop}%
\bibitem [{\citenamefont {Dell'Anna}\ \emph {et~al.}(2016)\citenamefont
  {Dell'Anna}, \citenamefont {Salberger}, \citenamefont {Barbiero},
  \citenamefont {Trombettoni},\ and\ \citenamefont
  {Korepin}}]{dellAnna2016_fredkin}%
  \BibitemOpen
  \bibfield  {author} {\bibinfo {author} {\bibfnamefont {L.}~\bibnamefont
  {Dell'Anna}}, \bibinfo {author} {\bibfnamefont {O.}~\bibnamefont
  {Salberger}}, \bibinfo {author} {\bibfnamefont {L.}~\bibnamefont {Barbiero}},
  \bibinfo {author} {\bibfnamefont {A.}~\bibnamefont {Trombettoni}},\ and\
  \bibinfo {author} {\bibfnamefont {V.~E.}\ \bibnamefont {Korepin}},\
  }\bibfield  {title} {\bibinfo {title} {Violation of cluster decomposition and
  absence of light cones in local integer and half-integer spin chains},\
  }\href {https://doi.org/10.1103/PhysRevB.94.155140} {\bibfield  {journal}
  {\bibinfo  {journal} {Phys. Rev. B}\ }\textbf {\bibinfo {volume} {94}},\
  \bibinfo {pages} {155140} (\bibinfo {year} {2016})}\BibitemShut {NoStop}%
\bibitem [{\citenamefont {Movassagh}\ and\ \citenamefont
  {Shor}(2016)}]{movassagh2016_supercrit}%
  \BibitemOpen
  \bibfield  {author} {\bibinfo {author} {\bibfnamefont {R.}~\bibnamefont
  {Movassagh}}\ and\ \bibinfo {author} {\bibfnamefont {P.~W.}\ \bibnamefont
  {Shor}},\ }\bibfield  {title} {\bibinfo {title} {Supercritical entanglement
  in local systems: Counterexample to the area law for quantum matter},\ }\href
  {https://doi.org/10.1073/pnas.1605716113} {\bibfield  {journal} {\bibinfo
  {journal} {Proceedings of the National Academy of Sciences}\ }\textbf
  {\bibinfo {volume} {113}},\ \bibinfo {pages} {13278} (\bibinfo {year}
  {2016})},\ \Eprint
  {https://arxiv.org/abs/https://www.pnas.org/doi/pdf/10.1073/pnas.1605716113}
  {https://www.pnas.org/doi/pdf/10.1073/pnas.1605716113} \BibitemShut {NoStop}%
\bibitem [{\citenamefont {Movassagh}(2018)}]{movassagh2018_fredkin}%
  \BibitemOpen
  \bibfield  {author} {\bibinfo {author} {\bibfnamefont {R.}~\bibnamefont
  {Movassagh}},\ }\bibfield  {title} {\bibinfo {title} {The gap of fredkin
  quantum spin chain is polynomially small},\ }\href
  {https://doi.org/10.4310/amsa.2018.v3.n2.a5} {\bibfield  {journal} {\bibinfo
  {journal} {Annals of Mathematical Sciences and Applications}\ }\textbf
  {\bibinfo {volume} {3}},\ \bibinfo {pages} {531–562} (\bibinfo {year}
  {2018})}\BibitemShut {NoStop}%
\bibitem [{\citenamefont {Haller}\ \emph {et~al.}(2023)\citenamefont {Haller},
  \citenamefont {Xu}, \citenamefont {Liu},\ and\ \citenamefont
  {Pollmann}}]{setqpt_tns}%
  \BibitemOpen
  \bibfield  {author} {\bibinfo {author} {\bibfnamefont {L.}~\bibnamefont
  {Haller}}, \bibinfo {author} {\bibfnamefont {W.-T.}\ \bibnamefont {Xu}},
  \bibinfo {author} {\bibfnamefont {Y.-J.}\ \bibnamefont {Liu}},\ and\ \bibinfo
  {author} {\bibfnamefont {F.}~\bibnamefont {Pollmann}},\ }\bibfield  {title}
  {\bibinfo {title} {Quantum phase transition between symmetry enriched
  topological phases in tensor-network states},\ }\href
  {https://doi.org/10.1103/PhysRevResearch.5.043078} {\bibfield  {journal}
  {\bibinfo  {journal} {Phys. Rev. Res.}\ }\textbf {\bibinfo {volume} {5}},\
  \bibinfo {pages} {043078} (\bibinfo {year} {2023})}\BibitemShut {NoStop}%
\bibitem [{\citenamefont {Albash}\ and\ \citenamefont
  {Lidar}(2018)}]{adiabaticQC2018}%
  \BibitemOpen
  \bibfield  {author} {\bibinfo {author} {\bibfnamefont {T.}~\bibnamefont
  {Albash}}\ and\ \bibinfo {author} {\bibfnamefont {D.~A.}\ \bibnamefont
  {Lidar}},\ }\bibfield  {title} {\bibinfo {title} {{Adiabatic quantum
  computation}},\ }\href {https://doi.org/10.1103/RevModPhys.90.015002}
  {\bibfield  {journal} {\bibinfo  {journal} {Rev. Mod. Phys.}\ }\textbf
  {\bibinfo {volume} {90}},\ \bibinfo {pages} {015002} (\bibinfo {year}
  {2018})}\BibitemShut {NoStop}%
\bibitem [{\citenamefont {Farhi}\ \emph {et~al.}(2000)\citenamefont {Farhi},
  \citenamefont {Goldstone}, \citenamefont {Gutmann},\ and\ \citenamefont
  {Sipser}}]{farhi2000_adiabatic}%
  \BibitemOpen
  \bibfield  {author} {\bibinfo {author} {\bibfnamefont {E.}~\bibnamefont
  {Farhi}}, \bibinfo {author} {\bibfnamefont {J.}~\bibnamefont {Goldstone}},
  \bibinfo {author} {\bibfnamefont {S.}~\bibnamefont {Gutmann}},\ and\ \bibinfo
  {author} {\bibfnamefont {M.}~\bibnamefont {Sipser}},\ }\href
  {https://arxiv.org/abs/quant-ph/0001106} {\bibinfo {title} {Quantum
  computation by adiabatic evolution}} (\bibinfo {year} {2000}),\ \Eprint
  {https://arxiv.org/abs/quant-ph/0001106} {arXiv:quant-ph/0001106 [quant-ph]}
  \BibitemShut {NoStop}%
\bibitem [{\citenamefont {Verstraete}\ \emph {et~al.}(2009)\citenamefont
  {Verstraete}, \citenamefont {Wolf},\ and\ \citenamefont
  {Ignacio~Cirac}}]{verstraete2009_dissip}%
  \BibitemOpen
  \bibfield  {author} {\bibinfo {author} {\bibfnamefont {F.}~\bibnamefont
  {Verstraete}}, \bibinfo {author} {\bibfnamefont {M.~M.}\ \bibnamefont
  {Wolf}},\ and\ \bibinfo {author} {\bibfnamefont {J.}~\bibnamefont
  {Ignacio~Cirac}},\ }\bibfield  {title} {\bibinfo {title} {Quantum computation
  and quantum-state engineering driven by dissipation},\ }\href
  {https://doi.org/10.1038/nphys1342} {\bibfield  {journal} {\bibinfo
  {journal} {Nature Physics}\ }\textbf {\bibinfo {volume} {5}},\ \bibinfo
  {pages} {633} (\bibinfo {year} {2009})}\BibitemShut {NoStop}%
\bibitem [{\citenamefont {Diehl}\ \emph {et~al.}(2008)\citenamefont {Diehl},
  \citenamefont {Micheli}, \citenamefont {Kantian}, \citenamefont {Kraus},
  \citenamefont {B{\"u}chler},\ and\ \citenamefont
  {Zoller}}]{diehl2008_dissip}%
  \BibitemOpen
  \bibfield  {author} {\bibinfo {author} {\bibfnamefont {S.}~\bibnamefont
  {Diehl}}, \bibinfo {author} {\bibfnamefont {A.}~\bibnamefont {Micheli}},
  \bibinfo {author} {\bibfnamefont {A.}~\bibnamefont {Kantian}}, \bibinfo
  {author} {\bibfnamefont {B.}~\bibnamefont {Kraus}}, \bibinfo {author}
  {\bibfnamefont {H.~P.}\ \bibnamefont {B{\"u}chler}},\ and\ \bibinfo {author}
  {\bibfnamefont {P.}~\bibnamefont {Zoller}},\ }\bibfield  {title} {\bibinfo
  {title} {Quantum states and phases in driven open quantum systems with cold
  atoms},\ }\href {https://doi.org/10.1038/nphys1073} {\bibfield  {journal}
  {\bibinfo  {journal} {Nature Physics}\ }\textbf {\bibinfo {volume} {4}},\
  \bibinfo {pages} {878} (\bibinfo {year} {2008})}\BibitemShut {NoStop}%
\bibitem [{\citenamefont {Kraus}\ \emph {et~al.}(2008)\citenamefont {Kraus},
  \citenamefont {B\"uchler}, \citenamefont {Diehl}, \citenamefont {Kantian},
  \citenamefont {Micheli},\ and\ \citenamefont {Zoller}}]{Kraus2008markov}%
  \BibitemOpen
  \bibfield  {author} {\bibinfo {author} {\bibfnamefont {B.}~\bibnamefont
  {Kraus}}, \bibinfo {author} {\bibfnamefont {H.~P.}\ \bibnamefont
  {B\"uchler}}, \bibinfo {author} {\bibfnamefont {S.}~\bibnamefont {Diehl}},
  \bibinfo {author} {\bibfnamefont {A.}~\bibnamefont {Kantian}}, \bibinfo
  {author} {\bibfnamefont {A.}~\bibnamefont {Micheli}},\ and\ \bibinfo {author}
  {\bibfnamefont {P.}~\bibnamefont {Zoller}},\ }\bibfield  {title} {\bibinfo
  {title} {Preparation of entangled states by quantum markov processes},\
  }\href {https://doi.org/10.1103/PhysRevA.78.042307} {\bibfield  {journal}
  {\bibinfo  {journal} {Phys. Rev. A}\ }\textbf {\bibinfo {volume} {78}},\
  \bibinfo {pages} {042307} (\bibinfo {year} {2008})}\BibitemShut {NoStop}%
\bibitem [{\citenamefont {Ticozzi}\ and\ \citenamefont
  {Viola}(2012)}]{ticozzi2012stabilize}%
  \BibitemOpen
  \bibfield  {author} {\bibinfo {author} {\bibfnamefont {F.}~\bibnamefont
  {Ticozzi}}\ and\ \bibinfo {author} {\bibfnamefont {L.}~\bibnamefont
  {Viola}},\ }\bibfield  {title} {\bibinfo {title} {Stabilizing entangled
  states with quasi-local quantum dynamical semigroups},\ }\href
  {https://doi.org/10.1098/rsta.2011.0485} {\bibfield  {journal} {\bibinfo
  {journal} {Philosophical Transactions of the Royal Society A: Mathematical,
  Physical and Engineering Sciences}\ }\textbf {\bibinfo {volume} {370}},\
  \bibinfo {pages} {5259} (\bibinfo {year} {2012})}\BibitemShut {NoStop}%
\bibitem [{\citenamefont {Mi}\ \emph {et~al.}(2024)\citenamefont {Mi},
  \citenamefont {Michailidis}, \citenamefont {Shabani}, \citenamefont {Miao},
  \citenamefont {Klimov}, \citenamefont {Lloyd}, \citenamefont {Rosenberg},
  \citenamefont {Acharya}, \citenamefont {Aleiner}, \citenamefont {Andersen}
  \emph {et~al.}}]{google2024_dissip}%
  \BibitemOpen
  \bibfield  {author} {\bibinfo {author} {\bibfnamefont {X.}~\bibnamefont
  {Mi}}, \bibinfo {author} {\bibfnamefont {A.~A.}\ \bibnamefont {Michailidis}},
  \bibinfo {author} {\bibfnamefont {S.}~\bibnamefont {Shabani}}, \bibinfo
  {author} {\bibfnamefont {K.~C.}\ \bibnamefont {Miao}}, \bibinfo {author}
  {\bibfnamefont {P.~V.}\ \bibnamefont {Klimov}}, \bibinfo {author}
  {\bibfnamefont {J.}~\bibnamefont {Lloyd}}, \bibinfo {author} {\bibfnamefont
  {E.}~\bibnamefont {Rosenberg}}, \bibinfo {author} {\bibfnamefont
  {R.}~\bibnamefont {Acharya}}, \bibinfo {author} {\bibfnamefont
  {I.}~\bibnamefont {Aleiner}}, \bibinfo {author} {\bibfnamefont {T.~I.}\
  \bibnamefont {Andersen}}, \emph {et~al.},\ }\bibfield  {title} {\bibinfo
  {title} {Stable quantum-correlated many-body states through engineered
  dissipation},\ }\href {https://doi.org/10.1126/science.adh9932} {\bibfield
  {journal} {\bibinfo  {journal} {Science}\ }\textbf {\bibinfo {volume}
  {383}},\ \bibinfo {pages} {1332} (\bibinfo {year} {2024})}\BibitemShut
  {NoStop}%
\bibitem [{\citenamefont {Lloyd}\ \emph {et~al.}(2025)\citenamefont {Lloyd},
  \citenamefont {Michailidis}, \citenamefont {Mi}, \citenamefont
  {Smelyanskiy},\ and\ \citenamefont {Abanin}}]{lloyd2025_cooling}%
  \BibitemOpen
  \bibfield  {author} {\bibinfo {author} {\bibfnamefont {J.}~\bibnamefont
  {Lloyd}}, \bibinfo {author} {\bibfnamefont {A.~A.}\ \bibnamefont
  {Michailidis}}, \bibinfo {author} {\bibfnamefont {X.}~\bibnamefont {Mi}},
  \bibinfo {author} {\bibfnamefont {V.}~\bibnamefont {Smelyanskiy}},\ and\
  \bibinfo {author} {\bibfnamefont {D.~A.}\ \bibnamefont {Abanin}},\ }\bibfield
   {title} {\bibinfo {title} {Quasiparticle cooling algorithms for quantum
  many-body state preparation},\ }\href
  {https://doi.org/10.1103/PRXQuantum.6.010361} {\bibfield  {journal} {\bibinfo
   {journal} {PRX Quantum}\ }\textbf {\bibinfo {volume} {6}},\ \bibinfo {pages}
  {010361} (\bibinfo {year} {2025})}\BibitemShut {NoStop}%
\bibitem [{\citenamefont {Gilyén}\ and\ \citenamefont
  {Sattath}(2017)}]{Gilyen2017_lovasz}%
  \BibitemOpen
  \bibfield  {author} {\bibinfo {author} {\bibfnamefont {A.~P.}\ \bibnamefont
  {Gilyén}}\ and\ \bibinfo {author} {\bibfnamefont {O.}~\bibnamefont
  {Sattath}},\ }\bibfield  {title} {\bibinfo {title} {{On Preparing Ground
  States of Gapped Hamiltonians: An Efficient Quantum Lovász Local Lemma}},\
  }in\ \href {https://doi.org/10.1109/FOCS.2017.47} {\emph {\bibinfo
  {booktitle} {2017 IEEE 58th Annual Symposium on Foundations of Computer
  Science (FOCS)}}}\ (\bibinfo {year} {2017})\ pp.\ \bibinfo {pages}
  {439--450}\BibitemShut {NoStop}%
\bibitem [{\citenamefont {Cubitt}(2023)}]{Cubitt2023_dissip}%
  \BibitemOpen
  \bibfield  {author} {\bibinfo {author} {\bibfnamefont {T.~S.}\ \bibnamefont
  {Cubitt}},\ }\href {https://arxiv.org/abs/2303.11962} {\bibinfo {title}
  {Dissipative ground state preparation and the dissipative quantum
  eigensolver}} (\bibinfo {year} {2023}),\ \Eprint
  {https://arxiv.org/abs/2303.11962} {arXiv:2303.11962 [quant-ph]} \BibitemShut
  {NoStop}%
\bibitem [{\citenamefont {Polla}\ \emph {et~al.}(2021)\citenamefont {Polla},
  \citenamefont {Herasymenko},\ and\ \citenamefont
  {O'Brien}}]{polla2021_cooling}%
  \BibitemOpen
  \bibfield  {author} {\bibinfo {author} {\bibfnamefont {S.}~\bibnamefont
  {Polla}}, \bibinfo {author} {\bibfnamefont {Y.}~\bibnamefont {Herasymenko}},\
  and\ \bibinfo {author} {\bibfnamefont {T.~E.}\ \bibnamefont {O'Brien}},\
  }\bibfield  {title} {\bibinfo {title} {Quantum digital cooling},\ }\href
  {https://doi.org/10.1103/PhysRevA.104.012414} {\bibfield  {journal} {\bibinfo
   {journal} {Phys. Rev. A}\ }\textbf {\bibinfo {volume} {104}},\ \bibinfo
  {pages} {012414} (\bibinfo {year} {2021})}\BibitemShut {NoStop}%
\bibitem [{\citenamefont {Roy}\ \emph {et~al.}(2020)\citenamefont {Roy},
  \citenamefont {Chalker}, \citenamefont {Gornyi},\ and\ \citenamefont
  {Gefen}}]{roy2020_steering}%
  \BibitemOpen
  \bibfield  {author} {\bibinfo {author} {\bibfnamefont {S.}~\bibnamefont
  {Roy}}, \bibinfo {author} {\bibfnamefont {J.~T.}\ \bibnamefont {Chalker}},
  \bibinfo {author} {\bibfnamefont {I.~V.}\ \bibnamefont {Gornyi}},\ and\
  \bibinfo {author} {\bibfnamefont {Y.}~\bibnamefont {Gefen}},\ }\bibfield
  {title} {\bibinfo {title} {Measurement-induced steering of quantum systems},\
  }\href {https://doi.org/10.1103/PhysRevResearch.2.033347} {\bibfield
  {journal} {\bibinfo  {journal} {Phys. Rev. Res.}\ }\textbf {\bibinfo {volume}
  {2}},\ \bibinfo {pages} {033347} (\bibinfo {year} {2020})}\BibitemShut
  {NoStop}%
\bibitem [{\citenamefont {Matthies}\ \emph {et~al.}(2024)\citenamefont
  {Matthies}, \citenamefont {Rudner}, \citenamefont {Rosch},\ and\
  \citenamefont {Berg}}]{Matthies2024_cooling}%
  \BibitemOpen
  \bibfield  {author} {\bibinfo {author} {\bibfnamefont {A.}~\bibnamefont
  {Matthies}}, \bibinfo {author} {\bibfnamefont {M.}~\bibnamefont {Rudner}},
  \bibinfo {author} {\bibfnamefont {A.}~\bibnamefont {Rosch}},\ and\ \bibinfo
  {author} {\bibfnamefont {E.}~\bibnamefont {Berg}},\ }\bibfield  {title}
  {\bibinfo {title} {Programmable adiabatic demagnetization for systems with
  trivial and topological excitations},\ }\href
  {https://doi.org/10.22331/q-2024-10-23-1505} {\bibfield  {journal} {\bibinfo
  {journal} {{Quantum}}\ }\textbf {\bibinfo {volume} {8}},\ \bibinfo {pages}
  {1505} (\bibinfo {year} {2024})}\BibitemShut {NoStop}%
\bibitem [{\citenamefont {Molpeceres}\ \emph {et~al.}(2025)\citenamefont
  {Molpeceres}, \citenamefont {Lu}, \citenamefont {Cirac},\ and\ \citenamefont
  {Kraus}}]{molpeceres2025_cooling}%
  \BibitemOpen
  \bibfield  {author} {\bibinfo {author} {\bibfnamefont {D.}~\bibnamefont
  {Molpeceres}}, \bibinfo {author} {\bibfnamefont {S.}~\bibnamefont {Lu}},
  \bibinfo {author} {\bibfnamefont {J.~I.}\ \bibnamefont {Cirac}},\ and\
  \bibinfo {author} {\bibfnamefont {B.}~\bibnamefont {Kraus}},\ }\href
  {https://arxiv.org/abs/2503.24330} {\bibinfo {title} {Quantum algorithms for
  cooling: a simple case study}} (\bibinfo {year} {2025}),\ \Eprint
  {https://arxiv.org/abs/2503.24330} {arXiv:2503.24330 [quant-ph]} \BibitemShut
  {NoStop}%
\bibitem [{\citenamefont {Herasymenko}\ \emph {et~al.}(2023)\citenamefont
  {Herasymenko}, \citenamefont {Gornyi},\ and\ \citenamefont
  {Gefen}}]{Herasymenko2023steering}%
  \BibitemOpen
  \bibfield  {author} {\bibinfo {author} {\bibfnamefont {Y.}~\bibnamefont
  {Herasymenko}}, \bibinfo {author} {\bibfnamefont {I.}~\bibnamefont
  {Gornyi}},\ and\ \bibinfo {author} {\bibfnamefont {Y.}~\bibnamefont
  {Gefen}},\ }\bibfield  {title} {\bibinfo {title} {Measurement-driven
  navigation in many-body hilbert space: Active-decision steering},\ }\href
  {https://doi.org/10.1103/PRXQuantum.4.020347} {\bibfield  {journal} {\bibinfo
   {journal} {PRX Quantum}\ }\textbf {\bibinfo {volume} {4}},\ \bibinfo {pages}
  {020347} (\bibinfo {year} {2023})}\BibitemShut {NoStop}%
\bibitem [{\citenamefont {Gily\'{e}n}\ \emph {et~al.}(2019)\citenamefont
  {Gily\'{e}n}, \citenamefont {Su}, \citenamefont {Low},\ and\ \citenamefont
  {Wiebe}}]{gilyen2019_qsvt}%
  \BibitemOpen
  \bibfield  {author} {\bibinfo {author} {\bibfnamefont {A.}~\bibnamefont
  {Gily\'{e}n}}, \bibinfo {author} {\bibfnamefont {Y.}~\bibnamefont {Su}},
  \bibinfo {author} {\bibfnamefont {G.~H.}\ \bibnamefont {Low}},\ and\ \bibinfo
  {author} {\bibfnamefont {N.}~\bibnamefont {Wiebe}},\ }\bibfield  {title}
  {\bibinfo {title} {Quantum singular value transformation and beyond:
  exponential improvements for quantum matrix arithmetics},\ }in\ \href
  {https://doi.org/10.1145/3313276.3316366} {\emph {\bibinfo {booktitle}
  {Proceedings of the 51st Annual ACM SIGACT Symposium on Theory of
  Computing}}},\ \bibinfo {series and number} {STOC 2019}\ (\bibinfo
  {publisher} {Association for Computing Machinery},\ \bibinfo {address} {New
  York, NY, USA},\ \bibinfo {year} {2019})\ p.\ \bibinfo {pages}
  {193–204}\BibitemShut {NoStop}%
\bibitem [{\citenamefont {Lin}\ and\ \citenamefont
  {Tong}(2020)}]{Lin2020_nearoptimal}%
  \BibitemOpen
  \bibfield  {author} {\bibinfo {author} {\bibfnamefont {L.}~\bibnamefont
  {Lin}}\ and\ \bibinfo {author} {\bibfnamefont {Y.}~\bibnamefont {Tong}},\
  }\bibfield  {title} {\bibinfo {title} {Near-optimal ground state
  preparation},\ }\href {https://doi.org/10.22331/q-2020-12-14-372} {\bibfield
  {journal} {\bibinfo  {journal} {{Quantum}}\ }\textbf {\bibinfo {volume}
  {4}},\ \bibinfo {pages} {372} (\bibinfo {year} {2020})}\BibitemShut {NoStop}%
\bibitem [{\citenamefont {Thibodeau}\ and\ \citenamefont
  {Clark}(2023)}]{Thibodeau2023_frust}%
  \BibitemOpen
  \bibfield  {author} {\bibinfo {author} {\bibfnamefont {M.}~\bibnamefont
  {Thibodeau}}\ and\ \bibinfo {author} {\bibfnamefont {B.~K.}\ \bibnamefont
  {Clark}},\ }\bibfield  {title} {\bibinfo {title} {Nearly-frustration-free
  ground state preparation},\ }\href
  {https://doi.org/10.22331/q-2023-08-16-1084} {\bibfield  {journal} {\bibinfo
  {journal} {{Quantum}}\ }\textbf {\bibinfo {volume} {7}},\ \bibinfo {pages}
  {1084} (\bibinfo {year} {2023})}\BibitemShut {NoStop}%
\bibitem [{\citenamefont {Alzetta}\ \emph {et~al.}(1976)\citenamefont
  {Alzetta}, \citenamefont {Gozzini}, \citenamefont {Moi},\ and\ \citenamefont
  {Orriols}}]{alzetta1976experimental}%
  \BibitemOpen
  \bibfield  {author} {\bibinfo {author} {\bibfnamefont {G.}~\bibnamefont
  {Alzetta}}, \bibinfo {author} {\bibfnamefont {A.}~\bibnamefont {Gozzini}},
  \bibinfo {author} {\bibfnamefont {L.}~\bibnamefont {Moi}},\ and\ \bibinfo
  {author} {\bibfnamefont {G.}~\bibnamefont {Orriols}},\ }\bibfield  {title}
  {\bibinfo {title} {An experimental method for the observation of rf
  transitions and laser beat resonances in oriented na vapour},\ }\href
  {https://doi.org/10.1007/BF02749417} {\bibfield  {journal} {\bibinfo
  {journal} {Il Nuovo Cimento B (1971-1996)}\ }\textbf {\bibinfo {volume}
  {36}},\ \bibinfo {pages} {5} (\bibinfo {year} {1976})}\BibitemShut {NoStop}%
\bibitem [{\citenamefont {Gray}\ \emph {et~al.}(1978)\citenamefont {Gray},
  \citenamefont {Whitley},\ and\ \citenamefont {Stroud}}]{Gray:78}%
  \BibitemOpen
  \bibfield  {author} {\bibinfo {author} {\bibfnamefont {H.~R.}\ \bibnamefont
  {Gray}}, \bibinfo {author} {\bibfnamefont {R.~M.}\ \bibnamefont {Whitley}},\
  and\ \bibinfo {author} {\bibfnamefont {C.~R.}\ \bibnamefont {Stroud}},\
  }\bibfield  {title} {\bibinfo {title} {Coherent trapping of atomic
  populations},\ }\href {https://doi.org/10.1364/OL.3.000218} {\bibfield
  {journal} {\bibinfo  {journal} {Opt. Lett.}\ }\textbf {\bibinfo {volume}
  {3}},\ \bibinfo {pages} {218} (\bibinfo {year} {1978})}\BibitemShut {NoStop}%
\bibitem [{\citenamefont {Gosset}\ and\ \citenamefont
  {Mozgunov}(2016)}]{gosset2016_gap}%
  \BibitemOpen
  \bibfield  {author} {\bibinfo {author} {\bibfnamefont {D.}~\bibnamefont
  {Gosset}}\ and\ \bibinfo {author} {\bibfnamefont {E.}~\bibnamefont
  {Mozgunov}},\ }\bibfield  {title} {\bibinfo {title} {Local gap threshold for
  frustration-free spin systems},\ }\href {https://doi.org/10.1063/1.4962337}
  {\bibfield  {journal} {\bibinfo  {journal} {Journal of Mathematical Physics}\
  }\textbf {\bibinfo {volume} {57}},\ \bibinfo {pages} {091901} (\bibinfo
  {year} {2016})},\ \Eprint
  {https://arxiv.org/abs/https://pubs.aip.org/aip/jmp/article-pdf/doi/10.1063/1.4962337/13685714/091901\_1\_online.pdf}
  {https://pubs.aip.org/aip/jmp/article-pdf/doi/10.1063/1.4962337/13685714/091901\_1\_online.pdf}
  \BibitemShut {NoStop}%
\bibitem [{\citenamefont {Ogunnaike}\ \emph {et~al.}(2023)\citenamefont
  {Ogunnaike}, \citenamefont {Feldmeier},\ and\ \citenamefont
  {Lee}}]{ogunnaike2023_lindblad}%
  \BibitemOpen
  \bibfield  {author} {\bibinfo {author} {\bibfnamefont {O.}~\bibnamefont
  {Ogunnaike}}, \bibinfo {author} {\bibfnamefont {J.}~\bibnamefont
  {Feldmeier}},\ and\ \bibinfo {author} {\bibfnamefont {J.~Y.}\ \bibnamefont
  {Lee}},\ }\bibfield  {title} {\bibinfo {title} {Unifying emergent
  hydrodynamics and lindbladian low-energy spectra across symmetries,
  constraints, and long-range interactions},\ }\href
  {https://doi.org/10.1103/PhysRevLett.131.220403} {\bibfield  {journal}
  {\bibinfo  {journal} {Phys. Rev. Lett.}\ }\textbf {\bibinfo {volume} {131}},\
  \bibinfo {pages} {220403} (\bibinfo {year} {2023})}\BibitemShut {NoStop}%
\bibitem [{\citenamefont {Masaoka}\ \emph {et~al.}(2024)\citenamefont
  {Masaoka}, \citenamefont {Soejima},\ and\ \citenamefont
  {Watanabe}}]{masaoka2024_gap}%
  \BibitemOpen
  \bibfield  {author} {\bibinfo {author} {\bibfnamefont {R.}~\bibnamefont
  {Masaoka}}, \bibinfo {author} {\bibfnamefont {T.}~\bibnamefont {Soejima}},\
  and\ \bibinfo {author} {\bibfnamefont {H.}~\bibnamefont {Watanabe}},\
  }\bibfield  {title} {\bibinfo {title} {Quadratic dispersion relations in
  gapless frustration-free systems},\ }\href
  {https://doi.org/10.1103/PhysRevB.110.195140} {\bibfield  {journal} {\bibinfo
   {journal} {Phys. Rev. B}\ }\textbf {\bibinfo {volume} {110}},\ \bibinfo
  {pages} {195140} (\bibinfo {year} {2024})}\BibitemShut {NoStop}%
\bibitem [{\citenamefont {Masaoka}\ \emph {et~al.}(2025)\citenamefont
  {Masaoka}, \citenamefont {Soejima},\ and\ \citenamefont
  {Watanabe}}]{masaoka2025oral}%
  \BibitemOpen
  \bibfield  {author} {\bibinfo {author} {\bibfnamefont {R.}~\bibnamefont
  {Masaoka}}, \bibinfo {author} {\bibfnamefont {T.}~\bibnamefont {Soejima}},\
  and\ \bibinfo {author} {\bibfnamefont {H.}~\bibnamefont {Watanabe}},\
  }\bibfield  {title} {\bibinfo {title} {Oral: Rigorous lower bound of dynamic
  critical exponents in critical frustration-free systems},\ }in\ \href@noop {}
  {\emph {\bibinfo {booktitle} {SMT 2025}}}\ (\bibinfo {organization} {APS},\
  \bibinfo {year} {2025})\BibitemShut {NoStop}%
\bibitem [{Note1()}]{Note1}%
  \BibitemOpen
  \bibinfo {note} {In our examples, the projectors $P_i$ are of rank 1, such
  that we can find local feedback that corrects $P_i=1$ to $P_i=0$. We note
  that general projectors $P_i$ can be written as a sum of rank 1 projectors,
  and the corresponding perfect feedback can be constructed accordingly. In
  general, we only require that $U_C(i)$ does not commute with $P_i$. We expect
  the protocol to still behave the same qualitatively, which we verified in the
  examples studied in this work.}\BibitemShut {Stop}%
\bibitem [{\citenamefont {Aharonov}\ \emph {et~al.}(2009)\citenamefont
  {Aharonov}, \citenamefont {Arad}, \citenamefont {Landau},\ and\ \citenamefont
  {Vazirani}}]{aharonov2009_DL}%
  \BibitemOpen
  \bibfield  {author} {\bibinfo {author} {\bibfnamefont {D.}~\bibnamefont
  {Aharonov}}, \bibinfo {author} {\bibfnamefont {I.}~\bibnamefont {Arad}},
  \bibinfo {author} {\bibfnamefont {Z.}~\bibnamefont {Landau}},\ and\ \bibinfo
  {author} {\bibfnamefont {U.}~\bibnamefont {Vazirani}},\ }\bibfield  {title}
  {\bibinfo {title} {The detectability lemma and quantum gap amplification},\
  }in\ \href {https://doi.org/10.1145/1536414.1536472} {\emph {\bibinfo
  {booktitle} {Proceedings of the Forty-First Annual ACM Symposium on Theory of
  Computing}}},\ \bibinfo {series and number} {STOC '09}\ (\bibinfo
  {publisher} {Association for Computing Machinery},\ \bibinfo {address} {New
  York, NY, USA},\ \bibinfo {year} {2009})\ p.\ \bibinfo {pages}
  {417–426}\BibitemShut {NoStop}%
\bibitem [{\citenamefont {Anshu}\ \emph {et~al.}(2016)\citenamefont {Anshu},
  \citenamefont {Arad},\ and\ \citenamefont {Vidick}}]{anshu2016_DL}%
  \BibitemOpen
  \bibfield  {author} {\bibinfo {author} {\bibfnamefont {A.}~\bibnamefont
  {Anshu}}, \bibinfo {author} {\bibfnamefont {I.}~\bibnamefont {Arad}},\ and\
  \bibinfo {author} {\bibfnamefont {T.}~\bibnamefont {Vidick}},\ }\bibfield
  {title} {\bibinfo {title} {Simple proof of the detectability lemma and
  spectral gap amplification},\ }\href
  {https://doi.org/10.1103/PhysRevB.93.205142} {\bibfield  {journal} {\bibinfo
  {journal} {Phys. Rev. B}\ }\textbf {\bibinfo {volume} {93}},\ \bibinfo
  {pages} {205142} (\bibinfo {year} {2016})}\BibitemShut {NoStop}%
\bibitem [{Note2()}]{Note2}%
  \BibitemOpen
  \bibinfo {note} {Via the DL, the spectral gap $\Delta $ of the
  frustration-free Hamiltonian is equivalent (up to constants) to the gap of
  $\protect \mathcal {P}$; in particular, the slow modes of $\protect \mathcal
  {P}$ are supported on low-lying excited states of $H$. We assume that the
  states $\mathinner {|{\psi (t)}\rangle }$ along the cooling dynamics
  generically carry a nonzero weight in this low-energy sector that is not
  parametrically smaller in system size than their weight in the ground-state
  subspace. We expect that these weights, which come from gapless excitations,
  cannot be removed by any finite-depth quantum circuits. Therefore, most
  trajectories which eventually reach the ground state must, at late times,
  enter a reset-free trajectory. The infidelity under a reset-free trajectory
  cannot decay faster than $\sim e^{-c t \Delta }$; this implies a convergence
  time $t \gtrsim \Delta ^{-1} |\log \epsilon |$ to reach small infidelity
  $\epsilon $ (if we only assume that the weight in the low-energy sector is
  non-zero, as done in the main text, that sets the minimal timescale to reach
  arbitrarily small infidelity). Under these assumptions, this also sets a
  minimal timescale for the full protocol to converge for any choices of
  $U_C(i)$.}\BibitemShut {Stop}%
\bibitem [{SM()}]{SM}%
  \BibitemOpen
  \href@noop {} {}\bibinfo {note} {See Supplemental Material for additional
  details, which includes
  Refs.~\cite{Shtanko2025boundsautonomous,Topo_quantum_memory_2002,nat:2023,Dengis:2014,wolf:2006}.}\BibitemShut
  {Stop}%
\bibitem [{\citenamefont {Cook}(1971)}]{cook:1971}%
  \BibitemOpen
  \bibfield  {author} {\bibinfo {author} {\bibfnamefont {S.~A.}\ \bibnamefont
  {Cook}},\ }\bibfield  {title} {\bibinfo {title} {The complexity of
  theorem-proving procedures},\ }in\ \href
  {https://doi.org/10.1145/800157.805047} {\emph {\bibinfo {booktitle}
  {Proceedings of the Third Annual ACM Symposium on Theory of Computing}}},\
  \bibinfo {series and number} {STOC '71}\ (\bibinfo  {publisher} {Association
  for Computing Machinery},\ \bibinfo {address} {New York, NY, USA},\ \bibinfo
  {year} {1971})\ p.\ \bibinfo {pages} {151–158}\BibitemShut {NoStop}%
\bibitem [{\citenamefont {Lichtenstein}(1982)}]{Lichtenstein:1982}%
  \BibitemOpen
  \bibfield  {author} {\bibinfo {author} {\bibfnamefont {D.}~\bibnamefont
  {Lichtenstein}},\ }\bibfield  {title} {\bibinfo {title} {Planar formulae and
  their uses},\ }\href {https://doi.org/10.1137/0211025} {\bibfield  {journal}
  {\bibinfo  {journal} {SIAM Journal on Computing}\ }\textbf {\bibinfo {volume}
  {11}},\ \bibinfo {pages} {329} (\bibinfo {year} {1982})}\BibitemShut
  {NoStop}%
\bibitem [{\citenamefont {Auerbach}(2012)}]{auerbach2012_interacting}%
  \BibitemOpen
  \bibfield  {author} {\bibinfo {author} {\bibfnamefont {A.}~\bibnamefont
  {Auerbach}},\ }\href@noop {} {\emph {\bibinfo {title} {Interacting electrons
  and quantum magnetism}}}\ (\bibinfo  {publisher} {Springer Science \&
  Business Media},\ \bibinfo {year} {2012})\BibitemShut {NoStop}%
\bibitem [{\citenamefont {Chen}\ \emph {et~al.}(2017)\citenamefont {Chen},
  \citenamefont {Fradkin},\ and\ \citenamefont
  {Witczak-Krempa}}]{Chen2017_fredkin}%
  \BibitemOpen
  \bibfield  {author} {\bibinfo {author} {\bibfnamefont {X.}~\bibnamefont
  {Chen}}, \bibinfo {author} {\bibfnamefont {E.}~\bibnamefont {Fradkin}},\ and\
  \bibinfo {author} {\bibfnamefont {W.}~\bibnamefont {Witczak-Krempa}},\
  }\bibfield  {title} {\bibinfo {title} {Gapless quantum spin chains: multiple
  dynamics and conformal wavefunctions},\ }\href
  {https://doi.org/10.1088/1751-8121/aa8dbc} {\bibfield  {journal} {\bibinfo
  {journal} {Journal of Physics A: Mathematical and Theoretical}\ }\textbf
  {\bibinfo {volume} {50}},\ \bibinfo {pages} {464002} (\bibinfo {year}
  {2017})}\BibitemShut {NoStop}%
\bibitem [{\citenamefont {Dell’Anna}\ \emph {et~al.}(2019)\citenamefont
  {Dell’Anna}, \citenamefont {Barbiero},\ and\ \citenamefont
  {Trombettoni}}]{DellAnna:2019}%
  \BibitemOpen
  \bibfield  {author} {\bibinfo {author} {\bibfnamefont {L.}~\bibnamefont
  {Dell’Anna}}, \bibinfo {author} {\bibfnamefont {L.}~\bibnamefont
  {Barbiero}},\ and\ \bibinfo {author} {\bibfnamefont {A.}~\bibnamefont
  {Trombettoni}},\ }\bibfield  {title} {\bibinfo {title} {Dynamics and
  correlations in motzkin and fredkin spin chains},\ }\href
  {https://doi.org/10.1088/1742-5468/ab5703} {\bibfield  {journal} {\bibinfo
  {journal} {Journal of Statistical Mechanics: Theory and Experiment}\ }\textbf
  {\bibinfo {volume} {2019}},\ \bibinfo {pages} {124025} (\bibinfo {year}
  {2019})}\BibitemShut {NoStop}%
\bibitem [{Note3()}]{Note3}%
  \BibitemOpen
  \bibinfo {note} {The adiabatic ramp is $H(t) = (1-r(t))\protect \,
  H_{\protect \mathrm {col}} + r(t)\protect \, H_{\protect \mathrm {RVB}}$,
  where $H_{\protect \mathrm {col}}$ has a columnar state as the exact ground
  state. For a linear ramp, $r(t) = t/ T_{\protect \mathrm {ramp}}$ with $0\leq
  t \leq T_{\protect \mathrm {ramp}}$. We have tested a smooth profile with
  $\partial _t r(t) = \protect \frac {1}{\protect \mathcal {N}}e^{-T_{\protect
  \mathrm {ramp}}\protect \bigl (1/t + 1/(T_{\protect \mathrm
  {ramp}}-t)\protect \bigr )}$, where $\protect \mathcal {N}$ is such that
  $r(T_{\protect \mathrm {ramp}})=1$.}\BibitemShut {Stop}%
\bibitem [{\citenamefont {Boixo}\ and\ \citenamefont
  {Somma}(2010)}]{boixo2010_adiab}%
  \BibitemOpen
  \bibfield  {author} {\bibinfo {author} {\bibfnamefont {S.}~\bibnamefont
  {Boixo}}\ and\ \bibinfo {author} {\bibfnamefont {R.~D.}\ \bibnamefont
  {Somma}},\ }\bibfield  {title} {\bibinfo {title} {Necessary condition for the
  quantum adiabatic approximation},\ }\href
  {https://doi.org/10.1103/PhysRevA.81.032308} {\bibfield  {journal} {\bibinfo
  {journal} {Phys. Rev. A}\ }\textbf {\bibinfo {volume} {81}},\ \bibinfo
  {pages} {032308} (\bibinfo {year} {2010})}\BibitemShut {NoStop}%
\bibitem [{\citenamefont {Berry}\ \emph {et~al.}(2007)\citenamefont {Berry},
  \citenamefont {Ahokas}, \citenamefont {Cleve},\ and\ \citenamefont
  {Sanders}}]{berry2007_trotter}%
  \BibitemOpen
  \bibfield  {author} {\bibinfo {author} {\bibfnamefont {D.~W.}\ \bibnamefont
  {Berry}}, \bibinfo {author} {\bibfnamefont {G.}~\bibnamefont {Ahokas}},
  \bibinfo {author} {\bibfnamefont {R.}~\bibnamefont {Cleve}},\ and\ \bibinfo
  {author} {\bibfnamefont {B.~C.}\ \bibnamefont {Sanders}},\ }\bibfield
  {title} {\bibinfo {title} {Efficient quantum algorithms for simulating sparse
  hamiltonians},\ }\href {https://doi.org/10.1007/s00220-006-0150-x} {\bibfield
   {journal} {\bibinfo  {journal} {Communications in Mathematical Physics}\
  }\textbf {\bibinfo {volume} {270}},\ \bibinfo {pages} {359} (\bibinfo {year}
  {2007})}\BibitemShut {NoStop}%
\bibitem [{\citenamefont {Acharya}\ \emph {et~al.}(2023)\citenamefont
  {Acharya}, \citenamefont {Aleiner}, \citenamefont {Allen}, \citenamefont
  {Andersen}, \citenamefont {Ansmann}, \citenamefont {Arute}, \citenamefont
  {Arya}, \citenamefont {Asfaw}, \citenamefont {Atalaya}, \citenamefont
  {Babbush} \emph {et~al.}}]{google2023_surface}%
  \BibitemOpen
  \bibfield  {author} {\bibinfo {author} {\bibfnamefont {R.}~\bibnamefont
  {Acharya}}, \bibinfo {author} {\bibfnamefont {I.}~\bibnamefont {Aleiner}},
  \bibinfo {author} {\bibfnamefont {R.}~\bibnamefont {Allen}}, \bibinfo
  {author} {\bibfnamefont {T.~I.}\ \bibnamefont {Andersen}}, \bibinfo {author}
  {\bibfnamefont {M.}~\bibnamefont {Ansmann}}, \bibinfo {author} {\bibfnamefont
  {F.}~\bibnamefont {Arute}}, \bibinfo {author} {\bibfnamefont
  {K.}~\bibnamefont {Arya}}, \bibinfo {author} {\bibfnamefont {A.}~\bibnamefont
  {Asfaw}}, \bibinfo {author} {\bibfnamefont {J.}~\bibnamefont {Atalaya}},
  \bibinfo {author} {\bibfnamefont {R.}~\bibnamefont {Babbush}}, \emph
  {et~al.},\ }\bibfield  {title} {\bibinfo {title} {Suppressing quantum errors
  by scaling a surface code logical qubit},\ }\href
  {https://doi.org/10.1038/s41586-022-05434-1} {\bibfield  {journal} {\bibinfo
  {journal} {Nature}\ }\textbf {\bibinfo {volume} {614}},\ \bibinfo {pages}
  {676} (\bibinfo {year} {2023})}\BibitemShut {NoStop}%
\bibitem [{\citenamefont {Bluvstein}\ \emph {et~al.}(2024)\citenamefont
  {Bluvstein}, \citenamefont {Evered}, \citenamefont {Geim}, \citenamefont
  {Li}, \citenamefont {Zhou}, \citenamefont {Manovitz}, \citenamefont {Ebadi},
  \citenamefont {Cain}, \citenamefont {Kalinowski}, \citenamefont {Hangleiter},
  \citenamefont {Bonilla~Ataides}, \citenamefont {Maskara}, \citenamefont
  {Cong}, \citenamefont {Gao}, \citenamefont {Sales~Rodriguez}, \citenamefont
  {Karolyshyn}, \citenamefont {Semeghini}, \citenamefont {Gullans},
  \citenamefont {Greiner}, \citenamefont {Vuleti{\'c}},\ and\ \citenamefont
  {Lukin}}]{Bluvstein2024_logic}%
  \BibitemOpen
  \bibfield  {author} {\bibinfo {author} {\bibfnamefont {D.}~\bibnamefont
  {Bluvstein}}, \bibinfo {author} {\bibfnamefont {S.~J.}\ \bibnamefont
  {Evered}}, \bibinfo {author} {\bibfnamefont {A.~A.}\ \bibnamefont {Geim}},
  \bibinfo {author} {\bibfnamefont {S.~H.}\ \bibnamefont {Li}}, \bibinfo
  {author} {\bibfnamefont {H.}~\bibnamefont {Zhou}}, \bibinfo {author}
  {\bibfnamefont {T.}~\bibnamefont {Manovitz}}, \bibinfo {author}
  {\bibfnamefont {S.}~\bibnamefont {Ebadi}}, \bibinfo {author} {\bibfnamefont
  {M.}~\bibnamefont {Cain}}, \bibinfo {author} {\bibfnamefont {M.}~\bibnamefont
  {Kalinowski}}, \bibinfo {author} {\bibfnamefont {D.}~\bibnamefont
  {Hangleiter}}, \bibinfo {author} {\bibfnamefont {J.~P.}\ \bibnamefont
  {Bonilla~Ataides}}, \bibinfo {author} {\bibfnamefont {N.}~\bibnamefont
  {Maskara}}, \bibinfo {author} {\bibfnamefont {I.}~\bibnamefont {Cong}},
  \bibinfo {author} {\bibfnamefont {X.}~\bibnamefont {Gao}}, \bibinfo {author}
  {\bibfnamefont {P.}~\bibnamefont {Sales~Rodriguez}}, \bibinfo {author}
  {\bibfnamefont {T.}~\bibnamefont {Karolyshyn}}, \bibinfo {author}
  {\bibfnamefont {G.}~\bibnamefont {Semeghini}}, \bibinfo {author}
  {\bibfnamefont {M.~J.}\ \bibnamefont {Gullans}}, \bibinfo {author}
  {\bibfnamefont {M.}~\bibnamefont {Greiner}}, \bibinfo {author} {\bibfnamefont
  {V.}~\bibnamefont {Vuleti{\'c}}},\ and\ \bibinfo {author} {\bibfnamefont
  {M.~D.}\ \bibnamefont {Lukin}},\ }\bibfield  {title} {\bibinfo {title}
  {Logical quantum processor based on reconfigurable atom arrays},\ }\href
  {https://doi.org/10.1038/s41586-023-06927-3} {\bibfield  {journal} {\bibinfo
  {journal} {Nature}\ }\textbf {\bibinfo {volume} {626}},\ \bibinfo {pages}
  {58} (\bibinfo {year} {2024})}\BibitemShut {NoStop}%
\bibitem [{\citenamefont {Reichardt}\ \emph {et~al.}(2024)\citenamefont
  {Reichardt}, \citenamefont {Aasen}, \citenamefont {Chao}, \citenamefont
  {Chernoguzov}, \citenamefont {van Dam}, \citenamefont {Gaebler},
  \citenamefont {Gresh}, \citenamefont {Lucchetti}, \citenamefont {Mills},
  \citenamefont {Moses} \emph {et~al.}}]{quantinuum2024_qec}%
  \BibitemOpen
  \bibfield  {author} {\bibinfo {author} {\bibfnamefont {B.~W.}\ \bibnamefont
  {Reichardt}}, \bibinfo {author} {\bibfnamefont {D.}~\bibnamefont {Aasen}},
  \bibinfo {author} {\bibfnamefont {R.}~\bibnamefont {Chao}}, \bibinfo {author}
  {\bibfnamefont {A.}~\bibnamefont {Chernoguzov}}, \bibinfo {author}
  {\bibfnamefont {W.}~\bibnamefont {van Dam}}, \bibinfo {author} {\bibfnamefont
  {J.~P.}\ \bibnamefont {Gaebler}}, \bibinfo {author} {\bibfnamefont
  {D.}~\bibnamefont {Gresh}}, \bibinfo {author} {\bibfnamefont
  {D.}~\bibnamefont {Lucchetti}}, \bibinfo {author} {\bibfnamefont
  {M.}~\bibnamefont {Mills}}, \bibinfo {author} {\bibfnamefont {S.~A.}\
  \bibnamefont {Moses}}, \emph {et~al.},\ }\href
  {https://arxiv.org/abs/2409.04628} {\bibinfo {title} {Demonstration of
  quantum computation and error correction with a tesseract code}} (\bibinfo
  {year} {2024}),\ \Eprint {https://arxiv.org/abs/2409.04628} {arXiv:2409.04628
  [quant-ph]} \BibitemShut {NoStop}%
\bibitem [{\citenamefont {Acharya}\ \emph {et~al.}(2025)\citenamefont
  {Acharya}, \citenamefont {Abanin}, \citenamefont {Aghababaie-Beni},
  \citenamefont {Aleiner}, \citenamefont {Andersen}, \citenamefont {Ansmann},
  \citenamefont {Arute}, \citenamefont {Arya}, \citenamefont {Asfaw} \emph
  {et~al.}}]{google2025_qec}%
  \BibitemOpen
  \bibfield  {author} {\bibinfo {author} {\bibfnamefont {R.}~\bibnamefont
  {Acharya}}, \bibinfo {author} {\bibfnamefont {D.~A.}\ \bibnamefont {Abanin}},
  \bibinfo {author} {\bibfnamefont {L.}~\bibnamefont {Aghababaie-Beni}},
  \bibinfo {author} {\bibfnamefont {I.}~\bibnamefont {Aleiner}}, \bibinfo
  {author} {\bibfnamefont {T.~I.}\ \bibnamefont {Andersen}}, \bibinfo {author}
  {\bibfnamefont {M.}~\bibnamefont {Ansmann}}, \bibinfo {author} {\bibfnamefont
  {F.}~\bibnamefont {Arute}}, \bibinfo {author} {\bibfnamefont
  {K.}~\bibnamefont {Arya}}, \bibinfo {author} {\bibfnamefont {A.}~\bibnamefont
  {Asfaw}}, \emph {et~al.},\ }\bibfield  {title} {\bibinfo {title} {Quantum
  error correction below the surface code threshold},\ }\href
  {https://doi.org/10.1038/s41586-024-08449-y} {\bibfield  {journal} {\bibinfo
  {journal} {Nature}\ }\textbf {\bibinfo {volume} {638}},\ \bibinfo {pages}
  {920} (\bibinfo {year} {2025})}\BibitemShut {NoStop}%
\bibitem [{\citenamefont {Dawson}\ and\ \citenamefont
  {Nielsen}(2005)}]{dawson2005_solvay}%
  \BibitemOpen
  \bibfield  {author} {\bibinfo {author} {\bibfnamefont {C.~M.}\ \bibnamefont
  {Dawson}}\ and\ \bibinfo {author} {\bibfnamefont {M.~A.}\ \bibnamefont
  {Nielsen}},\ }\href {https://arxiv.org/abs/quant-ph/0505030} {\bibinfo
  {title} {The solovay-kitaev algorithm}} (\bibinfo {year} {2005}),\ \Eprint
  {https://arxiv.org/abs/quant-ph/0505030} {arXiv:quant-ph/0505030 [quant-ph]}
  \BibitemShut {NoStop}%
\bibitem [{\citenamefont {Bluvstein}\ \emph {et~al.}(2025)\citenamefont
  {Bluvstein}, \citenamefont {Geim}, \citenamefont {Li}, \citenamefont
  {Evered}, \citenamefont {Ataides}, \citenamefont {Baranes}, \citenamefont
  {Gu}, \citenamefont {Manovitz}, \citenamefont {Xu}, \citenamefont
  {Kalinowski}, \citenamefont {Majidy}, \citenamefont {Kokail}, \citenamefont
  {Maskara}, \citenamefont {Trapp}, \citenamefont {Stewart}, \citenamefont
  {Hollerith}, \citenamefont {Zhou}, \citenamefont {Gullans}, \citenamefont
  {Yelin}, \citenamefont {Greiner}, \citenamefont {Vuletic}, \citenamefont
  {Cain},\ and\ \citenamefont {Lukin}}]{bluvstein2025_architect}%
  \BibitemOpen
  \bibfield  {author} {\bibinfo {author} {\bibfnamefont {D.}~\bibnamefont
  {Bluvstein}}, \bibinfo {author} {\bibfnamefont {A.~A.}\ \bibnamefont {Geim}},
  \bibinfo {author} {\bibfnamefont {S.~H.}\ \bibnamefont {Li}}, \bibinfo
  {author} {\bibfnamefont {S.~J.}\ \bibnamefont {Evered}}, \bibinfo {author}
  {\bibfnamefont {J.~P.~B.}\ \bibnamefont {Ataides}}, \bibinfo {author}
  {\bibfnamefont {G.}~\bibnamefont {Baranes}}, \bibinfo {author} {\bibfnamefont
  {A.}~\bibnamefont {Gu}}, \bibinfo {author} {\bibfnamefont {T.}~\bibnamefont
  {Manovitz}}, \bibinfo {author} {\bibfnamefont {M.}~\bibnamefont {Xu}},
  \bibinfo {author} {\bibfnamefont {M.}~\bibnamefont {Kalinowski}}, \bibinfo
  {author} {\bibfnamefont {S.}~\bibnamefont {Majidy}}, \bibinfo {author}
  {\bibfnamefont {C.}~\bibnamefont {Kokail}}, \bibinfo {author} {\bibfnamefont
  {N.}~\bibnamefont {Maskara}}, \bibinfo {author} {\bibfnamefont {E.~C.}\
  \bibnamefont {Trapp}}, \bibinfo {author} {\bibfnamefont {L.~M.}\ \bibnamefont
  {Stewart}}, \bibinfo {author} {\bibfnamefont {S.}~\bibnamefont {Hollerith}},
  \bibinfo {author} {\bibfnamefont {H.}~\bibnamefont {Zhou}}, \bibinfo {author}
  {\bibfnamefont {M.~J.}\ \bibnamefont {Gullans}}, \bibinfo {author}
  {\bibfnamefont {S.~F.}\ \bibnamefont {Yelin}}, \bibinfo {author}
  {\bibfnamefont {M.}~\bibnamefont {Greiner}}, \bibinfo {author} {\bibfnamefont
  {V.}~\bibnamefont {Vuletic}}, \bibinfo {author} {\bibfnamefont
  {M.}~\bibnamefont {Cain}},\ and\ \bibinfo {author} {\bibfnamefont {M.~D.}\
  \bibnamefont {Lukin}},\ }\href {https://arxiv.org/abs/2506.20661} {\bibinfo
  {title} {Architectural mechanisms of a universal fault-tolerant quantum
  computer}} (\bibinfo {year} {2025}),\ \Eprint
  {https://arxiv.org/abs/2506.20661} {arXiv:2506.20661 [quant-ph]} \BibitemShut
  {NoStop}%
\bibitem [{\citenamefont {Manetsch}\ \emph {et~al.}(2025)\citenamefont
  {Manetsch}, \citenamefont {Nomura}, \citenamefont {Bataille}, \citenamefont
  {Lv}, \citenamefont {Leung},\ and\ \citenamefont
  {Endres}}]{manetsch2025tweezer}%
  \BibitemOpen
  \bibfield  {author} {\bibinfo {author} {\bibfnamefont {H.~J.}\ \bibnamefont
  {Manetsch}}, \bibinfo {author} {\bibfnamefont {G.}~\bibnamefont {Nomura}},
  \bibinfo {author} {\bibfnamefont {E.}~\bibnamefont {Bataille}}, \bibinfo
  {author} {\bibfnamefont {X.}~\bibnamefont {Lv}}, \bibinfo {author}
  {\bibfnamefont {K.~H.}\ \bibnamefont {Leung}},\ and\ \bibinfo {author}
  {\bibfnamefont {M.}~\bibnamefont {Endres}},\ }\bibfield  {title} {\bibinfo
  {title} {A tweezer array with 6100 highly coherent atomic qubits},\
  }\href@noop {} {\bibfield  {journal} {\bibinfo  {journal} {Nature}\ ,\
  \bibinfo {pages} {1}} (\bibinfo {year} {2025})}\BibitemShut {NoStop}%
\bibitem [{\citenamefont {Li}\ \emph {et~al.}(2025)\citenamefont {Li},
  \citenamefont {Bao}, \citenamefont {Peper}, \citenamefont {Li},\ and\
  \citenamefont {Thompson}}]{li2025fast}%
  \BibitemOpen
  \bibfield  {author} {\bibinfo {author} {\bibfnamefont {Y.}~\bibnamefont
  {Li}}, \bibinfo {author} {\bibfnamefont {Y.}~\bibnamefont {Bao}}, \bibinfo
  {author} {\bibfnamefont {M.}~\bibnamefont {Peper}}, \bibinfo {author}
  {\bibfnamefont {C.}~\bibnamefont {Li}},\ and\ \bibinfo {author}
  {\bibfnamefont {J.~D.}\ \bibnamefont {Thompson}},\ }\bibfield  {title}
  {\bibinfo {title} {Fast, continuous and coherent atom replacement in a
  neutral atom qubit array},\ }\href@noop {} {\bibfield  {journal} {\bibinfo
  {journal} {arXiv preprint arXiv:2506.15633}\ } (\bibinfo {year}
  {2025})}\BibitemShut {NoStop}%
\bibitem [{\citenamefont {Chiu}\ \emph {et~al.}(2025)\citenamefont {Chiu},
  \citenamefont {Trapp}, \citenamefont {Guo}, \citenamefont {Abobeih},
  \citenamefont {Stewart}, \citenamefont {Hollerith}, \citenamefont
  {Stroganov}, \citenamefont {Kalinowski}, \citenamefont {Geim}, \citenamefont
  {Evered} \emph {et~al.}}]{chiu2025continuous}%
  \BibitemOpen
  \bibfield  {author} {\bibinfo {author} {\bibfnamefont {N.-C.}\ \bibnamefont
  {Chiu}}, \bibinfo {author} {\bibfnamefont {E.~C.}\ \bibnamefont {Trapp}},
  \bibinfo {author} {\bibfnamefont {J.}~\bibnamefont {Guo}}, \bibinfo {author}
  {\bibfnamefont {M.~H.}\ \bibnamefont {Abobeih}}, \bibinfo {author}
  {\bibfnamefont {L.~M.}\ \bibnamefont {Stewart}}, \bibinfo {author}
  {\bibfnamefont {S.}~\bibnamefont {Hollerith}}, \bibinfo {author}
  {\bibfnamefont {P.~L.}\ \bibnamefont {Stroganov}}, \bibinfo {author}
  {\bibfnamefont {M.}~\bibnamefont {Kalinowski}}, \bibinfo {author}
  {\bibfnamefont {A.~A.}\ \bibnamefont {Geim}}, \bibinfo {author}
  {\bibfnamefont {S.~J.}\ \bibnamefont {Evered}}, \emph {et~al.},\ }\bibfield
  {title} {\bibinfo {title} {Continuous operation of a coherent 3,000-qubit
  system},\ }\href@noop {} {\bibfield  {journal} {\bibinfo  {journal} {Nature}\
  ,\ \bibinfo {pages} {1}} (\bibinfo {year} {2025})}\BibitemShut {NoStop}%
\bibitem [{\citenamefont {Piroli}\ \emph {et~al.}(2024)\citenamefont {Piroli},
  \citenamefont {Styliaris},\ and\ \citenamefont {Cirac}}]{Piroli2024_Dicke}%
  \BibitemOpen
  \bibfield  {author} {\bibinfo {author} {\bibfnamefont {L.}~\bibnamefont
  {Piroli}}, \bibinfo {author} {\bibfnamefont {G.}~\bibnamefont {Styliaris}},\
  and\ \bibinfo {author} {\bibfnamefont {J.~I.}\ \bibnamefont {Cirac}},\
  }\bibfield  {title} {\bibinfo {title} {Approximating many-body quantum states
  with quantum circuits and measurements},\ }\href
  {https://doi.org/10.1103/PhysRevLett.133.230401} {\bibfield  {journal}
  {\bibinfo  {journal} {Phys. Rev. Lett.}\ }\textbf {\bibinfo {volume} {133}},\
  \bibinfo {pages} {230401} (\bibinfo {year} {2024})}\BibitemShut {NoStop}%
\bibitem [{\citenamefont {Yu}\ \emph {et~al.}(2024)\citenamefont {Yu},
  \citenamefont {Muleady}, \citenamefont {Wang}, \citenamefont {Schine},
  \citenamefont {Gorshkov},\ and\ \citenamefont {Childs}}]{Yu2024_Dicke}%
  \BibitemOpen
  \bibfield  {author} {\bibinfo {author} {\bibfnamefont {J.}~\bibnamefont
  {Yu}}, \bibinfo {author} {\bibfnamefont {S.~R.}\ \bibnamefont {Muleady}},
  \bibinfo {author} {\bibfnamefont {Y.-X.}\ \bibnamefont {Wang}}, \bibinfo
  {author} {\bibfnamefont {N.}~\bibnamefont {Schine}}, \bibinfo {author}
  {\bibfnamefont {A.~V.}\ \bibnamefont {Gorshkov}},\ and\ \bibinfo {author}
  {\bibfnamefont {A.~M.}\ \bibnamefont {Childs}},\ }\href
  {https://arxiv.org/abs/2411.03428} {\bibinfo {title} {Efficient preparation
  of dicke states}} (\bibinfo {year} {2024}),\ \Eprint
  {https://arxiv.org/abs/2411.03428} {arXiv:2411.03428 [quant-ph]} \BibitemShut
  {NoStop}%
\bibitem [{\citenamefont {Moessner}\ and\ \citenamefont
  {Sondhi}(2001)}]{moessner2001_rvb}%
  \BibitemOpen
  \bibfield  {author} {\bibinfo {author} {\bibfnamefont {R.}~\bibnamefont
  {Moessner}}\ and\ \bibinfo {author} {\bibfnamefont {S.~L.}\ \bibnamefont
  {Sondhi}},\ }\bibfield  {title} {\bibinfo {title} {Resonating valence bond
  phase in the triangular lattice quantum dimer model},\ }\href
  {https://doi.org/10.1103/PhysRevLett.86.1881} {\bibfield  {journal} {\bibinfo
   {journal} {Phys. Rev. Lett.}\ }\textbf {\bibinfo {volume} {86}},\ \bibinfo
  {pages} {1881} (\bibinfo {year} {2001})}\BibitemShut {NoStop}%
\bibitem [{\citenamefont {Misguich}\ \emph {et~al.}(2002)\citenamefont
  {Misguich}, \citenamefont {Serban},\ and\ \citenamefont
  {Pasquier}}]{misguich2002_rvb}%
  \BibitemOpen
  \bibfield  {author} {\bibinfo {author} {\bibfnamefont {G.}~\bibnamefont
  {Misguich}}, \bibinfo {author} {\bibfnamefont {D.}~\bibnamefont {Serban}},\
  and\ \bibinfo {author} {\bibfnamefont {V.}~\bibnamefont {Pasquier}},\
  }\bibfield  {title} {\bibinfo {title} {Quantum dimer model on the kagome
  lattice: Solvable dimer-liquid and ising gauge theory},\ }\href
  {https://doi.org/10.1103/PhysRevLett.89.137202} {\bibfield  {journal}
  {\bibinfo  {journal} {Phys. Rev. Lett.}\ }\textbf {\bibinfo {volume} {89}},\
  \bibinfo {pages} {137202} (\bibinfo {year} {2002})}\BibitemShut {NoStop}%
\bibitem [{\citenamefont {Salberger}\ and\ \citenamefont
  {Korepin}(2018)}]{salberger2018_fredkin}%
  \BibitemOpen
  \bibfield  {author} {\bibinfo {author} {\bibfnamefont {O.}~\bibnamefont
  {Salberger}}\ and\ \bibinfo {author} {\bibfnamefont {V.}~\bibnamefont
  {Korepin}},\ }\bibfield  {title} {\bibinfo {title} {Fredkin spin chain},\
  }in\ \href@noop {} {\emph {\bibinfo {booktitle} {Ludwig Faddeev Memorial
  Volume: A Life In Mathematical Physics}}}\ (\bibinfo  {publisher} {World
  Scientific},\ \bibinfo {year} {2018})\ pp.\ \bibinfo {pages}
  {439--458}\BibitemShut {NoStop}%
\bibitem [{\citenamefont {Zhu}\ and\ \citenamefont {Zhang}(2019)}]{zhu:2019}%
  \BibitemOpen
  \bibfield  {author} {\bibinfo {author} {\bibfnamefont {G.-Y.}\ \bibnamefont
  {Zhu}}\ and\ \bibinfo {author} {\bibfnamefont {G.-M.}\ \bibnamefont
  {Zhang}},\ }\bibfield  {title} {\bibinfo {title} {Gapless coulomb state
  emerging from a self-dual topological tensor-network state},\ }\href
  {https://doi.org/10.1103/PhysRevLett.122.176401} {\bibfield  {journal}
  {\bibinfo  {journal} {Phys. Rev. Lett.}\ }\textbf {\bibinfo {volume} {122}},\
  \bibinfo {pages} {176401} (\bibinfo {year} {2019})}\BibitemShut {NoStop}%
\bibitem [{\citenamefont {Zhu}\ \emph {et~al.}(2023)\citenamefont {Zhu},
  \citenamefont {Chen}, \citenamefont {Ye},\ and\ \citenamefont
  {Trebst}}]{zhu:2023}%
  \BibitemOpen
  \bibfield  {author} {\bibinfo {author} {\bibfnamefont {G.-Y.}\ \bibnamefont
  {Zhu}}, \bibinfo {author} {\bibfnamefont {J.-Y.}\ \bibnamefont {Chen}},
  \bibinfo {author} {\bibfnamefont {P.}~\bibnamefont {Ye}},\ and\ \bibinfo
  {author} {\bibfnamefont {S.}~\bibnamefont {Trebst}},\ }\bibfield  {title}
  {\bibinfo {title} {Topological fracton quantum phase transitions by tuning
  exact tensor network states},\ }\href
  {https://doi.org/10.1103/PhysRevLett.130.216704} {\bibfield  {journal}
  {\bibinfo  {journal} {Phys. Rev. Lett.}\ }\textbf {\bibinfo {volume} {130}},\
  \bibinfo {pages} {216704} (\bibinfo {year} {2023})}\BibitemShut {NoStop}%
\bibitem [{\citenamefont {Xu}\ \emph {et~al.}(2020)\citenamefont {Xu},
  \citenamefont {Zhang},\ and\ \citenamefont {Zhang}}]{xu:2020}%
  \BibitemOpen
  \bibfield  {author} {\bibinfo {author} {\bibfnamefont {W.-T.}\ \bibnamefont
  {Xu}}, \bibinfo {author} {\bibfnamefont {Q.}~\bibnamefont {Zhang}},\ and\
  \bibinfo {author} {\bibfnamefont {G.-M.}\ \bibnamefont {Zhang}},\ }\bibfield
  {title} {\bibinfo {title} {Tensor network approach to phase transitions of a
  non-abelian topological phase},\ }\href
  {https://doi.org/10.1103/PhysRevLett.124.130603} {\bibfield  {journal}
  {\bibinfo  {journal} {Phys. Rev. Lett.}\ }\textbf {\bibinfo {volume} {124}},\
  \bibinfo {pages} {130603} (\bibinfo {year} {2020})}\BibitemShut {NoStop}%
\bibitem [{\citenamefont {Liu}\ \emph {et~al.}(2024)\citenamefont {Liu},
  \citenamefont {Shtengel},\ and\ \citenamefont {Pollmann}}]{set_isotns}%
  \BibitemOpen
  \bibfield  {author} {\bibinfo {author} {\bibfnamefont {Y.-J.}\ \bibnamefont
  {Liu}}, \bibinfo {author} {\bibfnamefont {K.}~\bibnamefont {Shtengel}},\ and\
  \bibinfo {author} {\bibfnamefont {F.}~\bibnamefont {Pollmann}},\ }\bibfield
  {title} {\bibinfo {title} {Simulating two-dimensional topological quantum
  phase transitions on a digital quantum computer},\ }\href
  {https://doi.org/10.1103/PhysRevResearch.6.043256} {\bibfield  {journal}
  {\bibinfo  {journal} {Phys. Rev. Res.}\ }\textbf {\bibinfo {volume} {6}},\
  \bibinfo {pages} {043256} (\bibinfo {year} {2024})}\BibitemShut {NoStop}%
\bibitem [{\citenamefont {Boesl}\ \emph
  {et~al.}(2025{\natexlab{a}})\citenamefont {Boesl}, \citenamefont {Liu},
  \citenamefont {Xu}, \citenamefont {Pollmann},\ and\ \citenamefont
  {Knap}}]{Boesl2025}%
  \BibitemOpen
  \bibfield  {author} {\bibinfo {author} {\bibfnamefont {J.}~\bibnamefont
  {Boesl}}, \bibinfo {author} {\bibfnamefont {Y.-J.}\ \bibnamefont {Liu}},
  \bibinfo {author} {\bibfnamefont {W.-T.}\ \bibnamefont {Xu}}, \bibinfo
  {author} {\bibfnamefont {F.}~\bibnamefont {Pollmann}},\ and\ \bibinfo
  {author} {\bibfnamefont {M.}~\bibnamefont {Knap}},\ }\bibfield  {title}
  {\bibinfo {title} {Quantum phase transitions between symmetry-enriched
  fracton phases},\ }\href {https://doi.org/10.1103/fsr8-xd4n} {\bibfield
  {journal} {\bibinfo  {journal} {Phys. Rev. B}\ }\textbf {\bibinfo {volume}
  {112}},\ \bibinfo {pages} {125152} (\bibinfo {year}
  {2025}{\natexlab{a}})}\BibitemShut {NoStop}%
\bibitem [{\citenamefont {Boesl}\ \emph
  {et~al.}(2025{\natexlab{b}})\citenamefont {Boesl}, \citenamefont {Liu},
  \citenamefont {Pollmann},\ and\ \citenamefont
  {Knap}}]{Boesl:2025_stringnetiso}%
  \BibitemOpen
  \bibfield  {author} {\bibinfo {author} {\bibfnamefont {J.}~\bibnamefont
  {Boesl}}, \bibinfo {author} {\bibfnamefont {Y.-J.}\ \bibnamefont {Liu}},
  \bibinfo {author} {\bibfnamefont {F.}~\bibnamefont {Pollmann}},\ and\
  \bibinfo {author} {\bibfnamefont {M.}~\bibnamefont {Knap}},\ }\bibfield
  {title} {\bibinfo {title} {Skeleton of isometric tensor network states for
  abelian string-net models},\ }\href {https://arxiv.org/abs/2511.13821}
  {\bibfield  {journal} {\bibinfo  {journal} {arXiv:2511.13821}\ } (\bibinfo
  {year} {2025}{\natexlab{b}})}\BibitemShut {NoStop}%
\bibitem [{\citenamefont {Fern{\'a}ndez-Gonz{\'a}lez}\ \emph
  {et~al.}(2015)\citenamefont {Fern{\'a}ndez-Gonz{\'a}lez}, \citenamefont
  {Schuch}, \citenamefont {Wolf}, \citenamefont {Cirac},\ and\ \citenamefont
  {Perez-Garcia}}]{uncle_h}%
  \BibitemOpen
  \bibfield  {author} {\bibinfo {author} {\bibfnamefont {C.}~\bibnamefont
  {Fern{\'a}ndez-Gonz{\'a}lez}}, \bibinfo {author} {\bibfnamefont
  {N.}~\bibnamefont {Schuch}}, \bibinfo {author} {\bibfnamefont {M.~M.}\
  \bibnamefont {Wolf}}, \bibinfo {author} {\bibfnamefont {J.~I.}\ \bibnamefont
  {Cirac}},\ and\ \bibinfo {author} {\bibfnamefont {D.}~\bibnamefont
  {Perez-Garcia}},\ }\bibfield  {title} {\bibinfo {title} {Frustration free
  gapless hamiltonians for matrix product states},\ }\href@noop {} {\bibfield
  {journal} {\bibinfo  {journal} {Communications in Mathematical Physics}\
  }\textbf {\bibinfo {volume} {333}},\ \bibinfo {pages} {299} (\bibinfo {year}
  {2015})}\BibitemShut {NoStop}%
\bibitem [{\citenamefont {Somma}\ and\ \citenamefont
  {Boixo}(2013)}]{somma2013_gapAmp}%
  \BibitemOpen
  \bibfield  {author} {\bibinfo {author} {\bibfnamefont {R.~D.}\ \bibnamefont
  {Somma}}\ and\ \bibinfo {author} {\bibfnamefont {S.}~\bibnamefont {Boixo}},\
  }\bibfield  {title} {\bibinfo {title} {{Spectral Gap Amplification}},\ }\href
  {https://doi.org/10.1137/120871997} {\bibfield  {journal} {\bibinfo
  {journal} {SIAM Journal on Computing}\ }\textbf {\bibinfo {volume} {42}},\
  \bibinfo {pages} {593} (\bibinfo {year} {2013})},\ \Eprint
  {https://arxiv.org/abs/https://doi.org/10.1137/120871997}
  {https://doi.org/10.1137/120871997} \BibitemShut {NoStop}%
\bibitem [{\citenamefont {Low}\ and\ \citenamefont
  {Chuang}(2017)}]{low2017_gapAmp}%
  \BibitemOpen
  \bibfield  {author} {\bibinfo {author} {\bibfnamefont {G.~H.}\ \bibnamefont
  {Low}}\ and\ \bibinfo {author} {\bibfnamefont {I.~L.}\ \bibnamefont
  {Chuang}},\ }\bibfield  {title} {\bibinfo {title} {Hamiltonian simulation by
  uniform spectral amplification},\ }\href@noop {} {\bibfield  {journal}
  {\bibinfo  {journal} {arXiv preprint arXiv:1707.05391}\ } (\bibinfo {year}
  {2017})}\BibitemShut {NoStop}%
\bibitem [{\citenamefont {Hauschild}\ and\ \citenamefont
  {Pollmann}(2018)}]{hauschild2018_tenpy}%
  \BibitemOpen
  \bibfield  {author} {\bibinfo {author} {\bibfnamefont {J.}~\bibnamefont
  {Hauschild}}\ and\ \bibinfo {author} {\bibfnamefont {F.}~\bibnamefont
  {Pollmann}},\ }\bibfield  {title} {\bibinfo {title} {{Efficient numerical
  simulations with Tensor Networks: Tensor Network Python (TeNPy)}},\ }\href
  {https://doi.org/10.21468/SciPostPhysLectNotes.5} {\bibfield  {journal}
  {\bibinfo  {journal} {SciPost Phys. Lect. Notes}\ ,\ \bibinfo {pages} {5}}
  (\bibinfo {year} {2018})}\BibitemShut {NoStop}%
\bibitem [{\citenamefont {D{\"o}rstel}\ \emph {et~al.}(2025)\citenamefont
  {D{\"o}rstel}, \citenamefont {Iadecola}, \citenamefont {Wilson},\ and\
  \citenamefont {Buchhold}}]{dorstel2025frustration}%
  \BibitemOpen
  \bibfield  {author} {\bibinfo {author} {\bibfnamefont {T.}~\bibnamefont
  {D{\"o}rstel}}, \bibinfo {author} {\bibfnamefont {T.}~\bibnamefont
  {Iadecola}}, \bibinfo {author} {\bibfnamefont {J.}~\bibnamefont {Wilson}},\
  and\ \bibinfo {author} {\bibfnamefont {M.}~\bibnamefont {Buchhold}},\
  }\bibfield  {title} {\bibinfo {title} {Frustration-free control and
  absorbing-state transport in entangled state preparation},\ }\href@noop {}
  {\bibfield  {journal} {\bibinfo  {journal} {arXiv preprint arXiv:2510.24845}\
  } (\bibinfo {year} {2025})}\BibitemShut {NoStop}%
\bibitem [{\citenamefont {Shtanko}\ \emph {et~al.}(2025)\citenamefont
  {Shtanko}, \citenamefont {Liu}, \citenamefont {Lieu}, \citenamefont
  {Gorshkov},\ and\ \citenamefont {Albert}}]{Shtanko2025boundsautonomous}%
  \BibitemOpen
  \bibfield  {author} {\bibinfo {author} {\bibfnamefont {O.}~\bibnamefont
  {Shtanko}}, \bibinfo {author} {\bibfnamefont {Y.-J.}\ \bibnamefont {Liu}},
  \bibinfo {author} {\bibfnamefont {S.}~\bibnamefont {Lieu}}, \bibinfo {author}
  {\bibfnamefont {A.~V.}\ \bibnamefont {Gorshkov}},\ and\ \bibinfo {author}
  {\bibfnamefont {V.~V.}\ \bibnamefont {Albert}},\ }\bibfield  {title}
  {\bibinfo {title} {Bounds on {A}utonomous {Q}uantum {E}rror {C}orrection},\
  }\href {https://doi.org/10.22331/q-2025-07-22-1804} {\bibfield  {journal}
  {\bibinfo  {journal} {{Quantum}}\ }\textbf {\bibinfo {volume} {9}},\ \bibinfo
  {pages} {1804} (\bibinfo {year} {2025})}\BibitemShut {NoStop}%
\bibitem [{\citenamefont {Dennis}\ \emph {et~al.}(2002)\citenamefont {Dennis},
  \citenamefont {Kitaev}, \citenamefont {Landahl},\ and\ \citenamefont
  {Preskill}}]{Topo_quantum_memory_2002}%
  \BibitemOpen
  \bibfield  {author} {\bibinfo {author} {\bibfnamefont {E.}~\bibnamefont
  {Dennis}}, \bibinfo {author} {\bibfnamefont {A.}~\bibnamefont {Kitaev}},
  \bibinfo {author} {\bibfnamefont {A.}~\bibnamefont {Landahl}},\ and\ \bibinfo
  {author} {\bibfnamefont {J.}~\bibnamefont {Preskill}},\ }\bibfield  {title}
  {\bibinfo {title} {{Topological quantum memory}},\ }\href
  {https://doi.org/10.1063/1.1499754} {\bibfield  {journal} {\bibinfo
  {journal} {Journal of Mathematical Physics}\ }\textbf {\bibinfo {volume}
  {43}},\ \bibinfo {pages} {4452} (\bibinfo {year} {2002})}\BibitemShut
  {NoStop}%
\bibitem [{\citenamefont {Tantivasadakarn}\ \emph {et~al.}(2023)\citenamefont
  {Tantivasadakarn}, \citenamefont {Vishwanath},\ and\ \citenamefont
  {Verresen}}]{nat:2023}%
  \BibitemOpen
  \bibfield  {author} {\bibinfo {author} {\bibfnamefont {N.}~\bibnamefont
  {Tantivasadakarn}}, \bibinfo {author} {\bibfnamefont {A.}~\bibnamefont
  {Vishwanath}},\ and\ \bibinfo {author} {\bibfnamefont {R.}~\bibnamefont
  {Verresen}},\ }\bibfield  {title} {\bibinfo {title} {Hierarchy of topological
  order from finite-depth unitaries, measurement, and feedforward},\ }\href
  {https://doi.org/10.1103/PRXQuantum.4.020339} {\bibfield  {journal} {\bibinfo
   {journal} {PRX Quantum}\ }\textbf {\bibinfo {volume} {4}},\ \bibinfo {pages}
  {020339} (\bibinfo {year} {2023})}\BibitemShut {NoStop}%
\bibitem [{\citenamefont {Dengis}\ \emph {et~al.}(2014)\citenamefont {Dengis},
  \citenamefont {König},\ and\ \citenamefont {Pastawski}}]{Dengis:2014}%
  \BibitemOpen
  \bibfield  {author} {\bibinfo {author} {\bibfnamefont {J.}~\bibnamefont
  {Dengis}}, \bibinfo {author} {\bibfnamefont {R.}~\bibnamefont {König}},\
  and\ \bibinfo {author} {\bibfnamefont {F.}~\bibnamefont {Pastawski}},\
  }\bibfield  {title} {\bibinfo {title} {An optimal dissipative encoder for the
  toric code},\ }\href {https://doi.org/10.1088/1367-2630/16/1/013023}
  {\bibfield  {journal} {\bibinfo  {journal} {New Journal of Physics}\ }\textbf
  {\bibinfo {volume} {16}},\ \bibinfo {pages} {013023} (\bibinfo {year}
  {2014})}\BibitemShut {NoStop}%
\bibitem [{\citenamefont {Wolf}\ \emph {et~al.}(2006)\citenamefont {Wolf},
  \citenamefont {Ortiz}, \citenamefont {Verstraete},\ and\ \citenamefont
  {Cirac}}]{wolf:2006}%
  \BibitemOpen
  \bibfield  {author} {\bibinfo {author} {\bibfnamefont {M.~M.}\ \bibnamefont
  {Wolf}}, \bibinfo {author} {\bibfnamefont {G.}~\bibnamefont {Ortiz}},
  \bibinfo {author} {\bibfnamefont {F.}~\bibnamefont {Verstraete}},\ and\
  \bibinfo {author} {\bibfnamefont {J.~I.}\ \bibnamefont {Cirac}},\ }\bibfield
  {title} {\bibinfo {title} {Quantum phase transitions in matrix product
  systems},\ }\href {https://doi.org/10.1103/PhysRevLett.97.110403} {\bibfield
  {journal} {\bibinfo  {journal} {Phys. Rev. Lett.}\ }\textbf {\bibinfo
  {volume} {97}},\ \bibinfo {pages} {110403} (\bibinfo {year}
  {2006})}\BibitemShut {NoStop}%
\end{thebibliography}%
\let\addcontentsline\oldaddcontentsline

\leavevmode
\newpage
\leavevmode
\newpage

\onecolumngrid
\begin{center}
\textbf{Supplemental Material:}\\
\textbf{Digital dissipative state preparation for frustration-free gapless quantum systems}\\ \vspace{10pt}
Johannes Feldmeier$^{1}$, Yu-Jie Liu$^{2}$, Mikhail Lukin$^{1}$, and Soonwon Choi$^{2}$ \\ \vspace{6pt}

$^1$\textit{\small{Physics Department, Harvard University, 17 Oxford St, Cambridge MA 02138, USA}} \\
$^2$\textit{\small{Center for Theoretical Physics - a Leinweber Institute, Massachusetts Institute of Technology, Cambridge, MA 02139, USA}}
\vspace{10pt}
\end{center}
\maketitle
\onecolumngrid


\tableofcontents

\section{Guaranteed convergence of the preparation protocol}

Consider a frustration-free Hamiltonian $H=\sum_iP_i$ given by local projectors. In this section, we show that a protocol consisting of parallel local measurements followed by global (or local) recovery generally converges to the ground state, provided the recovery step applies random Pauli operators. The upper bounds on the convergence time obtained from these general constructions can be exponentially large and are not intended to accurately estimate the faster scaling observed in the main text.

We first record a monotonicity property common to all the protocols. Let
\begin{equation}
    \Lambda[\rho]=\sum_{\textbf{v}}K_{\textbf{v}}\rho K_{\textbf{v}}^\dag
\end{equation}
be the CPTP map for one step of the protocol. To be general, for each step, we allow classical randomness in the choice of measurement layer (each layer consists of commuting projectors) and recovery. Let $\Psi$ be the projector onto the full ground-state subspace. An outcome string $\textbf{v}$ is called trivial if every measured projector has outcome zero. For a trivial outcome string,
\begin{equation}
    K_{\textbf{v}}^\dag\Psi K_{\textbf{v}}
    =|c_{\textbf{v}}|^2\Psi,
    \qquad
    \sum_{\textbf{v}\ {\rm trivial}}|c_{\textbf{v}}|^2=1,
\end{equation}
because every factor $1-P_i$ acts as the identity on the ground state; the coefficients $c_{\textbf{v}}$ denote the probabilities for selecting different layers to measure. We therefore have, for every density matrix $\rho$,
\begin{align}\label{sm:eq:nonincrease}
    \Tr(\Psi\Lambda[\rho])
    &=\Tr(\Psi\rho)
      +\sum_{\textbf{v}\ {\rm nontrivial}}
      \Tr\!\left(K_{\textbf{v}}^\dag\Psi K_{\textbf{v}}\rho\right)
      \nonumber\\
    &\geq\Tr(\Psi\rho).
\end{align}
The inequality is strict whenever at least one term in the sum is positive. If the protocol does not apply any recovery and involves only measurements of the projectors, each nontrivial Kraus operator annihilates the ground state, so the second term vanishes and the overlap remains constant under repeated application of $\Lambda$.

We will later repeatedly use the following observation. If a CPTP map $\Phi$ satisfies
\begin{equation}\label{sm:eq:generic_contraction}
    \Tr(\Psi\Phi[\rho])
    \geq\Tr(\Psi\rho)
    +q\bigl(1-\Tr(\Psi\rho)\bigr)
\end{equation}
for every $\rho$ and some $0<q\leq1$, then induction gives
\begin{equation}\label{sm:eq:generic_iteration}
    \Tr(\Psi\Phi^n[\rho_0])
    \geq1-\left(1-\Tr(\Psi\rho_0)\right)(1-q)^n.
\end{equation}
For $0<\epsilon<1-\Tr(\Psi\rho_0)$, a sufficient number of applications to reach infidelity $\epsilon$ is
\begin{equation}\label{sm:eq:generic_time}
    n=O\!\left(
    q^{-1}\log\frac{1-\Tr(\Psi\rho_0)}{\epsilon}\right).
\end{equation}
We will also use the following notation throughout the section. Let the projectors be partitioned into $\mathcal{A}$ layers $B_a$ such that all projectors within a layer $B_a$ commute. Set
\begin{equation}
    H_a=\sum_{i\in B_a}P_i,\qquad
    Q_a=\prod_{i\in B_a}(1-P_i),
\end{equation}
and define
\begin{equation}\label{sm:eq:layer_bound}
    G=\frac{1}{\mathcal{A}}\sum_a(1-Q_a),\qquad
    m_{\max}=\max_a|B_a|,\qquad
    c=\frac{1}{\mathcal{A}m_{\max}}.
\end{equation}
Note that $m_{\max}\leq N$, where $N$ is the total number of qubits.
Because the projectors within $B_a$ commute, they can be diagonalized simultaneously. On a common eigenvector, $1-Q_a$ is the indicator of at least one violation, whereas $H_a$ is the number of violations. Hence
\begin{equation}
    1-Q_a\ \succeq\ \frac{H_a}{|B_a|}
    \ \succeq\ \frac{H_a}{m_{\max}},
\end{equation}    
and therefore $G\succeq cH$.

\subsection{Convergence with a global random recovery}\label{sm:sec:global_random}
Consider the following protocol: at each step, choose one layer $B_a$ uniformly, measure all its projectors, and, following any nontrivial outcome, apply a uniformly random Pauli string to all $N$ qubits. This differs from a round of the main protocol, which applies all layers in a fixed order; here a step contains only one randomly selected layer. The two protocols, however, share the same qualitative behavior (see an explicit analysis in Sec.~\ref{sm:sec:single_particle}). Let $\mathcal{P}_{a,\textbf{v}}$ be the projector onto outcome string $\textbf{v}$ in layer $a$, so $\mathcal{P}_{a,\textbf{0}}=Q_a$. The Pauli-twirling identity is, for any $2^N\times 2^N$ matrix $B$,
\begin{equation}
    \frac{1}{4^N}\sum_\alpha \sigma_\alpha B\sigma_\alpha=\frac{\Tr(B)}{2^N}\mathbbm{1},
\end{equation}
where $\sigma_\alpha$ runs over all $N$-qubit Pauli strings. Let $\Psi$ be the projector onto the full ground-state subspace and let $r=\Tr(\Psi)$ be its dimension. Twirling each nontrivial branch gives the exact relation
\begin{align}
    \Tr(\Psi\Lambda[\rho])
    &=\Tr(\Psi\rho)+\sum_a \frac{1}{\mathcal{A}}\sum_{\textbf{v}\neq\textbf{0}}\frac{1}{4^N}\sum_\alpha
    \Tr\!\left(\Psi\sigma_\alpha\mathcal{P}_{a,\textbf{v}}\rho
    \mathcal{P}_{a,\textbf{v}}\sigma_\alpha\right) \nonumber\\
    &=\Tr(\Psi\rho)+\frac{r}{2^N}\Tr(G\rho)
      \geq\Tr(\Psi\rho)+\frac{rc}{2^N}\Tr(H\rho). \label{sm:eq:global_random}
\end{align}
The first term is the contribution of the no-violation branches, while $\sum_{\textbf{v}\neq\textbf{0}}\mathcal{P}_{a,\textbf{v}}=1-Q_a$ gives the second line. If $\Delta$ is the gap above the full ground-state subspace, since $H$ is a sum of projectors and is frustration-free,
\begin{equation}
    \Tr(H\rho)\geq\Delta\bigl(1-\Tr(\Psi\rho)\bigr). \label{sm:eq:ff_energy}
\end{equation}
Combining Eq.~\eqref{sm:eq:global_random} and Eq.~\eqref{sm:eq:ff_energy},
\begin{equation}
    \Tr(\Psi\Lambda[\rho])\geq
    \left(1-\frac{rc\Delta}{2^N}\right)\Tr(\Psi\rho)
    +\frac{rc\Delta}{2^N}.
\end{equation}
This is Eq.~\eqref{sm:eq:generic_contraction} with $q=rc\Delta/2^N$.
The factor $2^N/r$ makes the convergence time upper bound generally exponential unless the ground-space degeneracy compensates for it. 

\subsection{Convergence with a local random recovery}
Consider the following protocol: at each step, choose one layer $B_a$ uniformly, measure all its projectors, and, following any nontrivial outcome, independently apply uniformly random Pauli strings to the support of each violated local projector. Our goal is to show that this protocol converges under mild assumptions.

We present the argument for a periodic chain of $N$ qubits. The same argument extends to bounded-degree higher-dimensional lattices. We assume that $H=\sum_iP_i$ has a unique ground state $\ket{GS}$. We show convergence of the protocol under the following assumption: Let the two-qubit projectors $P_e$ live on the edges $e=\{u,v\}$ of a connected interaction graph (which is very simple in 1D, we frame it like this for straightforward generalization), and assume
\begin{equation}\label{sm:eq:general_local_full_rank}
    \kappa_*\equiv
    \min_{\substack{e=\{u,v\}\\w=u,v}}
    \lambda_{\min}\!\left(\Tr_w P_e\right)>0.
\end{equation}
Here $w$ denotes either endpoint of the edge $e$, $\Tr_w$ is the partial trace over the qubit at that endpoint, and $\lambda_{\min}(B)$ denotes the smallest eigenvalue of the remaining one-qubit operator $B$. Thus, Eq.~\eqref{sm:eq:general_local_full_rank} requires both one-site partial traces of every edge projector to be positive definite, with $\kappa_*$ their smallest eigenvalue over all edges and both orientations. 

We show that this assumption holds, for example, when the ground state $\ket{GS}$ is not a product state, and $P_i$ is translational invariant, specified by a fixed two-qubit projector $P$ acting on nearest neighbor sites.
We prove this by contradiction. Let
\begin{equation}
    O_R\equiv\Tr_1(P),\qquad
    O_L\equiv\Tr_2(P),\qquad
    \kappa\equiv\min\{\lambda_{\min}(O_R),\lambda_{\min}(O_L)\}.
\end{equation}
If $\kappa=0$, we may exchange the two sites and assume without loss of generality that $O_R$ is singular. Choose a single-qubit state $\ket{\phi}$ in its kernel. Positivity of $P$ implies
\begin{equation}
    0=\bra{\phi}O_R\ket{\phi}
      =\sum_{a=0,1}\bra{a,\phi}P\ket{a,\phi}
      \quad\Longrightarrow\quad
      P\ket{a,\phi}=0\quad(a=0,1).
\end{equation}
In particular, $P\ket{\phi,\phi}=0$, so translation invariance makes $\ket{\phi}^{\otimes N}$ a ground state of $H$. By uniqueness, $\ket{GS}=\ket{\phi}^{\otimes N}$, which contradicts the assumption that the ground state is not a product state. So $\kappa>0$ and a translation-invariant chain with a nontrivial ground state naturally satisfies the assumption Eq.~\eqref{sm:eq:general_local_full_rank}.


Now we proceed to prove the convergence under the assumption Eq~\eqref{sm:eq:general_local_full_rank}. 
Set $Q_e=1-P_e$. The local measurement-and-recovery channel has Kraus operators
\begin{equation}\label{sm:eq:local_kraus}
    L_{e,0}=Q_e,
    \qquad
    L_{e,\alpha}=\frac{1}{4}\sigma_\alpha^{(e)}P_e,
    \quad \alpha=1,\ldots,16,
\end{equation}
where $\sigma_\alpha^{(e)}$ runs over all two-qubit Pauli strings on edge $e$. The label $0$ denotes the trivial measurement outcome, whereas $\alpha=1,\ldots,16$ denotes the nontrivial outcome followed by recovery $\sigma_\alpha^{(e)}$. Thus,
\begin{equation}\label{sm:eq:local_channel}
    \mathcal{K}_e[\rho]
    =\sum_{s=0}^{16}L_{e,s}\rho L_{e,s}^\dagger
    =Q_e\rho Q_e+\frac{1}{16}\sum_{\alpha=1}^{16}
    \sigma_\alpha^{(e)}P_e\rho P_e\sigma_\alpha^{(e)}.
\end{equation}
Because the edges within a layer have disjoint supports, a Kraus operator for one step, conditioned on choosing layer $a$, is
\begin{equation}\label{sm:eq:layer_kraus}
    K_{a,\mathbf{s}}
    =\frac{1}{\sqrt{\mathcal A}}
      \prod_{e\in B_a}L_{e,s_e},
    \qquad
    \Lambda[\rho]
    =\sum_{a,\mathbf{s}}K_{a,\mathbf{s}}\rho
      K_{a,\mathbf{s}}^\dagger.
\end{equation}
Here $\mathbf{s}=(s_e)_{e\in B_a}$ specifies the measurement and recovery result on every edge in the selected layer. The factor $1/\sqrt{\mathcal A}$ accounts for the uniform random choice of the layer.

For a block of $T$ rounds, let $\{W_\mu\}_\mu$ be the Kraus operators of $\Lambda^T$. A length-$T$ history is a sequence
\begin{equation}
    \mu=\bigl((a_1,\mathbf{s}_1),\ldots,
    (a_T,\mathbf{s}_T)\bigr),
\end{equation}
and its Kraus operator is the product
\begin{equation}\label{sm:eq:history_kraus}
    W_\mu
    =K_{a_T,\mathbf{s}_T}\cdots K_{a_1,\mathbf{s}_1}.
\end{equation}
Thus $\Lambda^T[\rho]=\sum_\mu W_\mu\rho W_\mu^\dagger$. A history is called nontrivial if at least one measured projector has the nontrivial outcome. We claim that one can choose $T=O(N)$ such that, for every edge $i$, there is a family $\mathfrak H_i$ of nontrivial histories and a number $\eta_N>0$, which may depend on the system size but not on $i$, for which
\begin{equation}\label{sm:eq:local_propagation}
    S_i\equiv
    \sum_{\mu\in\mathfrak H_i}W_\mu^\dagger\Psi W_\mu
    \succeq \eta_N P_i,
    \qquad \Psi=\ketbra{GS},
\end{equation}
where $\succeq$ denotes operator inequality: for semi-positive definite $A$ and $B$, $A\succeq B$ means $A-B$ is semi-positive definite.

We next define the family $\mathfrak H_i$ and use a tensor-network diagram to represent it and explain the proof. In the diagram below, each vertical black line is the worldline of one qubit, time runs from bottom to top, and each rectangle is one of the two-site operators $L_{e,s}$ in Eq.~\eqref{sm:eq:local_kraus}. The black line at the bottom is an arbitrary input on which $W_\mu$ acts. The bottom orange block is placed on the distinguished edge $i$. Moving upward, the orange blocks follow neighboring edges and form two propagation fronts that eventually cover the ring. For a general connected graph, one instead follows a walk on a spanning tree that visits every vertex, which requires $O(N)$ steps.
\begin{equation}
    \includegraphics[scale = 1]{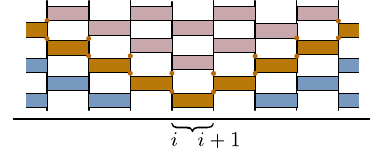}\nonumber
\end{equation}
The colors specify which histories are included in $\mathfrak H_i$. An orange block is constrained to have a nontrivial outcome, but all of its Pauli recovery labels are summed. For the convenience of explanation, we explicitly label the Pauli recovery following those nontrivial outcomes by orange circles, and the orange rectangular blocks represent the projectors $P$. A red block is constrained to have the Kraus operator $Q_e$ (i.e. trivial outcome). A blue block is unrestricted, and we sum over its complete set of measurement and recovery labels. We also fix a compatible sequence of layer choices; its nonzero probability $\mathcal A^{-T}$ is absorbed into $\eta_N$.

To evaluate $S_i$, attach $\Psi$ to the output of the diagram and contract the double-layer network for $W_\mu^\dagger\Psi W_\mu$. The red blocks adjacent to the output disappear because frustration freeness gives $Q_e\Psi Q_e=\Psi$. Now with the red blocks contracted, a subset of the Pauli recovery on top of the orange blocks directly acts on the ground state, summing over those Pauli recovery performs a global Pauli twirling and averages the ground-state projector $\Psi$ to the identity. This is diagrammatically shown as    
\begin{equation}
    \includegraphics[scale = 0.7]{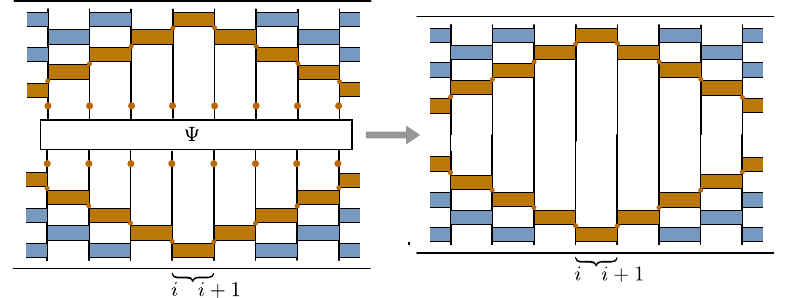}.\nonumber
\end{equation}
Now what is left is the Pauli recovery (orange circles) between the orange blocks. We then pull the orange projectors to the middle, and sum over the Pauli recovery sandwiching them one at a time, iterating this along the staircase. Note that summing the Pauli recovery at one site averages the operator
\begin{equation}
    \frac{1}{4}\sum_{\alpha=1}^{4}
    \sigma_\alpha B\sigma_\alpha
    =\frac{\mathbbm{1}_{i}}{2}
    \otimes\Tr_{i}(B).\label{sm:eq:one_site_twirl}
\end{equation}
With this relation, Eq.~\eqref{sm:eq:one_site_twirl} then replaces the projector on that edge by the appropriate one-site partial trace. The first two steps of this iterative procedure are depicted as follows:
\begin{equation}
    \includegraphics[scale = 0.7]{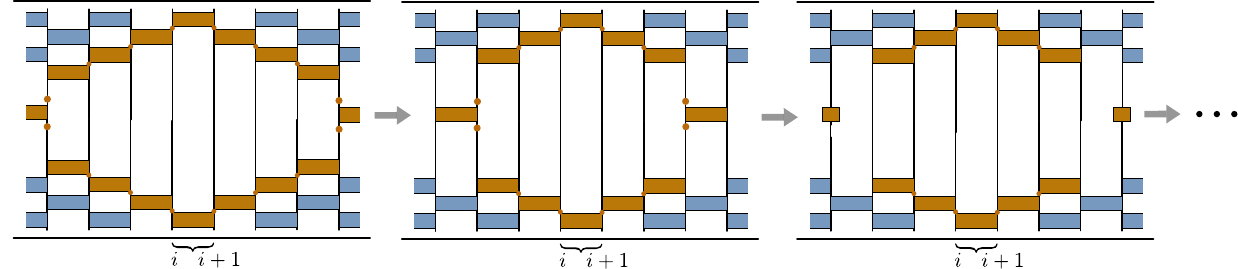}
\end{equation}
where the one-site orange box represents the one-site operator obtained after the partial trace. These one-site operators are either $O_R=\Tr_1(P)$ or $O_L=\Tr_2(P)$. After summing over all the orange Pauli recovery, we are left with a product of $P_i$, a string of those one-site operators and the blue blocks. More explicitly, the contribution for a fixed assignment $\gamma$ of the blue blocks has the form
\begin{equation}
    c_{i,N}\,B_\gamma^\dagger
    \left(P_i\prod_{v\in\mathcal V_i}R_v\right)B_\gamma,
    \qquad R_v\in\{O_L,O_R\},
\end{equation}
where $c_{i,N}>0$ contains the layer-selection, Kraus-normalization, and twirling factors, and the one-site operators $R_v$ have support away from the two endpoints of $i$. Hence
\begin{equation}
    P_i\prod_{v\in\mathcal V_i}R_v
    \succeq \kappa_*^{|\mathcal V_i|}P_i.
\end{equation}
The unrestricted blue blocks can then be contracted successively using $\sum_sL_{e,s}^\dagger L_{e,s}=\mathbbm{1}$, leaving only the operator $P_i$ multiplied by a non-zero prefactor. On a general graph, one performs the analogous contraction by following a spanning tree; at each step the new factor is one of the operators $\Tr_w(P_e)$ in Eq.~\eqref{sm:eq:general_local_full_rank}, and it may immediately be lower bounded by $\kappa_*\mathbbm{1}$ before the next contraction. Taking the smallest of the finitely many positive coefficients obtained for the different choices of $i$ defines an $i$-independent $\eta_N>0$ and proves Eq.~\eqref{sm:eq:local_propagation}. The coefficient contains $O(N)$ branch probabilities and powers of $\kappa_*$, so it can be exponentially small in $N$.

It remains to turn the propagation property Eq.~\eqref{sm:eq:local_propagation} into convergence. Summing $W_\mu^\dagger\Psi W_\mu$ over all histories with trivial outcome gives exactly $\Psi$: every $Q_e$ acts as the identity on the ground state, and the probabilities of all layer-choice sequences sum to one. Every nontrivial history contributes a positive-semidefinite operator. Therefore, Eq.~\eqref{sm:eq:local_propagation} implies, separately for every edge $i$,
\begin{equation}
    (\Lambda^T)^\dagger[\Psi]-\Psi
    \succeq\eta_NP_i.
\end{equation}
Averaging these $N$ inequalities gives
\begin{equation}\label{sm:eq:local_random}
    (\Lambda^T)^\dagger[\Psi]
    \succeq\Psi+\frac{\eta_N}{N}H.
\end{equation}
Combining Eq.~\eqref{sm:eq:local_random} and Eq.~\eqref{sm:eq:ff_energy}, we have for every density matrix $\rho$,
\begin{equation}
    \Tr(\Psi\Lambda^T[\rho])
    \geq\Tr(\Psi\rho)+\frac{\eta_N}{N}\Tr(H\rho)
    \geq\Tr(\Psi\rho)
      +\frac{\eta_N\Delta}{N}\bigl(1-\Tr(\Psi\rho)\bigr),
\end{equation}
where $\Delta$ is the spectral gap of $H$. Applying Eqs.~\eqref{sm:eq:generic_iteration} and \eqref{sm:eq:generic_time} to the block channel $\Phi=\Lambda^T$, with $q=\eta_N\Delta/N$, proves convergence under Eq.~\eqref{sm:eq:general_local_full_rank}. The resulting convergence-time bound is generally exponential in $N$, because $\eta_N$ is exponentially small.

As a separate locality consequence, each round of this protocol can be recast as a finite-depth local channel evolution. A Lieb--Robinson argument therefore gives a lower bound $T_c=\Omega(N^{1/d})$ for preparing a $d$-dimensional state with power-law or nonvanishing connected correlations from a product state.

As a side remark, we note that although the digital protocol bears similarities to the analog dissipative Lindbladian approach~\cite{verstraete2009_dissip}, the inherent temporal stochasticity of Lindbladian jumps can sometimes lead to subtly different dynamics compared to a digital implementation of the same jumps~\cite{Shtanko2025boundsautonomous}. Moreover, the digital protocol allows for additional optimization of the recovery procedure through classical communication of measurement outcomes, which in some cases enables the establishment of long-range entanglement at speeds unattainable by the Lindbladian approach with local jumps~\cite{Topo_quantum_memory_2002,nat:2023,Dengis:2014}. It therefore remains an interesting open question how the two approaches compare in general.

\subsection{Convergence of the random protocol with $U(1)$ symmetry}

A similar symmetry-preserving global recovery can be shown to converge for the RK-type examples considered in the main text. To state the required assumptions precisely, fix a $U(1)$ sector of $M=\sum_i Z_i$ and let $\mathcal{C}$ be a connected component of computational-basis configurations under the local moves generated by the $P_i$. We assume that every $P_i$ preserves
\begin{equation}
    \mathcal{H}_{\mathcal{C}}\equiv
    \operatorname{span}\{\ket{C}:C\in\mathcal{C}\},
\end{equation}
and that the restriction of $H$ to this subspace has the unique ground state
\begin{equation}
    \ket{GS}=\frac{1}{\sqrt{|\mathcal{C}|}}
    \sum_{C\in\mathcal{C}}\ket{C}.
\end{equation}
These are the properties of the connected RK sectors used in the examples. Let
\begin{equation}
    \Pi_{\mathcal{C}}=\sum_{C\in\mathcal{C}}\ketbra{C}
\end{equation}
denote the projector onto $\mathcal{H}_{\mathcal{C}}$. Since the $P_i$ are Hermitian and preserve this subspace, $[P_i,\Pi_{\mathcal{C}}]=0$.

Consider the local dephasing channels $\mathcal{Z}_i[\rho]=\frac{1}{2}(\rho+Z_i\rho Z_i)$ and their product $\mathcal{D}=\prod_i\mathcal{Z}_i$. Complete dephasing of the RK state gives
\begin{equation}
    \mathcal{D}[\Psi]
    =\frac{1}{|\mathcal{C}|}\sum_{C\in\mathcal{C}}\ketbra{C}
    =\frac{\Pi_{\mathcal{C}}}{|\mathcal{C}|},
    \qquad \Psi=\ketbra{GS}.
\end{equation}
We now use the protocol of Sec.~\ref{sm:sec:global_random}, but apply $\mathcal{D}$, rather than a uniformly random Pauli string, after every nontrivial outcome. Explicitly,
\begin{equation}
    \Lambda[\rho]=\frac{1}{\mathcal{A}}\sum_a\left(
    Q_a\rho Q_a+
    \sum_{\textbf{v}\neq\textbf{0}}
    \mathcal{D}\!\left[\mathcal{P}_{a,\textbf{v}}\rho
    \mathcal{P}_{a,\textbf{v}}\right]\right).
\end{equation}
The operators $G$ and the constant $c$ are those defined in Eq.~\eqref{sm:eq:layer_bound}. The dephasing channel is self-adjoint with respect to the Hilbert--Schmidt inner product. Therefore, for any $\rho$ supported on $\mathcal{H}_{\mathcal{C}}$,
\begin{align}
    \Tr(\Psi\Lambda[\rho])
    &=\Tr(\Psi\rho)
      +\frac{1}{\mathcal{A}}\sum_a\sum_{\textbf{v}\neq\textbf{0}}
      \Tr\!\left(\mathcal{D}[\Psi]
      \mathcal{P}_{a,\textbf{v}}\rho
      \mathcal{P}_{a,\textbf{v}}\right)\nonumber\\
    &=\Tr(\Psi\rho)+\frac{\Tr(G\rho)}{|\mathcal{C}|}
      \geq\Tr(\Psi\rho)
      +\frac{c\,\Tr(H\rho)}{|\mathcal{C}|}.
\end{align}
In the second line we used $[\Pi_{\mathcal{C}},\mathcal{P}_{a,\textbf{v}}]=0$, the support condition $\Pi_{\mathcal{C}}\rho=\rho$, and $\sum_{\textbf{v}\neq\textbf{0}}\mathcal{P}_{a,\textbf{v}}=1-Q_a$.
Let $\Delta_{\mathcal{C}}$ be the spectral gap of $H$ restricted to $\mathcal{H}_{\mathcal{C}}$. Uniqueness of the ground state in this subspace gives
\begin{equation}
    \Tr(H\rho)\geq
    \Delta_{\mathcal{C}}\bigl(1-\Tr(\Psi\rho)\bigr).
\end{equation}
Thus Eq.~\eqref{sm:eq:generic_contraction} holds on $\mathcal{H}_{\mathcal{C}}$ with $q=c\Delta_{\mathcal{C}}/|\mathcal{C}|$. The protocol therefore converges to $\ket{GS}$ from every initial state supported on this subspace.

\subsection{A rigorous bound for the single-particle case under deterministic correction}\label{sm:sec:single_particle}

We now keep the deterministic feedback $Z_i$ used in the main text and work in the single-particle sector of the one-dimensional ferromagnetic Heisenberg model. We first analyze the randomized-layer protocol, as considered in the earlier sections, that chooses the even or odd layer with equal probability at each step. We then relate it to fixed alternating layers for a translationally invariant initial state. The resulting bound is rigorous but is loose by a factor of $N$ compared with the observed convergence time. Let
\begin{align}
    \Lambda[\rho]&=\frac{1}{2}\bigl(\mathcal{E}[\rho]+\mathcal{O}[\rho]\bigr),\nonumber\\
    \mathcal{E}[\rho]&=\sum_{\textbf{v}_e}K_{e,\textbf{v}_e}\rho K_{e,\textbf{v}_e}^{\dag},
    \qquad
    \mathcal{O}[\rho]=\sum_{\textbf{v}_o}K_{o,\textbf{v}_o}\rho K_{o,\textbf{v}_o}^{\dag},\nonumber\\
    K_{a,\textbf{v}_a}&=\prod_{i\in a}R_i^{(v_i)},
    \qquad
    R_i^{(0)}=1-P_i,
    \qquad
    R_i^{(1)}=Z_iP_i,
\end{align}
where $a=e,o$ labels the even and odd layers and $\textbf{v}_a$ is the corresponding outcome string. The map $\Lambda$ is CPTP and invariant under translation by one site. Since $Z_i$ maps the singlet on bond $i$ to the orthogonal triplet, $P_iZ_iP_i=0$. The local Kraus operators therefore obey
\begin{equation}
    R_i^{(0)}R_i^{(0)}=R_i^{(0)},\qquad
    R_i^{(0)}R_i^{(1)}=R_i^{(1)},\qquad
    R_i^{(1)}R_i^{(0)}=R_i^{(1)}R_i^{(1)}=0.
\end{equation}
Consequently, the layer channels are idempotent, $\mathcal{E}^2=\mathcal{E}$ and $\mathcal{O}^2=\mathcal{O}$, and
\begin{equation}
    \Lambda^t = \frac{1}{2^t}(\mathcal{E}+\mathcal{O})^t = \frac{1}{2^t}\sum_{m = 1}^{t}\binom{t-1}{m-1}(\underbrace{\mathcal{E}\mathcal{O}\mathcal{E}\cdots}_m + \underbrace{\mathcal{O}\mathcal{E}\mathcal{O}\cdots}_m).
\end{equation}
This expansion groups all length-$t$ products according to the length $m$ of the alternating product obtained after contracting adjacent identical layers. For $t\in2\mathbb{N}$, either alternating product of length $m\leq t$ can be extended on the left to one of the two alternating products of length $t$. Monotonicity of the ground-state overlap under either layer, Eq.~\eqref{sm:eq:nonincrease}, therefore implies
\begin{equation}
    \Tr\!\left(\Psi(\mathcal{E}\mathcal{O})^{t/2}[\rho]\right)
    +\Tr\!\left(\Psi(\mathcal{O}\mathcal{E})^{t/2}[\rho]\right)
    \geq 2\Tr\!\left(\Psi\Lambda^t[\rho]\right).
\end{equation}
If $\rho$ is translationally invariant---for example, if it is obtained by averaging an initial state over all lattice translations---then translation by one site interchanges $\mathcal{E}$ and $\mathcal{O}$, while leaving $\Psi$ invariant. The two terms on the left-hand side are therefore equal, and
\begin{equation}
    \Tr\!\left(\Psi(\mathcal{E}\mathcal{O})^{t/2}[\rho]\right)
    \geq \Tr\!\left(\Psi\Lambda^t[\rho]\right),
\end{equation}
where $\mathcal{E}\mathcal{O}$ is one full round of a fixed alternating-layer protocol. 

It remains to bound the convergence of $\Lambda^t$. Let $\ket{j}$ denote the state with the single particle at site $j$. Within this sector,
\begin{equation}
    P_i=\ketbra{s_i},\qquad
    \ket{s_i}=\frac{\ket{i}-\ket{i+1}}{\sqrt{2}},\qquad
    Z_i\ket{s_i}=\ket{t_i}\equiv\frac{\ket{i}+\ket{i+1}}{\sqrt{2}},
\end{equation}
up to an irrelevant overall phase. Projectors on distinct bonds within the same layer have disjoint support and satisfy $P_iP_j=0$ in the single-particle sector. Thus, at most one projector in a measured layer can be violated. For an outcome with bond $i$ violated in layer $a$, the Kraus operator reduces to $Z_iP_i$, because $P_i\prod_{j\in a,\,j\neq i}(1-P_j)=P_i$. Moreover, let $\Psi = \ketbra{GS}$,
\begin{equation}
    P_iZ_i\Psi Z_i P_i=\Tr(P_iZ_i\Psi Z_i)P_i, = \left|\braket{GS|t_i}\right|^2P_i=\frac{2}{N}P_i,
\end{equation}
where $\ket{GS}=N^{-1/2}\sum_j\ket{j}$. As the even and odd layers are each selected with probability $1/2$, the increase in ground-state overlap is therefore exactly
\begin{align}
    \Tr(\Psi\Lambda[\rho])
    &=\Tr(\Psi\rho)
      +\frac{1}{2}\sum_i\Tr(\Psi Z_iP_i\rho P_iZ_i)\nonumber\\
    &=\Tr(\Psi\rho)+\frac{1}{N}\sum_i\Tr(P_i\rho)
     =\Tr(\Psi\rho)+\frac{1}{N}\Tr(H\rho).
\end{align}
Since,
\begin{equation}
    \Delta (1- \Tr(\Psi\rho)) \leq \Tr(H\rho),
\end{equation}
where $\Delta$ is the gap of $H$ in this sector. Hence,
\begin{equation}
     \Tr(\Psi\Lambda[\rho] )  \geq   \left(1-\frac{\Delta}{N}\right)\Tr(\Psi\rho) + \frac{\Delta}{N}.
\end{equation}
This is Eq.~\eqref{sm:eq:generic_contraction} with $q=\Delta/N$. Hence, for every initial state $\rho_0$, the randomized-layer channel obeys
\begin{equation}
    \Tr(\Psi\Lambda^t[\rho_0])
    \geq1-\left(1-\Tr(\Psi\rho_0)\right)
    \left(1-\frac{\Delta}{N}\right)^t.
\end{equation}
If $t$ is even and $\rho_0$ is translationally invariant, the fixed alternating protocol follows
\begin{equation}
     \Tr\!\left(\Psi(\mathcal{E
     }\mathcal{O})^{t/2}[\rho_0]\right)
     \geq \Tr(\Psi\Lambda^t[\rho_0])
     \geq 1-\left(1-\Tr(\Psi\rho_0)\right)
     \left(1-\frac{\Delta}{N}\right)^t.
\end{equation}
Thus, for the randomized-layer protocol, or for the fixed alternating protocol under the translation-invariance condition above, a target infidelity $\epsilon$ is reached after $O\left(\frac{N}{\Delta}
    \log\frac{1-\Tr(\Psi\rho_0)}{\epsilon}\right)$
layer applications. Again, we note that the resulting bound is rigorous but is loose by a factor of $N$ compared with the observed actual convergence time.

\section{Projection round dynamics: Symmetrized projection operator} \label{sec:AppProjectorTilde}

We recall the general definition of the projection round operator $\mathcal{P}$:
\begin{equation} \label{eq:proof1}
\mathcal{P} = \prod_{a=1}^{\mathcal{A}} \mathcal{P}_a = \prod_{a=1}^{\mathcal{A}} \prod_{i\in B_a} (1-P_{i}),
\end{equation}
where $P_{i}$ are the local projectors that make up the Hamiltonian, $H=\sum_i P_i$. The projectors are arranged into different sets $B_{a=1,\ldots,\mathcal{A}}$, within each of which $[P_{i},P_{i^\prime}]=0$ for $P_i,P_{i^\prime} \in B_a$. We assume in the following that $\mathcal{A}=2$, such that $\mathcal{P} = \mathcal{P}_1\mathcal{P}_2$ consists of two layers. This applies to the one-dimensional Heisenberg model (where the $P_i$ are singlet projectors on neighboring sites $i,i+1$) and the two-dimensional quantum dimer model considered in the main text. We show that the results of this section also apply to the single-particle sector of the Heisenberg model in dimensions $d>1$ (where $\mathcal{A}>2$) in Sec.~\ref{sec:SMsingleParticle}. We comment on a similar relation for the Fredkin chain (where $\mathcal{A}=3$) in Sec.~\ref{sec:AppFredkin}.

Ultimately, we are interested in the state $\mathcal{P}^\tau \ket{\psi}$ that results from repeated application of $\mathcal{P}$ to states $\ket{\psi}$ that are orthogonal to the ground state. In practice, however, we find that it is more convenient to work instead with a symmetrized projection round operator 
\begin{equation}
\tilde{\mathcal{P}} \equiv \frac{1}{2} \bigl( \mathcal{P} + \mathcal{P}^\dagger \bigr) = \frac{1}{2} \bigl( \mathcal{P}_1 \mathcal{P}_2 + \mathcal{P}_2 \mathcal{P}_1 \bigr),
\end{equation}
which, crucially, can be related to the Hamiltonian $H$ of the system at low energies (see Sec.~\ref{sec:SMsingleParticle} of this SM). Our goal is therefore to relate $\mathcal{P}^\tau \ket{\psi}$ to the properties of $\tilde{\mathcal{P}}^\tau \ket{\psi}$ instead. Under the assumption that $\mathcal{P}$ is diagonalizable, we show that $\mathcal{P}^\tau \approx \tilde{\mathcal{P}}^{4\tau/3}$, which we define in the following way:

Let $\ket{\tilde{\psi}}$ be an eigenstate of $\tilde{\mathcal{P}}$ with eigenvalue $\tilde{\lambda}_{\tilde{\psi}}$. We assume $\tilde{\lambda}_{\tilde{\psi}}$ is close to one, i.e., $0 < 1-\tilde{\lambda}_{\tilde{\psi}} \ll 1$. Intuitively, this means $\ket{\tilde{\psi}}$ is a low-energy state.
Then there is an eigenstate $\ket{\psi}$ of $\mathcal{P}$ with eigenvalue $\lambda_{\psi}$ such that:

\begin{enumerate}
\item \textit{Correspondence of eigenvalues:} $\lambda_{\psi} = \tilde{\lambda}_{\tilde{\psi}}^{4/3}$, up to small corrections of order $\mathcal{O}(|\log \tilde{\lambda}_{\tilde{\psi}}|^2)$.

\item \textit{Correspondence of eigenstates:} The overlap $\braket{\tilde{\psi}|\psi} = 1 -\mathcal{O}(|\log \tilde{\lambda}_{\tilde{\psi}}|)$ is large. 
\end{enumerate}
After rewriting the operator $\tilde{\mathcal{P}}^\tau$ in a convenient form in the following, we will subsequently show these points separately. We further confirm them numerically in \fig{fig:opDev}.

\subsection{Expanding $\tilde{\mathcal{P}}^\tau$ as a sum over operator strings}
We consider the repeated application of $\tilde{\mathcal{P}}$,
\begin{equation} \label{eq:symmProjOp2}
\tilde{\mathcal{P}}^\tau = \frac{1}{2^\tau} \sum_{ \bigl\{Q_n \in \{\mathcal{P},\mathcal{P}^\dagger\} \bigr\} } \; \prod_{n=1}^\tau Q_n,
\end{equation}
which is an average over all $2^\tau$ operator strings of length $\tau$ that can be formed by $\mathcal{P}$ and $\mathcal{P}^\dagger$.
Using the definition $\mathcal{P}=\mathcal{P}_1 \mathcal{P}_2$, $\mathcal{P}_a^2 = \mathcal{P}_a$, as well as $(\mathcal{P}^\dagger)^n = (\mathcal{P}_2\mathcal{P}_1)^n = \mathcal{P}_2 (\mathcal{P}_1\mathcal{P}_2)^{n-1} \mathcal{P}_1 = \mathcal{P}_2 \mathcal{P}^{n-1} \mathcal{P}_1$, we find the relation
\begin{equation} \label{eq:opStrings}
\mathcal{P} (\mathcal{P}^\dagger)^n \, \mathcal{P} = \mathcal{P}_1 \underbrace{\mathcal{P}_2 \, \mathcal{P}_2}_{=\mathcal{P}_2} \, \mathcal{P}^{n-1} \, \underbrace{\mathcal{P}_1 \, \mathcal{P}_1}_{=\mathcal{P}_1} \mathcal{P}_2 = \mathcal{P}^{n+1}.
\end{equation}
Following \eq{eq:opStrings}, when $\mathcal{P}$ and $\mathcal{P}^\dagger$ are neighbors in an operator string $\prod_{n=1}^\tau Q_n$ from \eq{eq:symmProjOp2}, they form a `domain wall', which can be removed in pairs. For each such pair, the length of the operator string is reduced by one according to \eq{eq:opStrings}. If there are $n_{DW}$ such domain walls in the string $\prod_{n=1}^\tau Q_n$, then
\begin{equation} \label{eq:opStringeff}
\begin{split}
\text{if } Q_1=Q_\tau=\mathcal{P}:& \quad \prod_{n=1}^\tau Q_n = \mathcal{P}^{\tau_{\mathrm{eff}}}, \quad \text{with } \tau_{\mathrm{eff}} = \tau - n_{DW}/2 \\ 
\text{if } Q_1=\mathcal{P}^\dagger,\; Q_\tau=\mathcal{P}:& \quad \prod_{n=1}^\tau Q_n = \mathcal{P}_2 \mathcal{P}^{\tau_{\mathrm{eff}}}, \quad \text{with } \tau_{\mathrm{eff}} = \tau - n_{DW}/2 - 1/2 \\ 
\text{if } Q_1=\mathcal{P},\; Q_\tau=\mathcal{P}^\dagger:& \quad \prod_{n=1}^\tau Q_n = \mathcal{P}^{\tau_{\mathrm{eff}}} \mathcal{P}_1, \quad \text{with } \tau_{\mathrm{eff}} = \tau - n_{DW}/2 - 1/2 \\
\text{if } Q_1=Q_\tau=\mathcal{P}^\dagger:& \quad \prod_{n=1}^\tau Q_n = \mathcal{P}_2\mathcal{P}^{\tau_{\mathrm{eff}}} \mathcal{P}_1, \quad \text{with } \tau_{\mathrm{eff}} = \tau - n_{DW}/2 - 1.
\end{split}
\end{equation}
\eq{eq:opStringeff} shows that up to a potential initial and final layer, each operator string can be expressed purely in terms of $\mathcal{P}$, where $\tau_{\mathrm{eff}}$ is the length of this $\mathcal{P}$-string.

We now consider $\tau \gg 1$, i.e., many repeated applications of $\tilde{\mathcal{P}}$. 
We note that each of the four possibilities in \eq{eq:opStringeff} appears an equal number of times, such that we can write the operator $\tilde{\mathcal{P}}^\tau$ as
\begin{equation} \label{eq:operatorID}
\tilde{\mathcal{P}}^\tau = \frac{\mathbb{1} + \mathcal{P}_2}{2}\, \biggl(\sum_{\tau_{\mathrm{eff}}=\tau/2}^\tau \rho(\tau_{\mathrm{eff}})\, \mathcal{P}^{\tau_{\mathrm{eff}}}\biggr) \, \frac{\mathbb{1} + \mathcal{P}_1}{2},
\end{equation}
where $\rho(\tau_{\mathrm{eff}})$ is the density of operator strings that result in a given $\tau_{\mathrm{eff}}$. Here, we can neglect a difference in $\tau_{\mathrm{eff}}$ by $\delta \tau_{\mathrm{eff}}=1$ for the different possibilities in \eq{eq:opStringeff}, since we will be interested in the scaling as $\tau \rightarrow \infty$. 
To determine $\rho(\tau_{\mathrm{eff}})$, we note that the number of operator strings in $\tilde{\mathcal{P}}^\tau$ that contain exactly $n_{DW}\in [0,\tau]$ domain walls is given by the binomial coefficient $\binom{\tau}{n_{DW}}$. Since $n_{DW}=2(\tau-\tau_{\mathrm{eff}})$ according to \eq{eq:opStringeff} (again ignoring an $\mathcal{O}(1)$ difference in $\tau_{\mathrm{eff}}$), $\rho(\tau_{\mathrm{eff}})$ is given by
\begin{equation} \label{eq:taudistribution}
\rho(\tau_{\mathrm{eff}}) = \frac{1}{2^\tau}\, \binom{\tau}{2(\tau-\tau_{\mathrm{eff}})} \xrightarrow{\tau \gg 1} \sqrt{\frac{8}{\tau\pi}} \exp\biggl\{-\frac{8}{\tau} \biggl(\tau_{\mathrm{eff}}-\frac{3}{4}\tau\biggr)^2\biggr\},
\end{equation}
which becomes a Gaussian distribution peaked at $\tau_{\mathrm{eff}}=\frac{3}{4}\tau$ as $\tau \gg 1$. This already suggests that $\tilde{\mathcal{P}}^\tau \sim \mathcal{P}^{3\tau/4}$; however, the growing variance $\sigma^2 \sim \tau$ of this distribution will lead to corrections as detailed below.

\subsection{Correspondence of eigenvalues}
Let us now take the eigenstate $\ket{\tilde{\psi}}$ of $\tilde{\mathcal{P}}$ with $\tilde{\mathcal{P}}\ket{\tilde{\psi}} = \tilde{\lambda}_{\tilde{\psi}} \ket{\tilde{\psi}}$ and $0<1-\tilde{\lambda}_{\tilde{\psi}} \ll 1$, and apply it to \eq{eq:operatorID}.
On the right hand side, the state $\frac{\mathbb{1}+\mathcal{P}_1}{2}\ket{\tilde{\psi}}$ can be expanded in terms of the eigenstates of $\mathcal{P}$.  Let us denote by $\ket{\psi}$ the eigenstate of $\mathcal{P}$ which, out of all eigenstates of $\mathcal{P}$ that have non-zero overlap with $\frac{\mathbb{1}+\mathcal{P}_1}{2}\ket{\tilde{\psi}}$, has the largest eigenvalue $\lambda_\psi < 1$.
Thus, in order for \eq{eq:operatorID} to hold as $\tau \rightarrow \infty$, there must be the following relation between $\tilde{\lambda}_{\tilde{\psi}}$ and $\lambda_\psi$:
\begin{equation} \label{eq:EVscaling}
\tilde{\lambda}^\tau_{\tilde{\psi}} = 
\sum_{\tau_{\mathrm{eff}}=0}^\tau \rho(\tau_{\mathrm{eff}})\, \lambda_\psi^{\tau_{\mathrm{eff}}} \; \xrightarrow{\tau \gg 1, \; |\log \lambda_\psi| \ll 1} \; \int_{-\infty}^{\infty} d\tau_{\mathrm{eff}} \, \rho(\tau_{\mathrm{eff}}) \lambda_\psi^{\tau_{\mathrm{eff}}} = \lambda_{\psi}^{\tau \,\bigl(\frac{3}{4}-\frac{1}{32}|\log(\lambda_\psi)|\bigr)},
\end{equation}
where we carried out the integral over $\tau_{\mathrm{eff}}$ upon inserting \eq{eq:taudistribution} in the last step. 
We may invert this relation, solving for $\lambda_\psi$ via
\begin{equation} \label{eq:EVinverse}
\tilde{\lambda}_{\tilde{\psi}} = \lambda_\psi^{\frac{3}{4} - \frac{1}{32}|\log \lambda_\psi|} \quad \Rightarrow \quad \lambda_\psi = \exp\biggl\{ -12\biggl(1-\sqrt{1-\frac{2}{9}|\log \tilde{\lambda}_{\tilde{\psi}}|}\biggr) \biggr\} = \tilde{\lambda}^{4/3}_{\tilde{\psi}} \cdot \exp\biggl\{ -\frac{2}{27}|\log\tilde{\lambda}_{\tilde{\psi}}|^2 + \mathcal{O}\bigl(|\log \tilde{\lambda}_{\tilde{\psi}}|^3\bigr) \biggr\},
\end{equation}
where the last equality was obtained upon expanding the square root.
Crucially, \eq{eq:EVinverse} relates the rate at which excited states decay under repeated application of $\mathcal{P}$ to the decay rate under repeated application of $\tilde{\mathcal{P}}$. 
We emphasize that the relation of \eq{eq:EVinverse} is independent of the underlying model, and, in particular, does not rely on an exact single-particle description.

\begin{figure}[t]
\centering
\includegraphics[trim={0cm 0cm 0cm 0cm},clip,width=0.99\linewidth]{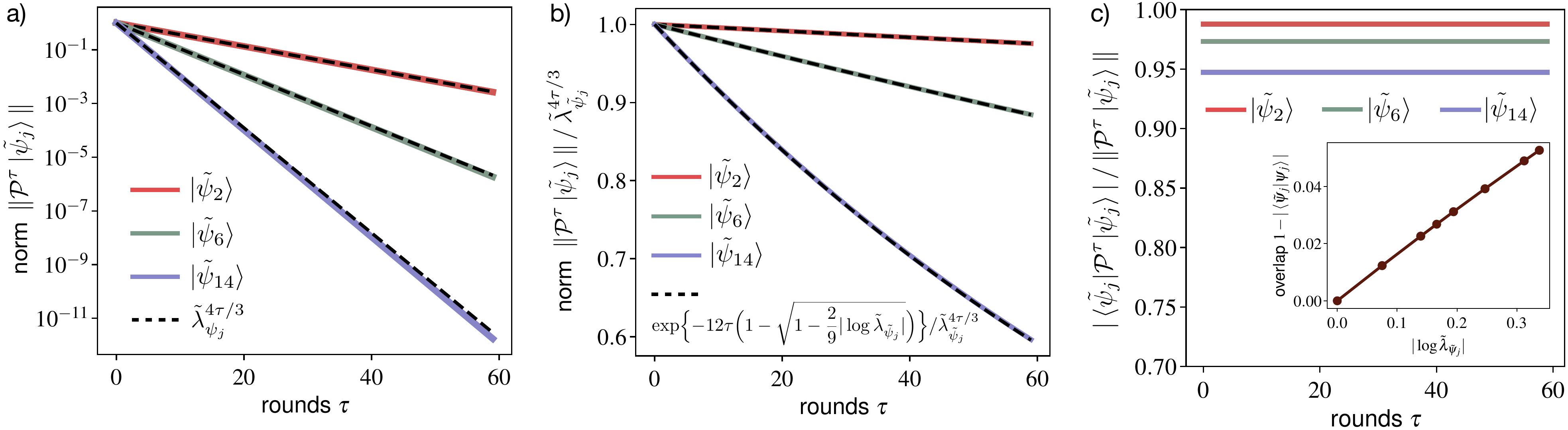}
\caption{\textbf{Numerical verification of $\mathcal{P}^\tau \approx \tilde{\mathcal{P}}^{4\tau/3}$.} \textbf{a)} Numerical evaluation of the norm $\|\mathcal{P}^\tau \ket{\tilde{\psi}_j}\|$ for selected eigenstates $\ket{\tilde{\psi}_j}$ of the operator $\tilde{\mathcal{P}}$ in the Heisenberg chain with $N=20$ and magnetization $\frac{1}{2}\sum_i (1-Z_i)=3$. 
The eigenstates with $\tilde{\mathcal{P}}\ket{\tilde{\psi}_j}=\tilde{\lambda}_{\tilde{\psi}_j} \ket{\tilde{\psi}_j}$ are organized such that $\tilde{\lambda}_{\tilde{\psi}_1}=1>\tilde{\lambda}_{\tilde{\psi}_2}\geq \cdots$.
The exponential decay of the norm follows the leading-order behavior $\sim \tilde{\lambda}_{\tilde{\psi}_j}^{4\tau/3}$ predicted by \eq{eq:EVinverse}. For the smallest eigenvalue $\tilde{\lambda}_{\tilde{\psi}_{14}}\approx 0.71$ displayed here, a very small deviation from the leading behavior becomes visible at late times.
\textbf{b)} Upon dividing out the leading-order scaling behavior, we find that these corrections are in excellent agreement with the predicted form of \eq{eq:EVinverse}. \textbf{c)} Overlap of the evolved state $\mathcal{P}^\tau \ket{\tilde{\psi}_j}$ with the initial eigenstate $\ket{\tilde{\psi}_j}$ of $\tilde{\mathcal{P}}$, relative to its total norm. This relative overlap takes on an $\mathcal{O}(1)$ value. Inset: Deviations from unity scale approximately linearly in $|\log \tilde{\lambda}_{\tilde{\psi}_j}|$ as predicted in \eq{eq:stateOverlap}.
}
\label{fig:opDev}
\end{figure}

To demonstrate this, we provide a numerical verification of \eq{eq:EVinverse}.
We consider the one-dimensional Heisenberg model with system size $N=20$ in the magnetization sector $\frac{1}{2}\sum_i (1-Z_i)=3$. We numerically compute the eigenstates $\ket{\tilde{\psi}_j}$ and associated eigenvalues $\tilde{\lambda}_{\tilde{\psi}_1} > \tilde{\lambda}_{\tilde{\psi}_2} \geq \cdots$ of $\tilde{\mathcal{P}}$. The largest eigenvalue $\tilde{\lambda}_{\tilde{\psi}_1} = 1$ corresponds to the Dicke state that is the unique ground state in this magnetization sector. We select a number of specific eigenstates (here: $j=2,6,14$) and numerically evaluate $\|\mathcal{P}^\tau \ket{\tilde{\psi}_j}\|$ in \figc{fig:opDev}{a}. The leading behavior $\|\mathcal{P}^\tau \ket{\tilde{\psi}_j}\| \sim \tilde{\lambda}^{4\tau/3}_{\tilde{\psi}_j}$ is clearly observed. 
We may further demonstrate that corrections to this scaling are predicted correctly by \eq{eq:EVinverse}. For this purpose, we divide out the leading behavior and consider the quantity $\|\mathcal{P}^\tau \ket{\tilde{\psi}_j}\| \, / \, \tilde{\lambda}^{4\tau/3}_{\tilde{\psi}_j}$, which we compare with \eq{eq:EVinverse} in \figc{fig:opDev}{b}. The numerical data show excellent agreement with our analytical prediction.

\subsection{Correspondence of eigenstates}
We now show that the states $\ket{\tilde{\psi}}$ and $\ket{\psi}$ have large overlap,
\begin{equation} \label{eq:stateOverlap}
\braket{\tilde{\psi}|\psi} = 1 - \mathcal{O}(|\log \tilde{\lambda}_{\tilde{\psi}}|).
\end{equation}
This may be seen by going back to \eq{eq:operatorID} and rewriting it as
\begin{equation} \label{eq:opIDinv}
2(\mathbb{1} + \mathcal{P}_2)^{-1} \, \tilde{\mathcal{P}}^\tau =  \biggl(\sum_{\tau_{\mathrm{eff}}=\tau/2}^\tau \rho(\tau_{\mathrm{eff}})\, \mathcal{P}^{\tau_{\mathrm{eff}}}\biggr) \, \frac{\mathbb{1} + \mathcal{P}_1}{2},
\end{equation}
where $\mathbb{1} + \mathcal{P}_2$ is Hermitian with strictly positive spectrum and thus invertible. 
We first note that the application of single layers $\mathcal{P}_a$ in \eq{eq:opIDinv} rotates $\ket{\tilde{\psi}}$ only by a small amount. To see this, consider
\begin{equation} \label{eq:singleLayer1}
\bra{\tilde{\psi}} \mathcal{P} \ket{\tilde{\psi}} = \bra{\tilde{\psi}} \mathcal{P}_1 \mathcal{P}_2 \ket{\tilde{\psi}} = \sum_{\tilde{\phi}} \bra{\tilde{\psi}} \mathcal{P}_1 \ket{\tilde{\phi}} \bra{\tilde{\phi}} \mathcal{P}_2 \ket{\tilde{\psi}},
\end{equation}
where we inserted a resolution of the identity in terms of the eigenstates of $\tilde{\mathcal{P}}$. 
The operator $\tilde{\mathcal{P}}$ is invariant under the $\mathbb{Z}_2$ operation $T_{\mathbb{Z}_2}$ that exchanges $\mathcal{P}_1 \xleftrightarrow{T_{\mathbb{Z}_2}} \mathcal{P}_2$ (for example, $T_{\mathbb{Z}_2}$ corresponds to site-centered inversion in the Heisenberg chain). As a result, the eigenstates $\ket{\tilde{\psi}}, \ket{\tilde{\phi}}$ of $\tilde{\mathcal{P}}$ in \eq{eq:singleLayer1} can simultaneously be chosen as eigenstates of $T_{\mathbb{Z}_2}$ with eigenvalues $\pm 1$. Consequently, $\bra{\tilde{\phi}} \mathcal{P}_2 \ket{\tilde{\psi}} = \bra{\tilde{\phi}} T_{\mathbb{Z}_2} \mathcal{P}_1 T_{\mathbb{Z}_2} \ket{\tilde{\psi}} = \pm  \bra{\tilde{\phi}} \mathcal{P}_1 \ket{\tilde{\psi}}$. Inserting into \eq{eq:singleLayer1} we obtain
\begin{equation} \label{eq:smallP1Rot}
|\bra{\tilde{\psi}} \mathcal{P} \ket{\tilde{\psi}}| \leq \sum_{\tilde{\phi}} |\bra{\tilde{\phi}}\mathcal{P}_1 \ket{\tilde{\psi}}|^2 = \bra{\tilde{\psi}} \mathcal{P}_1\ket{\tilde{\psi}}.
\end{equation}
We now note that $\braket{\tilde{\psi}|\mathcal{P}|\tilde{\psi}} = \braket{\tilde{\psi}|\tilde{\mathcal{P}}|\tilde{\psi}} = \tilde{\lambda}_{\tilde{\psi}} = 1 - \mathcal{O}(|\log \tilde{\lambda}_{\tilde{\psi}}|)$, since $\mathcal{P}$ and $\tilde{\mathcal{P}}$ have the same diagonal matrix elements. Inserting into \eq{eq:smallP1Rot} we find 
\begin{equation}
\braket{\tilde{\psi}|\mathcal{P}_1|\tilde{\psi}} \geq 1 - \mathcal{O}(|\log \tilde{\lambda}_{\tilde{\psi}}|),
\end{equation}
which shows that $\mathcal{P}_1$ rotates $\ket{\tilde{\psi}}$ only slightly; an analogous statement holds for $\mathcal{P}_2$. 
We now apply \eq{eq:opIDinv} to the state $\ket{\tilde{\psi}}$: Since $\ket{\tilde{\psi}}$ is an eigenstate of $\tilde{\mathcal{P}}$ and is modified only slightly by $(\mathbb{1} + \mathcal{P}_2)^{-1}$, the state on the left-hand side of this equation points in the direction of $\ket{\tilde{\psi}}$ up to a small rotation of $\mathcal{O}(|\log\tilde{\lambda}_{\tilde{\psi}}|)$. On the right-hand side, according to the arguments of the previous section, after application of $\sum_{\tau_{\mathrm{eff}}=\tau/2}^\tau \rho(\tau_{\mathrm{eff}})\, \mathcal{P}^{\tau_{\mathrm{eff}}}$ the state points in the direction $\ket{\psi}$. Comparing the left- and right-hand sides, \eq{eq:stateOverlap} follows.

To demonstrate \eq{eq:stateOverlap} numerically, we again consider the $N=20$ Heisenberg chain with $\frac{1}{2}\sum_i (1-Z_i)=3$ and the numerically computed eigenstates $\ket{\tilde{\psi}_j}$ of $\tilde{\mathcal{P}}$. We then evaluate the quantity
\begin{equation}
\frac{|\bra{\tilde{\psi}_j}\mathcal{P}^\tau \ket{\tilde{\psi}_j}|}{\|\mathcal{P}^\tau \ket{\tilde{\psi}_j}\|} \xrightarrow{\tau \rightarrow \infty} |\braket{\tilde{\psi}_j|\psi_j}|,
\end{equation}
which measures the overlap $|\braket{\tilde{\psi}_j|\psi_j}|$. As shown in \figc{fig:opDev}{c}, $|\braket{\tilde{\psi}_j|\psi_j}|$ assumes a finite value, in accordance with the prediction of \eq{eq:stateOverlap}. Moreover, we show the dependence of $|\braket{\tilde{\psi}_j|\psi_j}|$ on $|\log \tilde{\lambda}_{\tilde{\psi}_j}|$ in the inset of \figc{fig:opDev}{c}, which confirms the predicted linear relation.

\section{Single-particle projection round dynamics as imaginary time evolution} \label{sec:SMsingleParticle}
In this supplement, we show Eq.~(4) and  Eq.~(5) of the main text, which state that in the single-particle sector of the ferromagnetic Heisenberg model, repeated application of the local ground-state projector
\begin{equation} \label{eq:projectorApp}
\mathcal{P} = \prod_{a=1}^{2d} \mathcal{P}_a = \prod_{a=1}^{2d} \prod_{i\in B_a} (1-P_{i})
\end{equation}
acts as imaginary time evolution.

\subsection{Symmetric projection round operator}
As a first step, we relate the symmetrized projection round operator $\tilde{\mathcal{P}}$ introduced in the previous section to the Hamiltonian for systems with $\mathcal{A}=2$ layers.
In particular, the layer operators $\mathcal{P}_a = \prod_{i \in B_a} (1-P_i)$ can be expanded as 
\begin{equation}
\mathcal{P}_a = \mathbb{1} - \sum_{i \in B_a} P_i + \mathcal{O}(P_{i \in B_a} P_{j \in B_a}),
\end{equation}
where $\mathcal{O}(P_{i \in B_a} P_{j \in B_a})$ denotes operator strings containing at least two consecutive projectors $P_i$ from the same layer $B_a$. Accordingly, the projection round operator $\mathcal{P}$ can be expressed as
\begin{equation}
\mathcal{P} = \mathcal{P}_1 \mathcal{P}_2 = \mathbb{1} - \sum_{i \in B_1 \cup B_2} P_i + \sum_{\substack{i \in B_1 \\ j \in B_2}} P_i P_j + \mathcal{O}(P_{i \in B_a}P_{j \in B_a}).
\end{equation}
Given this result, the symmetric projection round operator $\tilde{\mathcal{P}} = \frac{1}{2}(\mathcal{P} + \mathcal{P}^\dagger)$, where $\mathcal{P}^\dagger = \mathcal{P}_2\mathcal{P}_1$, is given by
\begin{equation}
\tilde{\mathcal{P}} = \mathbb{1} - \sum_{i \in B_1 \cup B_2} P_i + \frac{1}{2} \sum_{\substack{i \in B_1 \\ j \in B_2}} (P_i P_j + P_j P_i) + \mathcal{O}(P_{i \in B_a}P_{j \in B_a}) = 
\mathbb{1} - \sum_{i \in B_1 \cup B_2} P_i + \frac{1}{2}\left[\sum_{\substack{i \in B_1 \cup B_2 \\ j \in B_1 \cup B_2}} P_{i}P_{j} - \sum_{i \in B_1 \cup B_2} P_{i} \right] + \mathcal{O}(P_{i \in B_a}P_{j \in B_a}).
\end{equation}
Since $\sum_{i \in B_1 \cup B_2} P_i = H$ is the Hamiltonian, we obtain
\begin{equation} \label{eq:symmProjApp}
\tilde{\mathcal{P}} = \mathbb{1} - \frac{3}{2}H + \frac{1}{2}H^2 + \mathcal{O}(P_{i \in B_a} P_{j\in B_a}).
\end{equation}
We note that this relation holds for all systems with $\mathcal{A}=2$ layers, beyond the specific example of the single-particle Heisenberg model.

\subsection{Dimension $d=1$}
We now focus specifically on the single-particle sector of the $d=1$ Heisenberg model, where the $P_{i}=\frac{1}{2}(\ket{01}-\ket{10})_{i,i+1}\, (\bra{01}-\bra{10})$ are singlet projectors on neighboring lattice sites.
The single-particle sector is specified by $ \biggl( \sum_i \frac{1}{2}(1-Z_i) \biggr)\ket{\psi} = \ket{\psi}$, i.e., all but one qubit are in state $\ket{0}$.
The key property we use is that for any state $\ket{\psi}$ in the single-particle sector, 
\begin{equation} \label{eq:projectorApp2}
P_{i}P_{j}\ket{\psi}=0, \;\; \text{if } [P_{i},P_{j}]=0. 
\end{equation}
This relation holds because $P_{j}\ket{\psi} = a_\psi \, (\ket{01}-\ket{10})_{j,j+1}$ with some $a_\psi \in \mathbb{C}$, i.e., the particle is localized to sites $j$ and $j+1$. If $[P_{i},P_{j}]=0$, the pair of sites $i,i+1$ has no overlap with the pair of sites $j,j+1$. Consequently, $P_{i}P_{j}\ket{\psi} \sim P_{i}(\ket{01}-\ket{10})_{j,j+1} = 0$, since the single-particle wave function strictly localized on $j,j+1$ has vanishing amplitude on sites $i,i+1$. 
Due to this property, \eq{eq:symmProjApp} becomes
\begin{equation} \label{eq:symmProjHeis}
\tilde{\mathcal{P}} = \mathbb{1} - \frac{3}{2}H + \frac{1}{2}H^2,
\end{equation}
when acting on the single-particle sector of the Heisenberg model $H = H_{\mathrm{Heis}}$.

As a result, $\tilde{\mathcal{P}}\ket{k} = \tilde{\lambda}_k \ket{k}$ with $\tilde{\lambda}_k = 1-\frac{3}{2}\varepsilon(k)+\frac{1}{2}\varepsilon(k)^2$, where $\ket{k}$ are the momentum eigenstates of the single-particle sector of $H_{\mathrm{Heis}}$.
To obtain Eq.~(4) of the main text, we insert \eq{eq:symmProjHeis} into the relation $\mathcal{P}^\tau \approx \tilde{\mathcal{P}}^{4\tau/3}$ we showed in Sec.~\ref{sec:AppProjectorTilde}. We recall that corrections to this relation are controlled by the eigenvalues $\tilde{\lambda}_k$ of $\tilde{\mathcal{P}}$ via the parameter $|\log \tilde{\lambda}_k|$. At low energies $\varepsilon(k) \ll 1$, $|\log \tilde{\lambda}_k| \approx \frac{3}{2} \varepsilon(k)$ and corrections will be small. We thus obtain 
\begin{equation} \label{eq:imagTApp1}
\mathcal{P}^\tau \ket{k} \approx \tilde{\mathcal{P}}^{4\tau/3} \ket{k} = \biggl(1-\frac{3}{2}\varepsilon(k)+\frac{1}{2}\varepsilon(k)^2\biggr)^{4\tau/3} \ket{k} \approx \exp\bigl\{ -2\varepsilon(k)\tau \bigr\} \ket{k}.
\end{equation}

\subsection{Dimension $d>1$}
For concreteness, we will show Eq.~(4) of the main text for $d=2$. Generalization to arbitrary dimensions will be straightforward. In two dimensions, we write the Heisenberg model as
\begin{equation} \label{eq:projectorApp7}
H_{\mathrm{Heis}} = H^{(x)}_{\mathrm{Heis}} + H^{(y)}_{\mathrm{Heis}},
\end{equation}
where $H^{(x)}_{\mathrm{Heis}}$ is the sum of all terms acting on bonds along the $x$-direction and analogously for $H^{(y)}_{\mathrm{Heis}}$. Moreover,
$\mathcal{P} = \mathcal{P}_1 \mathcal{P}_2 \mathcal{P}_3 \mathcal{P}_4$, where we assume that $\mathcal{P}_1$ and $\mathcal{P}_2$ correspond to products of local triplet projectors on bonds along the $x$-direction, while $\mathcal{P}_3$ and $\mathcal{P}_4$ correspond to those along the $y$-direction. Defining $\mathcal{P}_x = \mathcal{P}_1 \mathcal{P}_2$ and $\mathcal{P}_y = \mathcal{P}_3 \mathcal{P}_4$, we thus have $\mathcal{P}=\mathcal{P}_x \mathcal{P}_y$. We now define a new operator $\tilde{\mathcal{P}}$ as 
\begin{equation} \label{eq:projectorApp8}
\tilde{\mathcal{P}} = \frac{1}{2}\bigl(\mathcal{P}_x + \mathcal{P}_x^\dagger \bigr) \, \frac{1}{2}\bigl(\mathcal{P}_y + \mathcal{P}_y^\dagger \bigr) = \biggl( \mathbb{1} - \frac{3}{2} H^{(x)}_{\mathrm{Heis}} + \frac{1}{2}(H^{(x)}_{\mathrm{Heis}})^2 \biggr)\, \biggl( \mathbb{1} - \frac{3}{2} H^{(y)}_{\mathrm{Heis}} + \frac{1}{2}(H^{(y)}_{\mathrm{Heis}})^2 \biggr),
\end{equation}
which follows directly from the one-dimensional case of \eq{eq:symmProjHeis}. At small energies, $\tilde{\mathcal{P}} \approx \mathbb{1} - \frac{3}{2} H_{\mathrm{Heis}}$. 

In order to show Eq.~(4) of the main text, it remains to demonstrate that $\tilde{\mathcal{P}}^\tau \ket{k} \approx \mathcal{P}^{3\tau / 4} \ket{k}$, as in \eq{eq:imagTApp1} for $d=1$. 
We recall that the proof of $\tilde{\mathcal{P}}^\tau \approx \mathcal{P}^{3\tau/4}$ in Appendix~\ref{sec:AppProjectorTilde} was given for rounds of projector evolution consisting of only two layers, and does not immediately apply to the 2D Heisenberg case, where each round consists of four layers. This is because $\mathcal{P}_x$ and $\mathcal{P}_y$ do not generally commute, and we cannot immediately remove domain walls $\mathcal{P}_x \mathcal{P}_x^\dagger$ and $\mathcal{P}_y \mathcal{P}_y^\dagger$ of $x$- and $y$-type from the operator strings in $\tilde{\mathcal{P}}^\tau$.
To make progress, let us denote by $T_x$ and $T_y$ the one-site translation operators in the $x$- and $y$-directions, respectively. We now assume that the operators $\mathcal{P}_x$ and $\mathcal{P}_y$ are chosen such that
\begin{equation} \label{eq:translInv}
[T_y,\mathcal{P}_x] = [T_x,\mathcal{P}_y] = 0.
\end{equation}
\eq{eq:translInv} assumes that the operator $\mathcal{P}_x$ corresponds to a brickwork of triplet projectors along the $x$-direction which preserves translation invariance in the $y$-direction.
Explicitly, this condition is fulfilled by choosing $\mathcal{P}_1 = \prod_{i=(x,y): \, x \, \text{even}} (1-P_{i})$, where $P_i = \frac{1}{2}(\ket{01}-\ket{10})_{i,i+\hat{e}_x}(\bra{01}-\bra{10})$, and analogously for $\mathcal{P}_{2}$, $\mathcal{P}_{3}$, $\mathcal{P}_{4}$.
Let us now consider the two-dimensional single-particle basis states $\ket{i} = \ket{x} \otimes\ket{y}$. Because of \eq{eq:translInv},
\begin{equation}
\mathcal{P}_x \;\ket{x}\otimes\ket{y} = \sum_{x^\prime} a_x(x^\prime) \ket{x^\prime} \otimes \ket{y},
\end{equation}
where, crucially, the amplitude $a_x(x^\prime)$ does not depend on the $y$-coordinate of the particle. Therefore, $\mathcal{P}_x$ acts non-trivially only on the $x$-sector of the Hilbert space and analogously for $\mathcal{P}_y$. As a result, $[\mathcal{P}_x,\mathcal{P}_y]=0$ for the single-particle case and $\tilde{\mathcal{P}}^\tau \ket{k} \approx \mathcal{P}^{3\tau / 4} \ket{k}$ follows as a consequence.

\subsection{Energy dynamics during projection rounds}
Let us apply \eq{eq:imagTApp1} to the dynamics of the initial state $\ket{\psi(0)} = \frac{1}{\sqrt{2}}\bigl(\ket{i}+\ket{i+\hat{e}_x}\bigr)$ considered in the main text. Our goal is to derive Eq.~(5) of the main text for the energy 
\begin{equation} \label{eq:projectionEnergyApp}
e(\tau) = \frac{\braket{\psi(0)|(\mathcal{P}^\dagger)^\tau H_{\mathrm{Heis}}\mathcal{P}^\tau|\psi(0)}}{\|\mathcal{P}^\tau\ket{\psi(0)}\|^2}.
\end{equation}
For this purpose, let us express $\ket{\psi(0)}$ in terms of the momentum basis $\ket{k}$,
\begin{equation} 
\ket{\psi(0)} = \frac{1}{\sqrt{2N}}\sum_k (1+e^{ik\cdot \hat{e}_x}) \ket{k},
\end{equation}
which, at low momenta $k\ll 1$, is an equal amplitude superposition due to the real-space localized nature of the initial state. At times $\tau \gg 1$ we may focus on the low energy/momentum contributions $\varepsilon(k) = v\,k^z$ to $e(\tau)$ in \eq{eq:projectionEnergyApp}, which yields
\begin{equation} \label{eq:AppEnergyDynProj}
e(\tau \gg 1) \approx \frac{\sum_k \varepsilon(k) e^{-4\varepsilon(k)\tau}}{\sum_k e^{-4\varepsilon(k)\tau}} \xrightarrow{N \gg 1} \frac{\int_{2\pi/L}^\infty dk \, k^{d-1} \, vk^z \, e^{-4vk^z \tau}}{\int_0^\infty dk \, k^{d-1} e^{-4vk^z \tau}} = \frac{1}{\tau} \frac{\Gamma(1+\frac{d}{2},4v (\frac{2\pi}{L})^z \, \tau)}{\Gamma(1+\frac{d}{2},0)} = \frac{1}{\tau} \frac{\Gamma(1+\frac{d}{2},4\Delta \tau)}{\Gamma(1+\frac{d}{2},0)},
\end{equation}
where $\Gamma(\cdot,\cdot)$ denotes the incomplete upper gamma function and we used in the last step that $v(\frac{2\pi}{L})^z = \varepsilon(2\pi/L) = \Delta$ corresponds to the gap. Using that $\Gamma(1+d/2,4\Delta \tau) \overset{\tau \gg 1}{\sim} \exp\{-4\Delta\tau\}$, \eq{eq:AppEnergyDynProj} leads to Eq.~(5) of the main text,
\begin{equation} \label{eq:AppEnergyDynPro2}
e(1\ll\tau \ll \Delta^{-1}) = \frac{1}{2} \frac{\beta}{\tau}, \qquad e(\tau\Delta \gtrsim 1) \sim \beta \Delta \exp\{ -4\Delta \tau \}.
\end{equation}
Here, we have introduced a parameter $\beta$, expected to be of order $\mathcal{O}(1)$, such that the relation on the left is asymptotically exact (for $1 \ll \tau \ll \Delta^{-1}$). 
As mentioned in the main text and shown in Sec.~\ref{sec:AppMarkov}, remarkably, the parameter $\beta = d/z$ turns out to be \textit{universal}. 

\begin{figure}[t]
\centering
\includegraphics[trim={0cm 0cm 0cm 0cm},clip,width=0.8\linewidth]{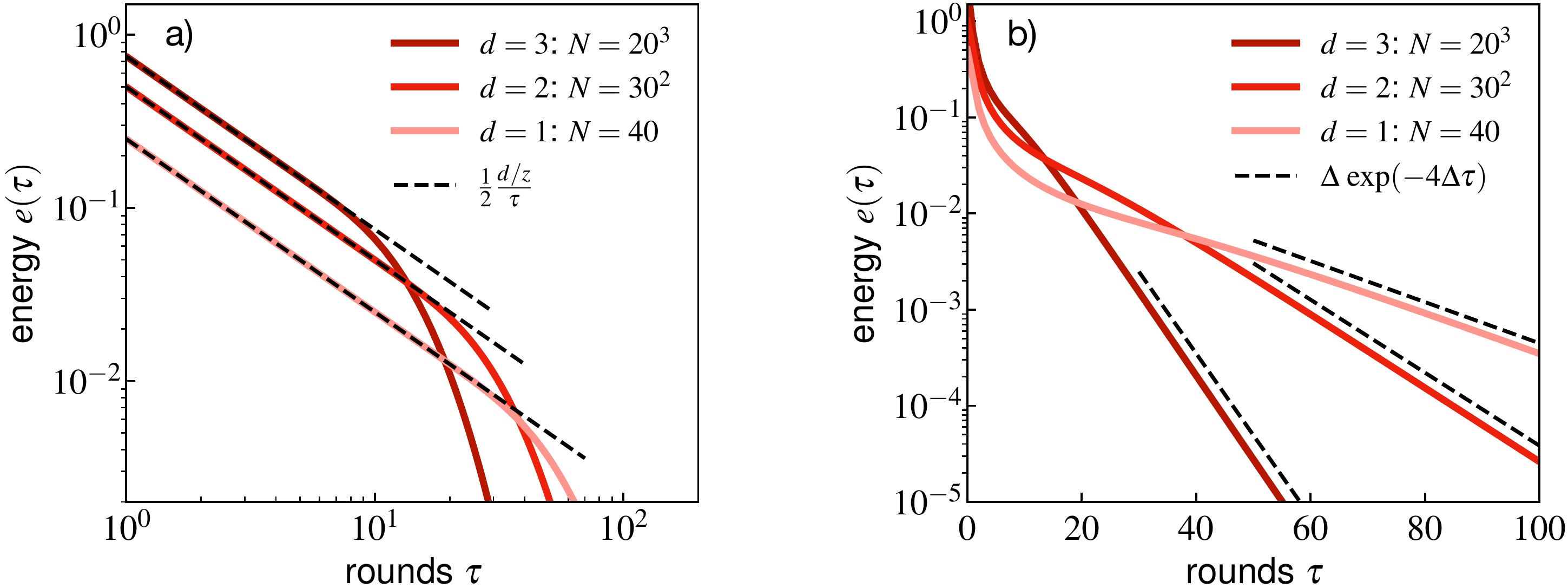}
\caption{\textbf{Single-particle energy dynamics under projection round evolution.} 
\textbf{a)} Early-time dynamics of the energy $e(\tau)$, see \eq{eq:projectionEnergyApp}, starting from an initially localized single-particle state $\ket{\psi(0)} = \frac{1}{\sqrt{2}}\bigl(\ket{i}+\ket{i+\hat{e}_x}\bigr)$ in different dimensions $d=1,2,3$. The early-time decay follows an algebraic form $e(\tau)=\frac{1}{2}\frac{\beta}{\tau}$, with $\beta=d/z=d/2$ as predicted by \eq{eq:AppEnergyDynPro2}.
\textbf{b)} Late-time exponential decay of $e(\tau)$, with rate $4\Delta$ set by four times the ground-state energy gap of the single-particle Heisenberg model, in agreement with \eq{eq:AppEnergyDynPro2}.
}
\label{fig:single_particle_projectorE}
\end{figure}

To conclude this section, we provide a numerical verification of \eq{eq:AppEnergyDynPro2} in \fig{fig:single_particle_projectorE}. Specifically, we evaluate the energy $e(\tau)$ defined in \eq{eq:projectionEnergyApp} numerically for the localized single-particle initial state $\ket{\psi(0)} = \frac{1}{\sqrt{2}}\bigl(\ket{i}+\ket{i+\hat{e}_x}\bigr)$ in different dimensions $d=1,2,3$. \figc{fig:single_particle_projectorE}{a} and \figc{fig:single_particle_projectorE}{b} show the early- and late-time behavior of $e(\tau)$, in excellent agreement with the prediction of \eq{eq:AppEnergyDynPro2}.

\section{The single-particle Markov process} \label{sec:AppMarkov}
We provide an analytical treatment of the single-particle Markov process introduced in the main text. While we rely on a number of approximations and scaling arguments, we find the analytical results derived below to be in excellent agreement with numerical studies. We restate the setup for completeness:

For a given trajectory, we start at time $t=0$ with $\tau(t=0)=0$ and energy $E(t=0) = e(\tau(0))$, where $e(\tau)$ is given by \eq{eq:AppEnergyDynPro2}. 
In each time step:
\begin{enumerate}
\item With probability $p(t) = 2E(t) = 2e(\tau(t))$, we set $\tau(t+1)=0$ and $E(t+1) = e(0)$. This corresponds to a reset.
\item With probability $1-p(t)$, we set $\tau(t+1) = \tau(t) + 1$ and $E(t+1) = e(\tau(t+1))$. This corresponds to a projection round.
\end{enumerate}
Our goal is to determine the trajectory-averaged energy $\overline{E(t)}$ as well as the trajectory-averaged ground state infidelity $\overline{\epsilon(t)} = 1 - \overline{|\braket{\Omega|\psi(t)}|^2}$.

\subsection{Number of resets per trajectory}
We define 
\begin{equation} \label{eq:AppQnoReset}
Q(\tau) \equiv \prod_{\tau^\prime=1}^\tau (1-p(\tau^\prime)),
\end{equation}
which is the probability to evolve for $\tau$ rounds without a reset. Since $p(\tau)=2e(\tau) \ll 1$ for $\tau\gg 1$, we can approximate
\begin{equation}
Q(\tau) \sim \prod_{\tau^\prime=1}^\tau \exp\{ -p(\tau^\prime) \} = \exp\Bigl\{- \sum_{\tau^\prime =1}^{\tau } p(\tau^\prime)\Bigr\} \approx \exp\Bigl\{ - \int_1^\tau d\tau^\prime \, p(\tau^\prime) \Bigr\}.
\end{equation}
Using the energy $e(\tau)$ from \eq{eq:AppEnergyDynPro2}, we obtain
\begin{equation}
\begin{split}
Q(\tau\Delta \ll 1) &\sim \exp\Bigl\{-\beta \int_1^\tau d\tau^\prime \, \frac{1}{\tau^\prime}\Bigr\} = \tau^{-\beta} \\ 
Q(\tau\Delta \gtrsim 1) &\sim \exp\Bigl\{-\beta \Bigl( \underbrace{\int_1^{\Delta^{-1}} d\tau^\prime \, \frac{1}{\tau^\prime}}_{\log(\Delta^{-1})} + \underbrace{\int_{\Delta^{-1}}^\infty d\tau^\prime\, \Delta \, e^{-4\Delta\tau^\prime}}_{e^{-4}/4 = \mathrm{const.}} \Bigr)\Bigr\} \sim \Delta^\beta.
\end{split}
\end{equation}
In particular, the probability to evolve for an infinite number of steps without reset is given by 
\begin{equation} \label{eq:AppProbNoReset}
Q_\infty \equiv Q(\tau \rightarrow \infty) = c\,\Delta^\beta,
\end{equation}
with some constant $c$.
Let us now consider the probability $P_{\# r}(N_r)$ to have \textit{exactly} $N_r$ resets in a given trajectory. We have already determined $P_{\# r}(N_r=0) = Q_\infty$ via \eq{eq:AppProbNoReset}. Accordingly, we obtain
\begin{equation}
\begin{split}
P_{\# r}(1) &= Q_\infty (1-Q_\infty) \\
P_{\# r}(2) &= Q_\infty (1-Q_\infty)^2\\
&\vdots \\
P_{\# r}(N_r) &= Q_\infty (1-Q_\infty)^{N_r},
\end{split}
\end{equation}
and note that $\sum_{N_r=0}^\infty P_{\# r}(N_r) = 1$. In the regime of small $Q_\infty \ll 1$, we obtain
\begin{equation} \label{eq:DistributionResets}
P_{\# r}(N_r) \approx Q_\infty \exp\{-Q_\infty \, N_r\} = c\,\Delta^\beta \exp\{-c\,\Delta^\beta N_r \}.
\end{equation}
The trajectory-averaged number of resets is then given by 
\begin{equation} \label{eq:AppAverageNrResets}
\overline{N_r} = \int_0^\infty dN_r \, N_r \, P_{\# r}(N_r) = c^{-1} \Delta^{-\beta} = 1/Q_\infty.
\end{equation}
Simultaneously, we know that the probability $Q_\infty$ to encounter no reset is given by the initial overlap with the ground state, $Q_\infty = |\braket{\Omega|\psi(0)}|^2 \sim 1/N = L^{-d}$.
Comparing with \eq{eq:AppProbNoReset}, we find $\Delta^{\beta} \sim L^{-z\,\beta} \sim L^{-d}$, and therefore
\begin{equation}
\beta = d/z,
\end{equation}
which is a universal exponent determined by spatial dimension and the dynamical exponent $z$. Moreover, the constant $c$ introduced in \eq{eq:AppProbNoReset} is proportional to the initial ground state overlap,
\begin{equation} \label{eq:AppConstantC}
c = Q_\infty \, \Delta^{-\beta} \sim \left|\braket{\Omega|\psi(0)}\right|^2 \, N,
\end{equation}
and corresponds to an $\mathcal{O}(1)$ number.

\subsection{Time between resets}
\begin{figure}[t]
\centering
\includegraphics[trim={0cm 0cm 0cm 0cm},clip,width=0.99\linewidth]{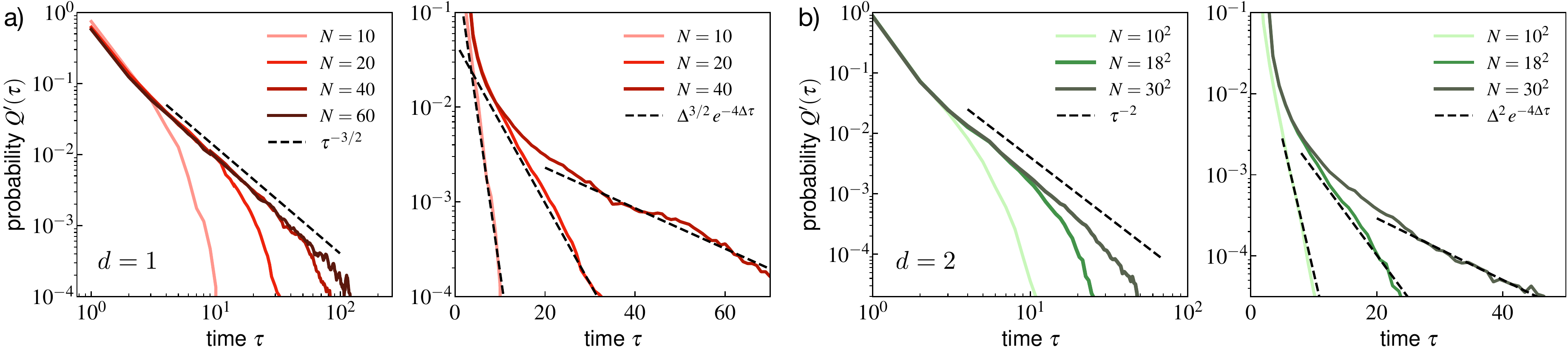}
\caption{\textbf{Distribution of time between resets.} \textbf{a)} Probability distribution $Q^\prime(\tau)$ defined in \eq{eq:timeResDisApp}, evaluated numerically for the single-particle Heisenberg chain in $d=1$. Early- and late-time behavior, shown on the left and right, respectively, follows the prediction \eq{eq:AppTbetweenResetsDistribution} for $\beta = d/z = 1/2$. \textbf{b)} Same as a) for a single-particle system in $d=2$ dimensions. The results agree with \eq{eq:AppTbetweenResetsDistribution} for $\beta = d/z = 1$. The times $\tau$ between resets were sampled from at least $1.4 \times 10^3$ trajectories for each system-size instance.
}
\label{fig:time_between_resets_supp}
\end{figure}

We have determined the probability distribution and average of the number of resets per trajectory. Let us now determine the average time between two resets in a given trajectory. Specifically, we define the quantity
\begin{equation} \label{eq:timeResDisApp}
Q^\prime (\tau) \equiv Q(\tau)\,p(\tau),
\end{equation}
which is the probability to evolve for $\tau$ rounds without a reset, followed by a subsequent reset. 
Inserting $Q(\tau)$ defined in \eq{eq:AppQnoReset}, we obtain
\begin{equation} \label{eq:AppTbetweenResetsDistribution}
Q^\prime(\tau\Delta\ll 1) \sim \tau^{-1-\beta}, \qquad Q^\prime(\tau\Delta\gtrsim 1) \sim  \Delta^{1+\beta} \exp\{-4\Delta\tau\}.
\end{equation}
Crucially, $Q^\prime (\tau)$ acts as a probability distribution for the time $\tau$ between two resets. 
We provide a numerical verification of \eq{eq:AppTbetweenResetsDistribution} for dimensions $d=1,2$ in \fig{fig:time_between_resets_supp}.
Following \eq{eq:AppTbetweenResetsDistribution}, the average time between two resets is given by
\begin{equation} \label{eq:AppTimeBetweenResets}
\overline{T_r} = \sum_{\tau=0}^\infty \tau \, Q^\prime(\tau) \sim \int_1^\infty d\tau\, \tau\,Q^\prime(\tau) \sim \int_1^{\Delta^{-1}} d\tau\, \tau^{-\beta} + \underbrace{\Delta^{1+\beta}\int_{\Delta^{-1}}^\infty d\tau\, \tau \exp\{-4\Delta\tau \}}_{\sim \Delta^{-(1-\beta)}} \sim 
\begin{cases}
\Delta^{-\max(1-\beta,0)}, \;\;\beta \neq 1 \\
|\log \Delta |, \;\; \beta = 1
\end{cases} \;.
\end{equation}
From \eq{eq:AppTimeBetweenResets}, we see that for $\beta<1$ the average time between resets grows with system size as $\overline{T_r}\sim \Delta^{\beta-1}$, and becomes a constant for $\beta > 1$. The case $\beta = 1$, corresponding to the $d=2$ single-particle system, is marginal and results in a logarithmically growing $\overline{T_r} \sim |\log\Delta| \sim \log N$.

\subsection{Time of last reset}
Given the distribution of the number of resets $P_{\# r}(N_r)$ per trajectory and the average time $\overline{T_r}$ between two resets, we can determine the distribution $P_{\text{last reset}}(t)$ of the time of the last reset in a given trajectory via the two relations
\begin{equation}
P_{\text{last reset}}(t) \, dt = P_{\# r}(N_r) \, dN_r \;, \qquad t = \overline{T_r}\, N_r \,.
\end{equation}
Inserting \eq{eq:DistributionResets} and \eq{eq:AppTimeBetweenResets} and solving for $P_{\text{last reset}}(t)$ for $t\Delta \gtrsim 1$ results in
\begin{equation} \label{eq:TimeLastReset}
P_{\text{last reset}}(t) \sim 
\begin{cases}
\Delta^{\max(1,\beta)}\, \exp\{- \lambda \Delta^{\max(1,\beta)}\, t\}\,, \;\; \beta \neq 1 \\
\Delta |\log\Delta|^{-1} \, \exp\{-\lambda \Delta t / |\log\Delta|\}, \;\; \beta = 1
\end{cases} \; ,
\end{equation}
where $\lambda>0$ is an $\mathcal{O}(1)$ constant. In particular, $\lambda$ is proportional to the constant $c$ of \eq{eq:AppConstantC} and thus to the initial ground state overlap,
\begin{equation} \label{eq:AppRater}
\lambda \sim \left|\braket{\Omega|\psi(0)}\right|^2 \, N.
\end{equation}
Moreover, we note that for $\beta < 1$ we must have $\lambda \leq 4$, since the probability to have a reset cannot decay faster than the energy $e(\tau)$ of the pure projector round evolution \eq{eq:AppEnergyDynPro2}.

\subsection{Average energy}
The distribution \eq{eq:TimeLastReset} of the last reset time is the crucial ingredient for determining the average energy and ground-state infidelity.
Specifically, let us consider a trajectory where the last reset occurs at time $t^\prime$. Let us further denote the energy of this trajectory at time $t \Delta \gtrsim 1$ as $E^{}_{t^\prime}(t)$. We now approximate
\begin{equation} \label{eq:EnergyLastResetTrajectory}
E^{}_{t^\prime}(t) \approx \theta(t^\prime - t)\, e\bigl(\overline{T_r}/2\bigr) + \theta(t-t^\prime)\, e(t-t^\prime),
\end{equation}
where $\theta(\cdot)$ denotes the Heaviside step function. \eq{eq:EnergyLastResetTrajectory} expresses that the energy at time $t<t^\prime$ before the last reset is given -- on average -- by the energy $e(\overline{T}_r/2)$ at an average time between resets. On the other hand, if $t>t^\prime$, the energy is given by projector round evolution since the last reset. We thus obtain for the trajectory-averaged energy
\begin{equation} \label{eq:MarkovAverageEnergy}
\overline{E(t)} \approx \int_0^\infty dt^\prime \, P_{\text{last reset}}(t^\prime) \, E^{}_{t^\prime}(t) = \int_0^t dt^\prime\, P_{\text{last reset}}(t^\prime)\, e(t-t^\prime)  + \int_t^\infty dt^\prime\, P_{\text{last reset}}(t^\prime) \, e(\overline{T_r}/2) \,.
\end{equation}
By inserting \eq{eq:TimeLastReset} for $P_{\text{last reset}}(t^\prime)$ and \eq{eq:AppEnergyDynPro2} for $e(t-t^\prime)$, we find that the first term in \eq{eq:MarkovAverageEnergy} scales as 
\begin{equation} \label{eq:MarkovEnergyTerm1}
\int_0^t dt^\prime\, P_{\text{last reset}}(t^\prime)\, e(t-t^\prime) \sim 
\begin{cases}
\Delta^{\max(1,\beta)}|\log\Delta| \, \exp\{-\lambda\Delta^{\max(1,\beta)}t\}, \;\; \beta \neq 1 \\
\Delta \, \exp\{-\lambda \Delta t / |\log\Delta|\}, \;\; \beta = 1
\end{cases} \;. 
\end{equation}
Moreover, using that
\begin{equation}
e(\overline{T_r}/2) \sim 1 / \overline{T_r} \sim 
\begin{cases}
\Delta^{\max(1-\beta,0)}, \;\; \beta \neq 1\\
|\log \Delta|^{-1}, \;\; \beta = 1
\end{cases} \; ,
\end{equation}
we obtain for the second term of \eq{eq:MarkovAverageEnergy},
\begin{equation} \label{eq:MarkovEnergyTerm2}
\int_t^\infty dt^\prime\, P_{\text{last reset}}(t^\prime) \, e(\overline{T_r}/2) \sim 
\begin{cases}
\Delta^{\max(1-\beta,0)}\, \exp\{-\lambda\Delta^{\max(1,\beta)}t\}, \;\; \beta \neq 1 \\
|\log \Delta|^{-1}\, \exp\{-\lambda \Delta t / |\log \Delta | \}, \;\; \beta = 1
\end{cases} \;.
\end{equation}
Comparing \eq{eq:MarkovEnergyTerm1} and \eq{eq:MarkovEnergyTerm2}, we see that for small energy gap $\Delta$, i.e., large systems, the second term of \eq{eq:MarkovEnergyTerm2} dominates, such that $\overline{E(t)}$ is asymptotically given by
\begin{equation} \label{eq:MarkovEnergyLateTime}
\overline{E(t)} \sim 
\begin{cases}
\Delta^{\max(1-\beta,0)}\, \exp\{-\lambda\Delta^{\max(1,\beta)}t\}, \;\; \beta \neq 1 \\
|\log \Delta|^{-1}\, \exp\{-\lambda \Delta t / |\log \Delta | \}, \;\; \beta = 1
\end{cases} \;.
\end{equation}
\eq{eq:MarkovEnergyLateTime} is the trajectory-averaged energy for times $t\Delta \gtrsim 1$ larger than the gap. 
In order to estimate the energy $\overline{E(t)}$ at early times, we note that the average time between resets for $t\Delta \ll 1$ is obtained by generalizing \eq{eq:AppTimeBetweenResets} to
\begin{equation}
\overline{T_r(t)}\sim \int_1^t d\tau\, \tau\, Q^\prime(\tau) \sim \int_1^t d\tau\,\tau\,\tau^{-1-\beta} \sim 
\begin{cases}
t^{\max(1-\beta,0)}, \;\; \beta \neq 1 \\
1/ |\log t|, \;\; \beta = 1
\end{cases} \;.
\end{equation}
As a consequence, we expect
\begin{equation} \label{eq:MarkovEnergyEarlyTime}
\overline{E(t)} \sim \frac{1}{\overline{T_r(t)}} \sim
\begin{cases}
1/ t^{\max(1-\beta,0)}, \;\; \beta \neq 1 \\
1/ |\log t|, \;\; \beta = 1
\end{cases} \; , 
\qquad \text{for}\quad t\Delta \ll 1.
\end{equation}
Importantly, the expression of \eq{eq:MarkovEnergyEarlyTime} for the early-time decay of the average energy matches \eq{eq:MarkovEnergyLateTime} for the late-time decay upon inserting $t=\Delta^{-1}$ (up to an $\mathcal{O}(1)$ prefactor). 

In addition, we see from \eq{eq:MarkovEnergyLateTime} that the rate $\lambda$ of \eq{eq:AppRater} effectively renormalizes the decay rate towards the ground state. According to \eq{eq:AppRater}, this rate can be tuned via the initial ground state overlap, and is thus directly connected to the quality of the correction step after a measurement yields $P_i = 1$. We reemphasize the marginal case $\beta=1$, which is attained for $z=2$ in spatial dimension $d=2$, where the decay rate exhibits a logarithmic correction according to \eq{eq:MarkovEnergyLateTime}. We verify the case $\beta = 1$ of \eq{eq:MarkovEnergyLateTime} numerically in \figc{fig:single_particle2D_dynamics_supp}{a}. \\

\subsection{Average ground state infidelity}
\begin{figure}[t]
\centering
\includegraphics[trim={0cm 0cm 0cm 0cm},clip,width=0.6\linewidth]{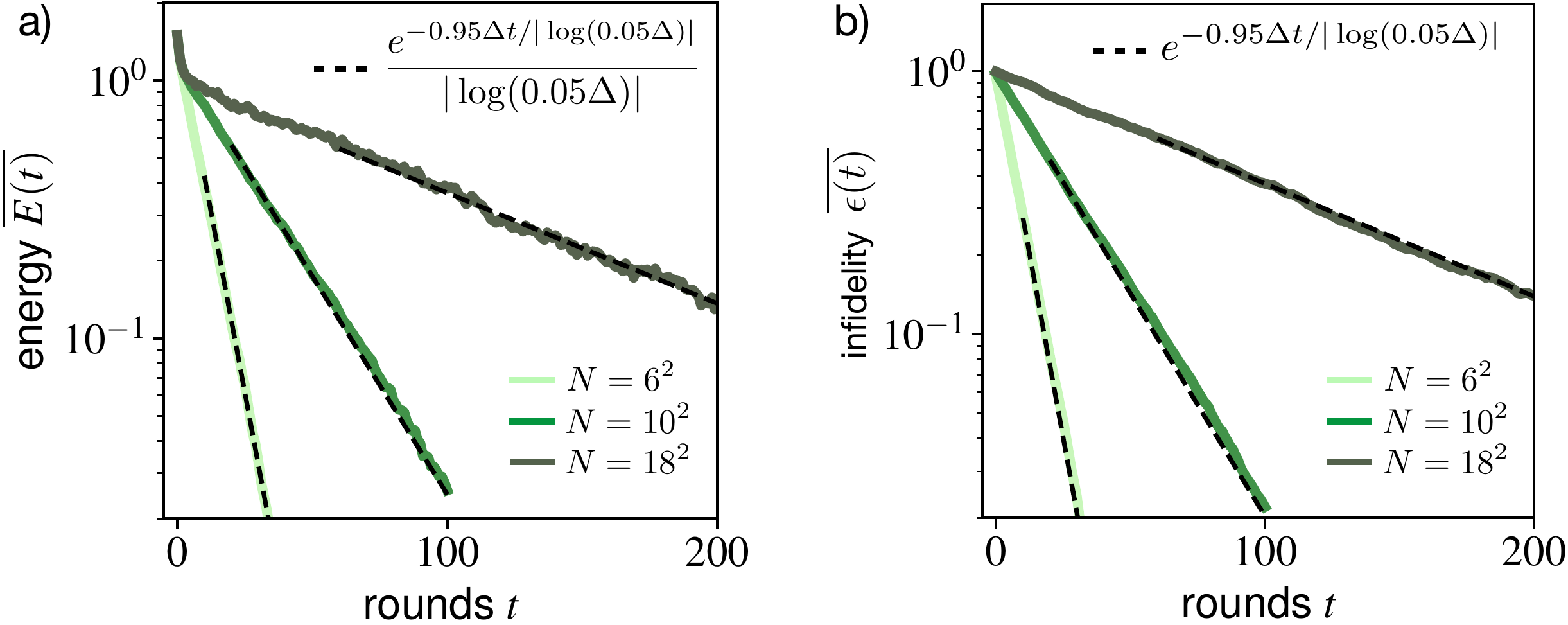}
\caption{\textbf{Dynamics of the $d=2$ single-particle system.} \textbf{a)} Average energy $\overline{E(t)}$ follows the prediction of \eq{eq:MarkovEnergyLateTime} for $\beta=d/z=1$ and $\lambda \approx 0.95$. 
\textbf{b)} Dynamics of the average infidelity $\overline{\epsilon(t)}$ follows \eq{eq:MarkovInfidelityLateTime} for $\beta = 1$.
}
\label{fig:single_particle2D_dynamics_supp}
\end{figure}

Finally, we determine the dynamics of the trajectory-averaged ground state infidelity $\overline{\epsilon(t)}$. For this purpose, we use an approach similar to \eqs{eq:EnergyLastResetTrajectory}{eq:MarkovAverageEnergy} for the energy. In particular, we denote by $\epsilon_{t^\prime}(t)$ the infidelity for a trajectory at time $t\Delta \gtrsim 1$ where the last reset happens at time $t^\prime$. Using \eq{eq:EnergyLastResetTrajectory} and the bound 
\begin{equation}
\epsilon_{t^\prime}(t) \leq \min(E_{t^\prime}(t)/\Delta,1),
\end{equation}
we obtain
\begin{equation}
\epsilon_{t^\prime}(t) \leq \theta(t^\prime + \Delta^{-1} - t)  + \theta(t-t^\prime - \Delta^{-1})\, \frac{e(t-t^\prime)}{\Delta},
\end{equation}
where the additional shift by $\Delta^{-1}$ in the theta-functions makes sure that $\frac{e(t-t^\prime)}{\Delta} < 1$. We thus conclude that
\begin{equation} \label{eq:MarkovAverageInfidelity}
\overline{\epsilon (t)} = \int_0^\infty dt^\prime\, P_{\text{last reset}}(t^\prime)\, \epsilon_{t^\prime}(t) \leq \int_0^{t-\Delta^{-1}} dt^\prime\, P_{\text{last reset}}(t^\prime)\, \frac{e(t-t^\prime)}{\Delta}\, + \int_{t-\Delta^{-1}}^\infty dt^\prime\, P_{\text{last reset}}(t^\prime).
\end{equation}
Inserting \eq{eq:TimeLastReset} for $P_{\text{last reset}}(t^\prime)$ and \eq{eq:AppEnergyDynPro2} for $e(t-t^\prime)$, we find, similar to the analysis of \eq{eq:MarkovAverageEnergy}, that the second term in 
\eq{eq:MarkovAverageInfidelity} dominates with
\begin{equation} \label{eq:MarkovInfidelityLateTime}
\overline{\epsilon (t)} \lesssim 
\begin{cases}
\exp\{-\lambda\Delta^{\max(1,\beta)}\,t\}\,, \;\; \beta \neq 1 \\
\exp\{-\lambda\Delta\,t / |\log \Delta|\}\,, \;\; \beta = 1
\end{cases}
\qquad \text{for}\quad t\Delta \gtrsim 1.
\end{equation}
From \eq{eq:MarkovInfidelityLateTime} we directly obtain the preparation time $t_c$ to reach a target average infidelity $\overline{\epsilon}$:
\begin{equation} \label{eq:AppAveragePreparationTime}
t_c \sim 
\begin{cases}
\Delta^{-\max(1,\beta)}\, \log(1/\overline{\epsilon}), \;\; \beta \neq 1 \\
|\log \Delta|\,\Delta^{-1}\, \log(1/\overline{\epsilon}), \;\; \beta = 1
\end{cases} \;.
\end{equation}
In addition to the numerical results presented in Fig.~2 of the main text, we verify \eq{eq:MarkovInfidelityLateTime} for $\beta=1$ in \figc{fig:single_particle2D_dynamics_supp}{b}.

\section{Details on the Fredkin chain} \label{sec:AppFredkin}
\subsection{Hamiltonian and gap}
The Hamiltonian of the Fredkin spin chain used in the main text is given by
\begin{equation} \label{eq:AppFredkin}
H_{\mathrm{Fred}} = \sum_{j=1}^{N-1} \frac{1}{2}\bigl( \ket{01}-\ket{10} \bigr)_{j,j+1} \bigl( \bra{01}-\bra{10} \bigr) \otimes \bigl( \mathbb{1} - \ket{10}\bra{10}_{j-1,j+2}  \bigr) \equiv \sum_j P_j
\end{equation}
Note that $P_j=P_j^2$ projects onto a spin singlet state at sites $j,j+1$, conditioned on the two neighboring sites $j-1,j+2$ being in a state orthogonal to $\ket{10}_{j-1,j+2}$. 
We consider the Hamiltonian \eq{eq:AppFredkin} on a spin-1/2 chain of even length $N$, where $Z_0 = 1$ and $Z_{N+1}=-1$ are fixed. These two boundary spins are left invariant under the dynamics of \eq{eq:AppFredkin}. We restrict to the sector connected to the Néel state $\ket{0101\cdots 01}$. The ground state $\ket{\Omega}$ is given by the equal-amplitude superposition of all computational basis states within this sector, so-called `Dyck paths', which fulfill the constraint $\sum_{i=0}^x Z_i \geq 0$ for all $x$. 
As in slight variations of this Fredkin Hamiltonian~\cite{Chen2017_fredkin}, we find numerically that the ground state gap decreases algebraically with system size as $\Delta(N) \sim N^{-z}$, where we find $z$ consistent with $z \approx 8/3$, see \figc{fig:fredkin_supp}{a}.

\subsection{Projection round evolution}
\begin{figure}[t]
\centering
\includegraphics[trim={0cm 0cm 0cm 0cm},clip,width=0.99\linewidth]{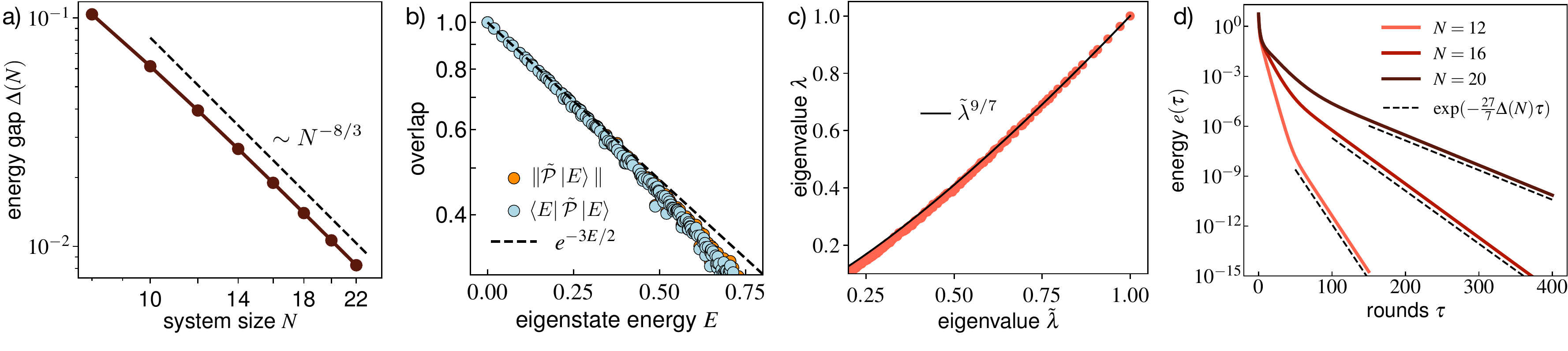}
\caption{\textbf{Fredkin spin chain.} \textbf{a)} Energy gap $\Delta(N)$ of the Fredkin Hamiltonian defined in \eq{eq:AppFredkin}. We find an algebraic decay $\Delta(N) \sim N^{-z}$, with dynamical exponent $z\approx 8/3$. 
\textbf{b)} Matrix elements of the operator $\tilde{\mathcal{P}}$ as defined in \eq{eq:tildeP_fredkin} with respect to the energy eigenstates $\ket{E}$ of $H_{\mathrm{Fred}}$. At low energies $E \ll 1$, the energy eigenstates $\ket{E}$ are also eigenstates of $\tilde{\mathcal{P}}$, as $\|\tilde{\mathcal{P}}\ket{E}\| = \braket{E|\tilde{\mathcal{P}}|E}$ (blue and orange dots coincide). The associated eigenvalues $\tilde{\lambda}_E$ of $\tilde{\mathcal{P}}$ asymptotically are in close agreement with the lowest-order prediction of \eqs{eq:tildeP_fredkin}{eq:tildeP_eigenvals} (black line). \textit{Inset:} Norm and diagonal matrix elements of $\tilde{\mathcal{P}}$ across the entire energy spectrum. 
\textbf{c)} Relation between the eigenvalues $\tilde{\lambda}$ of $\tilde{\mathcal{P}}$ and the eigenvalues $\lambda$ of $\mathcal{P}$. We find that the relation $\lambda = \tilde{\lambda}^{9/7}$ predicted in \eq{eq:lambda_relation_fredkin} holds at $1-\tilde{\lambda} \ll 1$.
\textbf{d)} Energy $e(\tau)$ defined in \eq{eq:projector_evolution_energy_fredkin} of a Néel product initial state $\ket{\psi}$ evolved under repeated application of the projector string $\mathcal{P}$. The asymptotic decay rate of $e(\tau)$ is proportional to the gap $\Delta(N)$, with a prefactor in excellent agreement with the estimate of \eq{eq:projector_evolution_energy_fredkin}. 
}
\label{fig:fredkin_supp}
\end{figure}

We consider the evolution of an initial state $\ket{\psi}$ under the local ground state projections $(1-P_j)$. The corresponding projection string operator $\mathcal{P}$ is defined as 
\begin{equation}
\mathcal{P} = \prod_{a=1}^{\mathcal{A}=3} \mathcal{P}_a = \prod_{a=1}^3 \prod_{j\in B_a} (1-P_{j}) = \prod_{a=1}^3 \;\; \prod_{j: \;a\leq 3j+a \leq N-2} (1-P_{3j+a}),
\end{equation}
where $\mathcal{P}_a^2 = \mathcal{P}_a$. We note that although the local $P_j$ are four-site operators, we only require three layers, since $[P_j,P_{j+3}]=0$.
As for evolution with only two layers, we seek to relate the convergence of $\mathcal{P}^\tau\ket{\psi}$ to the ground state (or, if $\braket{\Omega|\psi}=0$, its decay) to the energy eigenstates $\ket{E}$ of the Hamiltonian $H_{\mathrm{Fred}}$. For this purpose, similar to the two-layer case, we define a symmetrized projection string operator $\tilde{\mathcal{P}}$ via 
\begin{equation} \label{eq:tildeP_fredkin}
\tilde{\mathcal{P}} \equiv \frac{1}{6} \sum_{\pi \in \mathrm{Perm}(3)} \mathcal{P}_{\pi(1)}\, \mathcal{P}_{\pi(2)} \, \mathcal{P}_{\pi(3)} = \mathbb{1} - \frac{3}{2} H_{\mathrm{Fred}} + \frac{1}{2} (H_{\mathrm{Fred}})^2 + \mathcal{O}(P_jP_{j+3}),
\end{equation}
by summing over all permutations of the ordering of the three projection layers $\mathcal{P}_{a=1,2,3}$. The second equality in \eq{eq:tildeP_fredkin} follows in a manner analogous to the two-layer case.
Assuming that at low energies $E\ll 1$, corrections from the term $\mathcal{O}(P_jP_{j+3})$ in \eq{eq:tildeP_fredkin} are small, we expect that  
\begin{equation} \label{eq:tildeP_eigenvals}
\braket{E|\tilde{\mathcal{P}}|E} \approx \| \tilde{\mathcal{P}}\ket{E} \| \approx \exp\bigl\{ -\frac{3}{2}E \bigr\},
\end{equation}
which expresses that $\ket{E}$ asymptotically becomes an eigenstate of $\tilde{\mathcal{P}}$ at low energies. We confirm \eq{eq:tildeP_eigenvals} numerically for an $N=16$ system in \figc{fig:fredkin_supp}{b}. We emphasize that although one generally expects corrections to the factor $\frac{3}{2}$ on the right-hand side of \eq{eq:tildeP_eigenvals}, they appear to remain remarkably small in this system. 
Moreover, even at elevated energies $E=\mathcal{O}(1)$ the eigenstates $\ket{E}$ of $H_{\mathrm{Fred}}$ remain approximate eigenstates of $\tilde{\mathcal{P}}$, since $\braket{E|\tilde{\mathcal{P}}|E} \approx \| \tilde{\mathcal{P}}\ket{E} \|$ remains valid.
As a consequence of \eq{eq:tildeP_eigenvals}, the operator $\tilde{\mathcal{P}}$ is directly related to the Hamiltonian and its eigenstates at low energies, and effectively implements an imaginary time step as desired.

We now consider the repeated application of $\tilde{\mathcal{P}}$ and attempt to relate it to the repeated application of $\mathcal{P}$, similar to the two-layer case discussed in Sec.~\ref{sec:AppProjectorTilde}.
However, the relation of \eq{eq:operatorID} between $\mathcal{P}^\tau$ and $\tilde{\mathcal{P}}^\tau$ for the two-layer case does not apply in this three-layer example. Indeed, while we do not derive a corresponding relation for this three-layer case at the same level of rigor, in the following we provide a heuristic argument that we find to be in excellent agreement with numerical results. 
Specifically, we note that $\tilde{\mathcal{P}}^\tau$ may be expanded as a sum over operator strings using \eq{eq:tildeP_fredkin}. Within such an operator string, single round projectors $\mathcal{P}_{\pi(1)}\mathcal{P}_{\pi(2)}\mathcal{P}_{\pi(3)}$ with different layer permutations may interface, leading to substrings $\mathcal{P}_{\pi(1)}\mathcal{P}_{\pi(2)}\mathcal{P}_{\pi(3)}\, \mathcal{P}_{\pi^\prime(1)}\mathcal{P}_{\pi^\prime(2)}\mathcal{P}_{\pi^\prime(3)}$. Inspecting this substring, we note two points:
\begin{enumerate}
\item With probability $1/3$, we have $\pi(3) = \pi^\prime(1)$ and thus $\mathcal{P}_{\pi(3)}\mathcal{P}_{\pi^\prime(1)}=\mathcal{P}_{\pi(3)}$, which reduces the total length of the string by one layer, which corresponds to $1/3$ of a whole round.
\item Repeated application of only two layers $(\mathcal{P}_{a_1}\mathcal{P}_{a_2})^n$ cannot prepare the global ground state, since the system is effectively cut into separate three-site clusters. States at low global energy are close to the ground states of these local clusters already, and we thus expect $(\mathcal{P}_{a_1}\mathcal{P}_{a_2})^2 \approx \mathcal{P}_{a_1}\mathcal{P}_{a_2}$. This reduces the total length of the string by two layers and thus $2/3$ of a round. We encounter this situation with probability $1/6$, if $\pi(2) = \pi^\prime(1)$, $\pi(3) = \pi^\prime(2)$.
\end{enumerate}
Combining these two points, we expect the final contracted operator string to have a reduced number of effective rounds $\tau_{\mathrm{eff}} \approx \tau \bigl(1 - \frac{1}{3}\cdot \frac{1}{3}-\frac{1}{6}\cdot\frac{2}{3}\bigr) = \frac{7}{9}\tau$. We emphasize, however, that the remaining operator string is \textit{not} of the form $\mathcal{P}^{\tau_{\mathrm{eff}}}$ as in \eq{eq:operatorID} for the two-layer case. 
Nonetheless, we conjecture that there is still a relation between the eigenvalues $\tilde{\lambda}$ of $\tilde{\mathcal{P}}$ and $\lambda$ of $\mathcal{P}$ given by
\begin{equation} \label{eq:lambda_relation_fredkin}
\tilde{\lambda} = \lambda^{7/9} \quad \Rightarrow \quad \lambda = \tilde{\lambda}^{9/7} \approx \exp\Bigl\{-\frac{9}{7}\cdot\frac{3}{2} E\Bigr\},
\end{equation}
where we used \eq{eq:tildeP_eigenvals} at small energies $E \ll 1$ in the last step. We numerically confirm the validity of this relation at low energies for a system of size $N=16$ in \figc{fig:fredkin_supp}{c}.

Let us now consider the dynamics of an initial state $\ket{\psi}$ under repeated application of $\mathcal{P}$. We take $\ket{\psi}$ to be an initial Néel product state for concreteness. We then expect the energy $e(\tau)$ of the normalized state $\mathcal{P}^\tau \ket{\psi} / \|\mathcal{P}^\tau \ket{\psi}\|$ with respect to $H_{\mathrm{Fred}}$ to be asymptotically given by
\begin{equation} \label{eq:projector_evolution_energy_fredkin}
e(\tau) = \frac{\braket{\psi|(\mathcal{P}^\dagger)^\tau H_{\mathrm{Fred}}\mathcal{P}^\tau|\psi}}{\|\mathcal{P}^\tau\ket{\psi}\|^2} \xrightarrow{\tau \gg \Delta^{-1}} \exp\Bigl\{ - \frac{27}{7}\Delta \tau \Bigr\},
\end{equation}
where $\Delta$ is the gap and thus the energy of the first excited state. We confirm \eq{eq:projector_evolution_energy_fredkin} for different system sizes \figc{fig:fredkin_supp}{d}. We emphasize again that the prefactor $27/7$ that appears in the decay rate of \eq{eq:projector_evolution_energy_fredkin} is an estimate that generally exhibits corrections. However, as \figc{fig:fredkin_supp}{d} demonstrates, \eq{eq:projector_evolution_energy_fredkin} is remarkably accurate and corrections thus small.

\subsection{Additional numerical data}
\begin{figure}[t]
\centering
\includegraphics[trim={0cm 0cm 0cm 0cm},clip,width=0.65\linewidth]{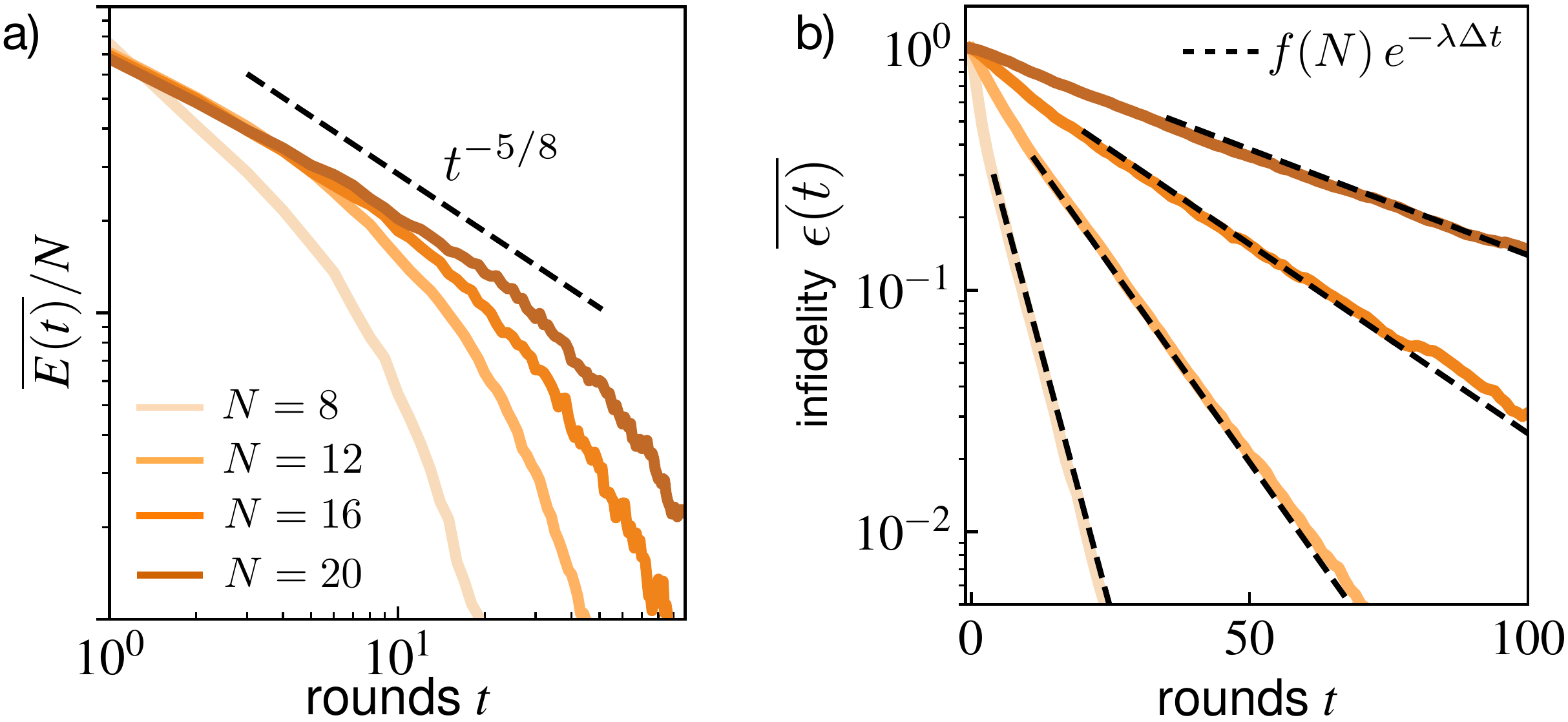}
\caption{\textbf{Fredkin chain: additional data.} \textbf{a)} Early-time energy dynamics $\overline{E(t)}$ is roughly consistent with a power-law decay $\sim t^{-5/8}$. \textbf{b)} The average infidelity decays with rate linear in the gap $\Delta$. We find $\lambda \approx 1.9$ and $f(N)$ consistent with $f(N) \sim \sqrt{N}$.
}
\label{fig:fredkin_supp2}
\end{figure}

We present additional data supporting the conclusions drawn from Fig.~4(a) of the main text. In particular, \figc{fig:fredkin_supp2}{a} shows the dynamics of the trajectory-averaged energy $\overline{E(t)}$ at early times. From our analysis and the critical exponent $z\approx 8/3$, we expect a decay $\sim t^{-5/8}$. Our numerical data are consistent with this scaling. Moreover, \figc{fig:fredkin_supp2}{b} shows the dynamics of the average ground state infidelity $\overline{\epsilon(t)}$, whose decay rate follows that expected linear-in-gap scaling $\overline{\epsilon(t)} \sim f(N) \, e^{-\lambda \Delta t}$. We find $f(N)$ to be consistent with $f(N)\sim \sqrt{N}$.

\section{Cooling dynamics without $U(1)$ symmetry: 1D critical cluster-Ising model}

\begin{figure}[b]
    \centering
    \includegraphics[width=0.8\linewidth]{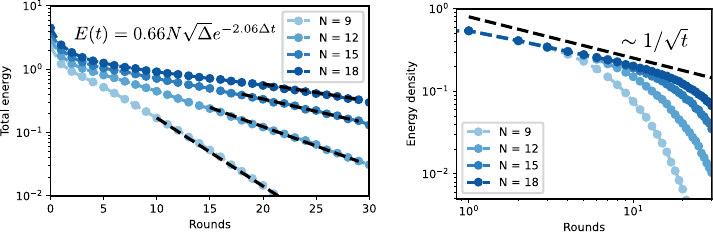}
    \caption{\textbf{Simulation of the 1D critical cluster-Ising model}. The dynamics of the average total energy and the energy density are obtained over $\geq 4000$ trajectories and are shown on log and log-log scales, respectively. The energy gaps $\Delta$ for different system sizes are obtained via exact diagonalization. The dynamics of the energy with increasing system size are consistent with the proposed universal form with $\beta = 1/2$, at late and early times. }
    \label{fig:sm_clusterIsing}
\end{figure}

The examples we have examined so far focus on RK-type Hamiltonians with $U(1)$ symmetry. Whether the universal behavior proposed in the main text extends beyond this subclass remains an intriguing open question. A systematic study is left for future work; here, we simply show that the same universal behavior emerges in the cluster-Ising model in 1D when lacking an explicit $U(1)$ symmetry.

The 1D cluster-Ising model is a frustration-free model. On a ring with length $N$, the Hamiltonian is a sum of local projectors $H_{CI}(g) = \sum_i P_i(g)$, where $i$ labels the site on the ring and each projector $P_i(g)$ is defined as~\cite{wolf:2006,setqpt_tns}
\begin{equation}
    P_i(g) = \frac{1}{2} + \frac{(1-g)^2}{4(1+g^2)} Z_{i-1}X_i Z_{i+1} - \frac{1-g^2}{4(1+g^2)}(Z_{i-1}Z_i + Z_iZ_{i+1}) - \frac{(1+g)^2}{4(1+g^2)} X_i,
\end{equation}
where the parameter $g\in [-1, 1]$. The model exhibits a quantum phase transition at $g = 0$ between symmetry-protected topological (SPT) phases protected by $\mathbb{Z}_2\times\mathbb{Z}_2^T$ symmetry generated by the global spin flip $\prod_i X_i$ and the complex conjugation. At $g = -1$, the ground state is the standard cluster state with a non-trivial $\mathbb{Z}_2\times\mathbb{Z}_2^T$ SPT order. At $g = 1$, the ground state is the paramagnetic state with a trivial SPT order. This particular path has the nice feature that the ground states for all $g$ can be exactly represented by an MPS with bond dimension 2.

Here we are interested in the critical point $g = 0$. At this point, a ground state of the system is the GHZ state $(\ket{00\cdots 0} + \ket{11\cdots 1})/\sqrt{2}$. The ground-state manifold for any finite system is two-fold degenerate, spanned by $\ket{00\cdots 0}$ and $\ket{11\cdots 1}$. The Hamiltonian $H_{CI}(0)$ is gapless with a dynamical critical exponent $z = 2$. This provides an interesting example of an “uncle’’ Hamiltonian for the GHZ state: in stark contrast to the usual gapped ferromagnetic parent Hamiltonian, it is gapless. Note that the model has an accidental $U(1)$ symmetry that only appears at $g = 0$; the symmetry is generated by $h = \sum_i Z_iZ_{i+1}$ that satisfies $[H_{CI}(0), h] = 0$. In our preparation protocol, we will choose the recovery operation that explicitly violates this $U(1)$ symmetry, ensuring this symmetry is removed.

To implement the protocol, we choose the initial state to be the product state $\ket{++\cdots +}$ where $\ket{+} = (\ket{0} + \ket{1})/\sqrt{2}$. The projectors $\{P_i\}$ are grouped into three layers of non-overlapping projectors. At each round, the projectors in each of the three layers are measured in parallel. If any of the projectors are measured to be $P_i = +1$, a single Pauli $X_i$ operator is implemented at site $i$. Unlike the previous examples, the local $X_i$ does not reset the projector $P_i$ back to 0, but it is chosen such that the total energy is lowered and the dynamics respect the $\mathbb{Z}_2$ symmetry. Under this setup, the unique fixed point state of the protocol is the GHZ state. We simulate the protocol using exact diagonalization for system sizes $N = 9, 12, 15$ and $18$. The results are summarized in Fig.~\ref{fig:sm_clusterIsing}. The dynamics of the average energy demonstrate good quantitative agreement with the universal form for $\beta = d/z = 1/2$.

\section{Supplementary data for the decay of infidelity}
\begin{figure}[t]
\centering
\includegraphics[trim={0cm 0cm 0cm 0cm},clip,width=0.65\linewidth]{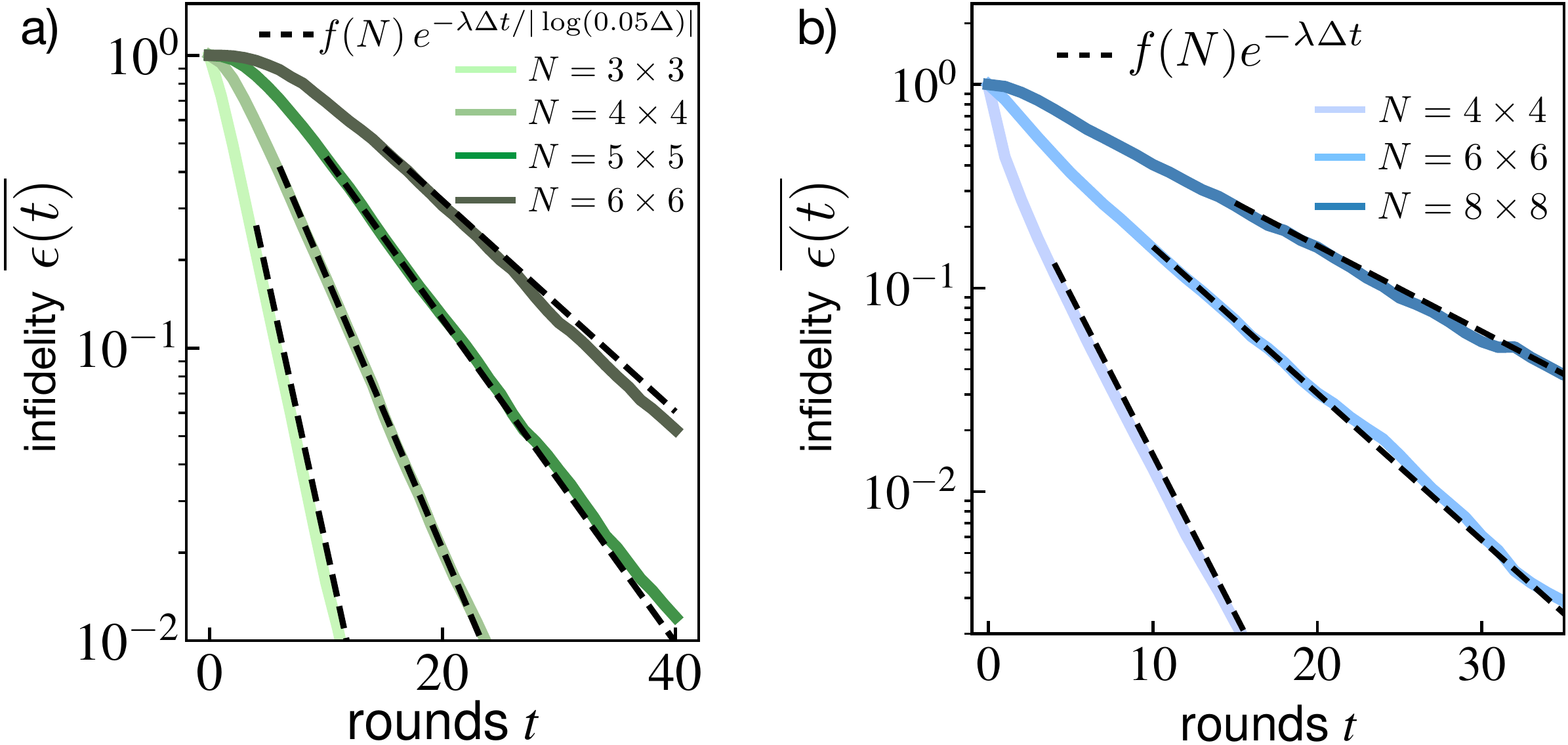}
\caption{\textbf{Supplemental data.} \textbf{a)} Dynamics of average infidelity $\overline{\epsilon(t)}$ in the 2D Heisenberg model, showing a decay consistent with a linear-in-gap rate up to a log-correction, as expected for $\beta=1/2$. We find the system-size dependence of the prefactor $f(N)$ to be consistent with $f(N) \lesssim N^{1/8}$. \textbf{b)} Dynamics of average infidelity for the 2D quantum dimer model. The decay is consistent with our scaling predictions for $\beta = 1/2$, showing a linear-in-gap behavior. We find $f(N)$ consistent with $f(N)\sim \sqrt{N}$.
}
\label{fig:supp_data}
\end{figure}

We show additional data on the dynamics of ground-state infidelity both for the 2D Heisenberg model and the 2D quantum dimer model for resonating valence bond states in \fig{fig:supp_data}. 

\leavevmode
\newline
\leavevmode
\newline
\leavevmode
\newline


\end{document}